\documentclass[twocolumn,aps,prx,groupedaddress]{revtex4}
\usepackage{graphicx,wrapfig,amsmath,amssymb,enumitem,upgreek,bbm,mathtools,amsfonts,physics,multirow,color,caption,subcaption,hyperref}
\usepackage[normalem]{ulem}

\renewcommand{\vec}[1]{\oldbm{#1}}

\renewcommand{\vec}[1]{\boldsymbol{#1}}

\def\bb{{\vec b}}
\def\bk{{\vec K}}
\def\bk{{\vec k}}

\def\bg{{\vec g}}

\def\bA{{\vec A}}
\def\bB{{\vec B}}
\def\bz{{\vec z}}

\def\ba{{\vec a}}
\def\bz{{\vec z}}
\def\bK{{\vec K}}
\def\bq{{\vec q}}

\def\bG{{\vec G}}

\def\bB{{\bf B}}

\def\bp{{\bf p}}

\def\bn{{\vec n}}

\def\br{{\vec r}}
\def\bu{{\vec u}}
\def\ad{\mathrm{ad}}

\def\bgamma{{\boldsymbol \gamma}}

\def\bsigma{{\boldsymbol \sigma}}
\def\bnabla{{\boldsymbol \nabla}}

\newcommand{\ov}[1]{\overline{#1}}

\def\sgn{\mathop{\mathrm{sgn}}}
\def\tr{\mathop{\mathrm{tr}}}

\def\T{\mathcal{T}}
\def\L{\mathcal{L}}

\def\P{\mathcal{P}}

\def\H{\mathcal{H}}
\def\K{\mathcal{K}}

\def\diag{{\rm diag}}
\def\ad{{\rm ad}}

\def\U{{\rm U}}

\newcommand{\beq}{\begin{equation}}
\newcommand{\eeq}{\end{equation}}
\newcommand{\beqarray}{\begin{eqnarray}}
\newcommand{\eeqarray}{\end{eqnarray}}

\begin{document}

\title{
TB or not TB? \\ Contrasting properties of twisted bilayer graphene and the alternating twist $n$-layer structures ($n=3, 4, 5, \dots$) 
}

%\author[1]{Eslam Khalaf}
%\author[1]{Ashvin Vishwanath}

%\affil[1]{\normalsize\it Department of Physics, Harvard University, Cambridge MA 02138, USA}

\author{Patrick J. Ledwith$^1$}
\author{Eslam Khalaf$^1$}
\author{Ziyan Zhu$^1$}
\author{Stephen Carr$^2$}
\author{Efthimios Kaxiras$^1$}
\author{Ashvin Vishwanath$^1$}
\affiliation{$^1$Department of Physics, Harvard University, Cambridge, MA 02138}
\affiliation{$^2$Brown Theoretical Physics Center and Department of Physics, Brown University, Providence, Rhode Island 02912, USA}

\begin{abstract}
The emergence of alternating twist multilayer graphene (ATMG) as a  generalization of twisted bilayer graphene (TBG) raises the question  - in what important ways do these systems differ? Here, we utilize a combination of techniques including ab-initio relaxation and single-particle theory, analytical strong coupling analysis, and Hartree-Fock to contrast ATMG with $n=3,4,5,\ldots$ layers and TBG.
\textbf{I}: We show how external fields enter in the decomposition of ATMG into twisted bilayer graphene and graphene subsystems. The parallel magnetic field is expected to have a much smaller effect when $n$ is odd due to mirror symmetry, but surprisingly also for any $n > 2$ if we are are the largest magic angle. \textbf{II}: We compute the lattice relaxation of the multilayers leading to the effective parameters for each TBG subsystem as well as small mixing between the subsystems. We find that the second magic angle for $n=5$, $\theta \approx 1.14$, provides the closest realization of the ``chiral" model and is protected from mixing by mirror symmetry. It may be an optimal host for fractional Chern insulators. \textbf{III}: When there are no external fields, we integrate out the non-magic subsystems and reduce ATMG to the magic angle TBG subsystem with a screened interaction. 
\textbf{IV}: We perform an analytic strong coupling analysis of the effect of external fields and corroborate our results with numerical Hartree Fock simulations.
For TBG itself, we find that an in-plane magnetic field can drive a phase transition to a valley Hall state or a gapless ``magnetic semimetal''
while having a weaker effect on $n\geq3$ ATMG at the first magic angle. In contrast, displacement field ($V$) has very little effect on TBG, but induces a gapped phase in ATMG for small $V$ for $n = 4$ and above a finite critical $V$ for $n = 3$.
For $n\geq 3$, we extract the superexchange coupling - believed to set the scale of superconductivity in the skyrmion mechanism - and show that it increases with $V$ at angles near and below the magic angle.
\textbf{V}: We complement our strong coupling approach with a phenomenological weak coupling theory of ATMG pair-breaking. While for $n=2$ orbital effects of the in-plane magnetic field can give a critical field of the same order as the Pauli field, for $n>2$ we expect the critical field to exceed the Pauli limit. 
\end{abstract}

\maketitle

\tableofcontents

\section{Introduction}

The discovery of superconductivity and correlated insulators in  twisted bilayer graphene (TBG) \cite{PabloSC,PabloMott} close to the theoretically predicted magic angle $\theta_{\rm TBG} \approx 1.1^o$ \cite{Bistritzer2011, dosSantos2012, Mele2010, dosSantos2007} has stimulated a huge amount of experimental \cite{Dean-Young, Efetov, YoungScreening, EfetovScreening,  CascadeShahal, CascadeYazdani, RozenEntropic2021, CaltechSTM, RutgersSTM, ColombiaSTM, PrincetonSTM} and theoretical work \cite{BalentsReview,AndreiMacdonaldReview,LecNotes,PoPRX, StiefelWhitney, Song, KangVafekPRL, KangVafekPRX, VafekRG,XieMacdonald, ShangNematic, Tarnopolsky,KIVCpaper,Khalaf2020, KhalafSoftMode,Parker,KwanIKS,wagner2021globalHF,TBGIII, TBGIV, TBGV, TBGVI,Potasz,Parker,VafekDMRG,QMC,PanQMC2021} on the system. In addition to the correlated insulators and superconductivity observed in the early works, more recent discoveries include the quantum anomalous Hall states with \cite{Zhang19,Bultinck2019,SharpeQAH, YoungQAH} and without \cite{EfetovQAH} an aligned hBN substrate, charge density waves \cite{PierceCDW}, as well as the prediction \cite{LedwithFCI,AbouelkomsanFCI,RepellinFCI,LauchliFCI} and subsequent observation \cite{XieFCI} of fractional Chern insulators (FCIs). Besides TBG, other graphene multilayers have been explored in search for similarly interesting phases. These include mono-bi graphene \cite{YankowitzMonoBi1, YankowitzMonoBi2, YoungMonoBi,Morell2013ElectronicPO}, twisted double bilayer graphene \cite{TDBG_Pablo,TDBGexp2019,IOP_TDBG,YankowitzTDBG,KimTDBG,he2021chiralitydependent,Lee2019,Zhang2018}, and ABC graphene aligned with hBN \cite{ABCMoire}. While all these systems have displayed interaction-induced phases such as correlated insulators, TBG remained the only Moir\'e system where robust superconductivity has been observed and reproduced by several groups.
Note that these other systems, like hBN-aligned twisted bilayer graphene, do not possess $C_2 \T$ symmetry which relates the two degenerate Chern bands in a single valley of TBG. This suggests that $C_2 \T$ symmetry  plays a significant role in stabilizing superconductivity,  consistent with the predictions of a topological mechanism for superconductivity based on skyrmions  \cite{Khalaf2020,chatterjee2020skyrmion,ChristosWZW} (alternative explanations for the importance of $C_2 \T$ symmetry were proposed in Refs.~\cite{phonondichotomy,MacdonaldFluctuations}). 

This raises the question of designing other graphene Moir\'e structures that retain $C_2 \T$ symmetry. Alternating twist multilayer graphene (ATMG) heterostructures proposed in Ref.~\cite{Khalaf2019}, where the twist angle alternates between $\pm \theta$ from interface to interface, fit the bill, since they have essentially the same symmetry as magic angle TBG. This makes them promising platforms to realize all the phases seen in TBG including superconductivity, if the symmetry ingredients have been correctly identified. Twisted trilayer graphene (TTG) is therefore a natural next step after twisted bilayer both experimentally \cite{Hao2021TTG,Park2021TTG,seaofmagic,caltechTrilayerSTM,TrilayerScreening} and theoretically \cite{ChristosTTG,LakeSenthil,MacdonaldInPlane,MacdonaldTri,TSTGI,TSTGII,guerci2021higherorder,Chou2021,assi2021floquet,li2021induced}. It is a structure that is symmetric under mirror symmetry $M_z$ and it decomposes into twisted bilayer graphene and ordinary graphene where the subsystems are in opposite mirror sectors \cite{Carr2020}. The TBG subsystem has tunneling matrix elements that are scaled by $\sqrt{2}$ which leads to a magic angle $\sqrt{2} \theta_{\rm TBG}\approx 1.56^o$. TTG was expected to host similar correlated physics to TBG due to this symmetry protected TBG subsystem, and the higher magic angle implies elevated energy scales and likely more structural uniformity. In addition, TTG has tunability advantages over TBG because external fields couple in different ways; due to the mirror symmetry $M_z$ out-of-plane electric and in-plane magnetic fields act purely as tunneling terms between the TBG and graphene subsystems. These predictions and advantages were borne out in recent experiments where TTG tuned close to the predicted new magic angle of $1.56^\circ$ was found to exhibit strong coupling superconductivity that is even stronger than TBG superconductivity and tunable via displacement field \cite{Hao2021TTG,Park2021TTG}. Furthermore, the critical in-plane magnetic field was found to exceed the Pauli limiting field \cite{PabloTrilayerInPlane}. 

Remarkably, the decomposition of TTG into TBG and graphene subsystems is actually a general property of ATMG for an arbitrary number of layers $n$ \cite{Khalaf2019}. For example, twisted quadrilayer graphene (TQG) has two TBG subsystems with magic angles $\varphi \theta_{\rm TBG}\approx 1.78^\circ $ and $\varphi^{-1} \theta_{\rm TBG} \approx 0.68^\circ$ where $\varphi =\frac{1+\sqrt{5}}{2}$ is the golden ratio  \cite{LecNotes,Khalaf2019}. Twisted pentalayer graphene (TPG) has two twisted bilayer graphene subsystems with magic angles $\sqrt{3} \theta_{\rm TBG}$ and $\theta_{\rm TBG}$.

In this paper we perform a comprehensive comparison between $n$-layer ATMG and $n=2$ TBG. The low energy physics of ATMG is largely controlled by its \emph{magic sector}, the TBG subsystem that has flat bands.  Nonetheless, there is much to be learned by studying ATMG and contrasting it with TBG. Indeed, we will quantitatively catalogue many differences between the systems. One important difference is the way external fields couple to ATMG versus TBG; we describe this along with the non-interacting band structure and symmetries in section \ref{sec:BandStructure}. While for twisted bilayer graphene, displacement fields (electric field applied perpendicular to the layers) and  magnetic fields, applied parallel to the layers, couple directly to the magic sector, for {\em odd} $n$ the external fields exclusively act as tunnelings between subsystems due to mirror symmetry $M_z$. In particular, the in-plane magnetic field does not, on its own, have a pair-breaking effect, as noticed in $n=3$ TTG \cite{MacdonaldInPlane,LakeSenthil}, because it is symmetric under the time reversal symmetry combined with in plane reflection $M_z \T$. Indeed, experimentally the in-plane critical field of TTG is found to vastly exceeds that of TBG and the Pauli limit \cite{PabloTrilayerInPlane}. Amazingly, we find that the same is very nearly true at the first magic angle for all $n>2$ {\em despite} the absence of a microscopic mirror symmetry for $n$ even which enforces this term to vanish; instead, the intra-magic-subsystem coupling of the external fields at the largest magic angle is suppressed by a factor $\pi^2/(2n^3)$ for large $n$. 

In section \ref{sec:relax} we present a detailed computation for the lattice relaxation effects for $n=2,\,3,\,4,\,5$ ATMG. It is known that in twisted bilayer graphene, the ratio between the same-sublattice and opposite-sublattice hopping $\kappa = w_0/w_1$ has a crucial role in the interacting physics. In particular small $\kappa$ is found to give rise to favorable quantum band geometry for realizing fractional Chern insulators.   \cite{LedwithFCI,RepellinFCI,XieFCI}. While nominally this parameter is $1$ for a fully-rigid lattice, both in-plane and out-of-plane lattice relaxation result in a lower effective $\kappa$. For $n \geq 4$ ATMG, we find that both $w_0$ and $w_1$ depend on the layers connected by tunneling. While the decoupling mapping may be extended to include non-constant $w_1$, leading to slightly different magic angles, the different ratios $w_0/w_1$ lead to small tunnelings between subsystems. In addition, we compute the effective $\kappa$ for the various TBG subsystems.  We generally find smaller values of $\kappa$ compared to what has been reported in earlier works \cite{Carr2018relax, Koshino2018}; for example, we find $\kappa \approx 0.51$ for $n = 2$ TBG. 
While these $\kappa$ values are sensitive to details of the relaxation model we employ, the angle and layer dependence of $\kappa$ that we find should be universal, making our predictions regarding the relative strength of relaxation in different multilayer systems compared to TBG robust. We find that more layers tend to increase the lattice relaxation and reduce $\kappa$ compared to its value in TBG at the same angle. However,  the increase in the value of the first magic angle with the number of layers ends up slightly outweighing this effect leading to a slight overall increase in the value $\kappa$ at the first magic angle with $n$. This does not apply to higher magic angles where we can exploit the increased relaxation without necessarily going to larger angle. A notable case that we highlight is the {\em second} magic angle for $n=5$ which has the same magic angle $\theta_{\rm TBG}$ as the {\em first} magic angle of $n=2$ TBG, but experiences enhanced relaxation due to the increased number of layers. This leads to a low effective $\kappa \approx 0.45$. We further note that the TBG sub-block where this magic angle is realized is protected by mirror symmetry from mixing with other sub-blocks. We therefore highlight the second magic angle of $n=5$ as a promising platform for hosting FCIs and other correlated phases.

We then turn to the low energy interacting physics of ATMG in section \ref{sec:screen}. When one of the TBG sectors is at its magic angle, the low energy bandstructure of ATMG consists of the flat magic angle TBG (MATBG) bands together with Dirac cones; the latter have much smaller density of states and the associated electrons move much faster than the MATBG electrons. Based on this perspective, we integrate out the non-magic ATMG electrons and argue that to a good approximation they only statically screen the Coulomb interaction. 

We then perturbatively include the effect of external fields in ATMG at integer filling of the magic sector in section \ref{sec:externalfields}. We build off of the analytic strong coupling approach to integer fillings of MATBG \cite{LecNotes,KIVCpaper,Khalaf2020,VafekRG,KhalafSoftMode,TBGIII,TBGIV}, which has been numerically supported first by Hartree-Fock \cite{XieMacdonald, ShangNematic, KIVCpaper, GuineaHF} and then by less biased methods like DMRG \cite{Tomo, Parker, VafekDMRG}, exact diagonalization \cite{TBGVI, Potasz}, and QMC \cite{QMC,PanQMC2021}. 

While much of this paper is devoted to $n>2$ due to the extensive theoretical literature for $n=2$ TBG, we first focus on TBG because the strong orbital effect of the in-plane magnetic field on TBG insulators has not been previously understood. In particular, we 
% We show that while a displacement field has little effect due to the tiny layer polarization, 
find that an in-plane magnetic field can drive a phase transition from the Kramers' intervalley coherent state to either a valley Hall state or a ``magnetic semimetal.'' The latter state is similar to the  nematic semimetal \cite{ShangNematic} except it spontaneously breaks $C_2$ and $\T$ while preserving $C_2 \T$. Either of these transitions would impede the skyrmion mechanism of superconductivity \cite{Khalaf2020}.

As discussed in section \ref{sec:BandStructure}, the external fields have little impact in the magic subsystem for $n>2$ and instead mostly act as tunneling terms between the different subsystems. We find that this tunneling indirectly affects the dispersion of the magic sector, depending on the angle. At small displacement fields, the dispersion increases (decreases) depending on whether the twist angle is smaller (larger) than the magic angle. Any theory of superconductivity that relies on dispersion - such as skyrmion superconductivity \cite{Khalaf2020} - will therefore see $T_c$ enhancement with displacement field for small angles and a reduction in $T_c$ for large angles for small displacement fields. 
This behavior is consistent with the observations in Refs. \cite{Hao2021TTG,Park2021TTG}, where the angles are slightly smaller than the magic angle and the superconductivity is enhanced by the displacement field. We also find that the displacement field tunneling can transfer symmetry breaking orders from the magic sector to non-magic subsystems. For $n=3$, this implies that the valley Hall state becomes fully gapped in the presence of a displacement field whereas the Kramers' intervalley coherent (KIVC) zero-field ground state \cite{KIVCpaper} remains gapless since the intervalley mass cannot gap out the Dirac cone. This favors the valley Hall (VH) state over the KIVC state which all together implies a transition from a gapless state at small displacement field to a fully gapped state at large field as previously noticed in Refs.~\onlinecite{ChristosTTG,TSTGII} and may correspond to resistance peaks observed in experiments \cite{Park2021TTG,Hao2021TTG}. Note that at similar fields superconductivity disappears, which is compatible with the skyrmion mechanism of superconductivity where pairing is disfavored by the easy-axis anisotropy that picks out the VH  state \cite{Khalaf2020,chatterjee2020skyrmion}. At the largest fields the system becomes fully symmetric and metallic. In contrast, for $n = 4$, the non-magic TBG subsystem can gap out from IVC order and thus the system can become gapped at charge neutrality as soon as $V \neq 0$.

Finally, to complement our strong coupling theory, we develop a phenomenological weak coupling theory of superconductivity based on a general mean-field pairing whose origin we leave unspecified. We find quite generally that for $n=2$ MATBG the orbital coupling of the in-plane magnetic field is pair breaking with a critical field that is of the same order as the Pauli limiting field. However for $n>2$ ATMG the critical field is expected to be much larger because the intra-subsystem magnetic field either doesn't exist ($n$-odd) or couples very weakly to the magic sector ($n > 2$ at the first magic angle). 

\section{Non-Interacting Band Structure and Symmetries}
\label{sec:BandStructure}

We begin with the Bistritzer-Macdonald (BM) model \cite{Bistritzer2011} generalized to alternating twist multilayer graphene (ATMG) with $n$ layers \cite{Khalaf2019}. The Hamiltonian for a single spin and a single graphene valley is given by 
\beq
H(\br) = \left(\begin{array}{cccc}
        -i v \bsigma_{+} \cdot \bnabla  & T(\br) & 0 & \cdots \\
        T^\dagger(\br) & -i v \bsigma_{-} \cdot \bnabla & T^\dagger(\br) &  \cdots \\
        0 & T(\br) & -i v \bsigma_{+} \cdot \nabla  & \cdots \\
\cdots & \cdots & \cdots & \cdots 
\end{array} \right).
\label{HATMG}
\eeq
Here $\bsigma_{\pm} = e^{\mp\frac{i}{4} \theta \sigma_z} (\sigma_x, \sigma_y ) e^{\pm\frac{i}{4} \theta \sigma_z}$ and 
\begin{equation}
    \begin{aligned}
        T(\br) & = \begin{pmatrix} w_0 U_0(\br) & w_1 U_1(\br) \\ w_1 U^*_1(-\br) & w_0 U_0(\br) \end{pmatrix}, \\
        U_0(\br) & = e^{-i \bq_1 \cdot \br } + e^{-i \bq_2 \cdot \br } + e^{-i \bq_3 \cdot \br },  \\
        U_1(\br) & = e^{-i \bq_1 \cdot \br } + e^{i \phi}e^{-i \bq_2 \cdot \br } + e^{-i \phi}e^{-i \bq_3 \cdot \br },
        \label{tunnel}
    \end{aligned}
\end{equation}
with $\phi = 2\pi/3$.  The vectors $\bq_i$ are $\bq_1 = k_\theta(0,-1)$ and $\bq_{2,3} =k_\theta(\pm \sqrt{3}/2,1/2)$.  The wavevector $k_\theta = 2k_D\sin \frac{\theta}{2}$ is the moir\'{e} version of the Dirac wavevector $k_D= 4\pi/3a_0$, where $a_0$ is the graphene lattice constant. 
% The quantity $E_0 = vk_\theta$ gives a natural energy scale for the band structure at an angle $\theta$.  
Note that we have assumed that all layers are aligned - for bilayers this does not matter since a shift may be removed by a gauge transformation. For $n >2$, the shifts are important and may alter the band structure \cite{Khalaf2019, MacdonaldTri}. For $n = 3$, it was argued that the zero shift configuration is energetically favorable \cite{Carr2020} and it is reasonable to assume that this also applies for larger $n$. Thus, in the following, we will assume the layers are aligned.

ATMG may be decoupled into $\lfloor \frac{n}{2} \rfloor$ copies of TBG where each copy has a tunneling scaled by a different number $\lambda_k$.
\begin{equation}
    H^{(n)}_k = \begin{pmatrix} -i v \bsigma_+ \cdot \bnabla & \lambda_k^{(n)} T(\br) \\
    \lambda_k T^\dag(\br) & -iv \bsigma_- \cdot \bnabla \end{pmatrix}.
    \label{tbgsector}
\end{equation}
If $n$ is odd there is another decoupled graphene layer $H_G = -iv \bsigma_{\theta/2} \cdot \bnabla$. The mapping is reviewed in detail in the supplemental material, section \ref{SuppSec:mappingreview}. We will order the TBG subsystems such that $\lambda_k^{(n)}$ is decreasing with $k = 1,\ldots,\lfloor \frac{n}{2} \rfloor$. We refer to the angle $\lambda^{(n)}_k \theta_{\rm TBG}$ as the \emph{$k$'th magic angle} of $n$ layer ATMG; this is really the ``first" magic angle for the $k$'th TBG subsystem, but in this paper we do not consider the higher magic angles of a given TBG subsystem.

\begin{figure*}
    \centering
    \includegraphics[width=\textwidth]{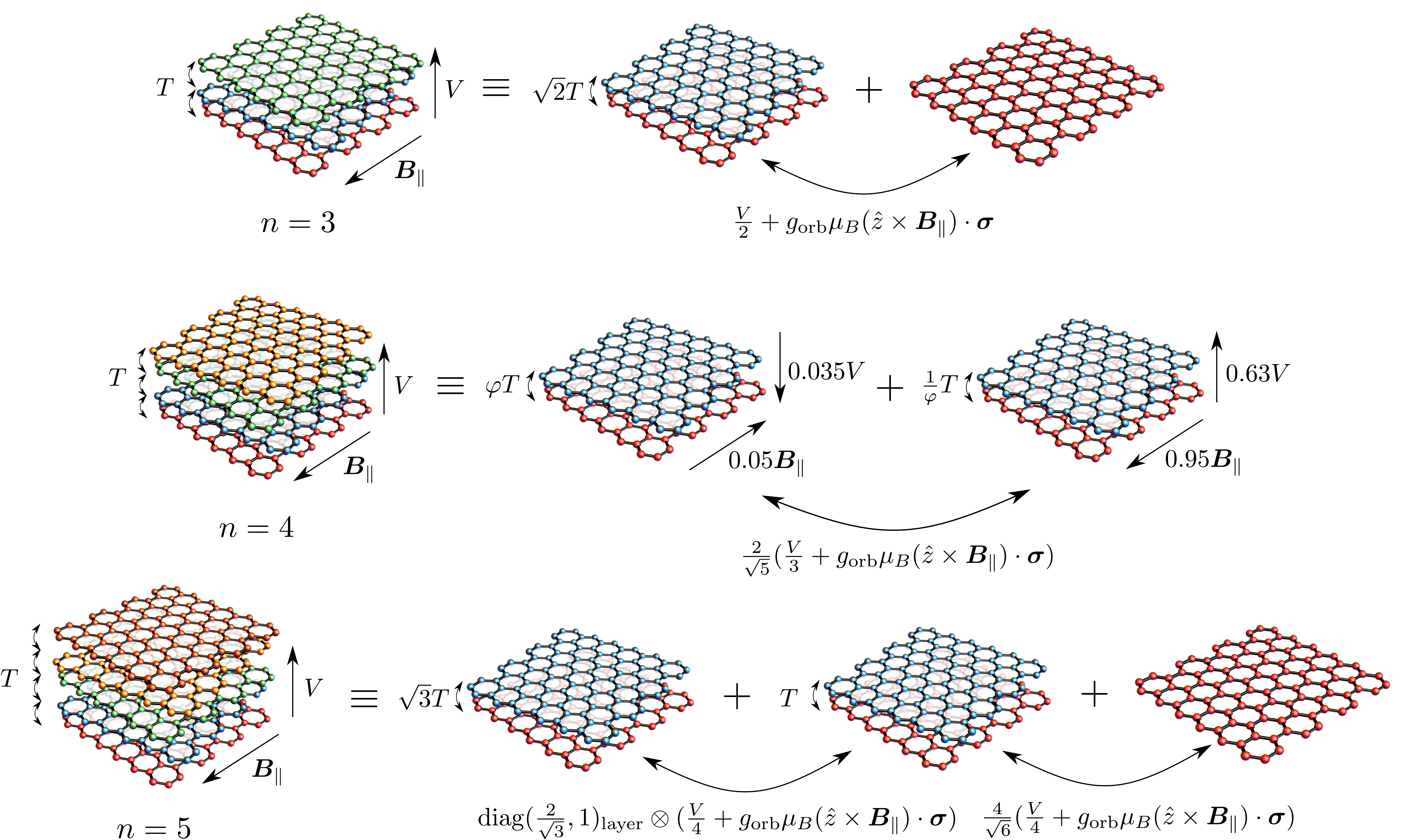}
    \caption{Illustration of the decompositon of ATMG into TBG and graphene subsystems for $n = 3,\,4,\,5$. Even $n$ ATMG has $n/2$ TBG sectors whereas odd $n$ ATMG has $(n-1)/2$ TBG sectors and one graphene sector.  The terms generated by  external fields $(V, \,{\bf B}_\parallel)$ can act both within subsystems and also as couplings between subsystems. For odd $n$, mirror symmetry $M_z$ implies that the intra-subsystem action of the external fields vanishes.} 
    \label{fig:decomposition}
\end{figure*}

ATMG has an approximate particle hole symmetry inherited from TBG and graphene. In particular, TBG has a well known approximate particle hole symmetry $\P = i \sigma_x \mu_y K$ \cite{BernevigTBGTopology, HejaziTopological}, where $\mu_{x,y,z}$ are the Pauli matrices in TBG layer space. In even-layer ATMG, particle-hole symmetry may be realized by applying it to each TBG subsystem \eqref{tbgsector}. For an odd number of layers, we additionally define $\P = \sigma_x \K$ in the graphene sector. It is an exact symmetry of \eqref{HATMG} if we drop the rotations on $\bsigma$ which is valid up to corrections proportional to $\theta^2 \ll 1$. There are other sources of particle-hole symmetry breaking including momentum-dependent moir\'{e} hoppings \cite{CarrPH} and next-nearest-layer ATMG tunneling \cite{Carr2020} but both of these terms are quite small.

Note however that when we move to $\bk$-space $\P$ does not have a simple action. This is because in each TBG sector the Bloch states have the form 
\begin{equation}
    \psi_{\bk}(\br) = u_{\bk}(\br)\begin{pmatrix} e^{i(\bk - \bK)\br} \\ e^{i(\bk - \bK')\br} \end{pmatrix},
    \label{tbgBloch}
\end{equation}
where $\bK$ and $\bK'$ are the wave-vectors at the the moir\'{e} $K_M$ and $K_M'$ points respectively. We choose the $\Gamma$ point to correspond to $\bk = 0$, such that $\bK = -\bq_1$ and $\bK' = \bq_1.$  Then we obtain that $\P$ takes $\bk \mapsto -\bk$. However the Bloch states for the graphene sector are proportional to $e^{i(\bk - \bK)\cdot \br}$, and so $\P$ inverts $\bk$ about the $K_M$ point, not the $\Gamma$ point. 

An important symmetry for an odd number of layers is a mirror reflection about the middle layer: $M_{zll'} = \delta_{l,n-l'}$.
The perturbations associated with an out of plane electric field and in plane magnetic field anticommute with this operation, whereas the TBG and graphene subsystems each have a specific eigenvalue under $M_z$. These external fields therefore act as tunneling terms between subsystems that have opposite $M_z$ eigenvalues. In contrast, for an even number of layers these external fields generically modify the intra-subsystem Hamiltonians
\begin{equation}
    H_k \to H_k - r^{(n)}_k \begin{pmatrix} \Gamma^{(n)}_+ & 0 \\ 0 & -\Gamma^{(n)}_- \end{pmatrix},
\end{equation}
where 
\begin{equation}
    \Gamma_{\pm}^{(n)} = \frac{V}{n-1} + g_{\rm orb} \mu_B \hat{\boldsymbol z} \times \bB \cdot \bsigma_{\pm}
\end{equation}
Here, $V$ is the voltage induced between the top and bottom layer after accounting for screening by high energy electrons and $\bB$ is the parallel magnetic field with $g_{\rm orb} = \frac{evd_\perp}{\mu_B} \approx 6.14$, where the bilayer graphene thickness is $d_\perp \approx 3.5$ \AA. 
The number $r^{(n)}_k$ is the external field coupling strength for the $k$'th TBG subsystem and is $1/2$ for $n=2$ TBG. Henceforth we will supress the superscript $(n)$ in $\lambda_k$, $\Gamma_k$ and $r_k$, since it will be clear from the multilayer case being considered.

For concreteness we illustrate the above points for $n=3, 4$ and $5$. For trilyer graphene the Hamiltonian is 
\begin{equation}
    H = \begin{pmatrix} -i v \bsigma_{+} \cdot \bnabla & T(\br ) & 0 \\
        T^\dag (\br ) & -iv \bsigma_{-} \cdot \bnabla & T^\dag(\br ) \\
    0 & T(\br ) & -i v \bsigma_{+} \cdot \bnabla \end{pmatrix},
    \label{ham}
\end{equation}
and $M_z$ acts by exchanging the first and third layers. 
An out of plane electric field and parallel magnetic field modify the Hamiltonian via
\begin{equation}
    H \mapsto H + \begin{pmatrix} \Gamma_{+} & 0 & 0 \\ 0 & 0 & 0 \\ 0 & 0 & -\Gamma_{+} \end{pmatrix},
    \label{efieldbfield}
\end{equation}
The Hamiltonian may be decoupled into a TBG sector, even under $M_z$, and a graphene sector, odd under $M_z$. The external fields act as tunneling terms between the subsystems.
\begin{equation}
    H_{\rm dec} = \begin{pmatrix} -i v \bsigma_{+} \cdot \bnabla & \sqrt{2} T(\br) & \Gamma_+ \\
    \sqrt{2} T^\dag(\br) & -iv \bsigma_{-} \cdot \bnabla & 0  \\
    \Gamma_+ & 0 & -iv \bsigma_{+} \cdot \bnabla \end{pmatrix}
    % \label{tbgsector}
\end{equation}

\begin{figure*}
\begin{center}
    \includegraphics[width=\textwidth]{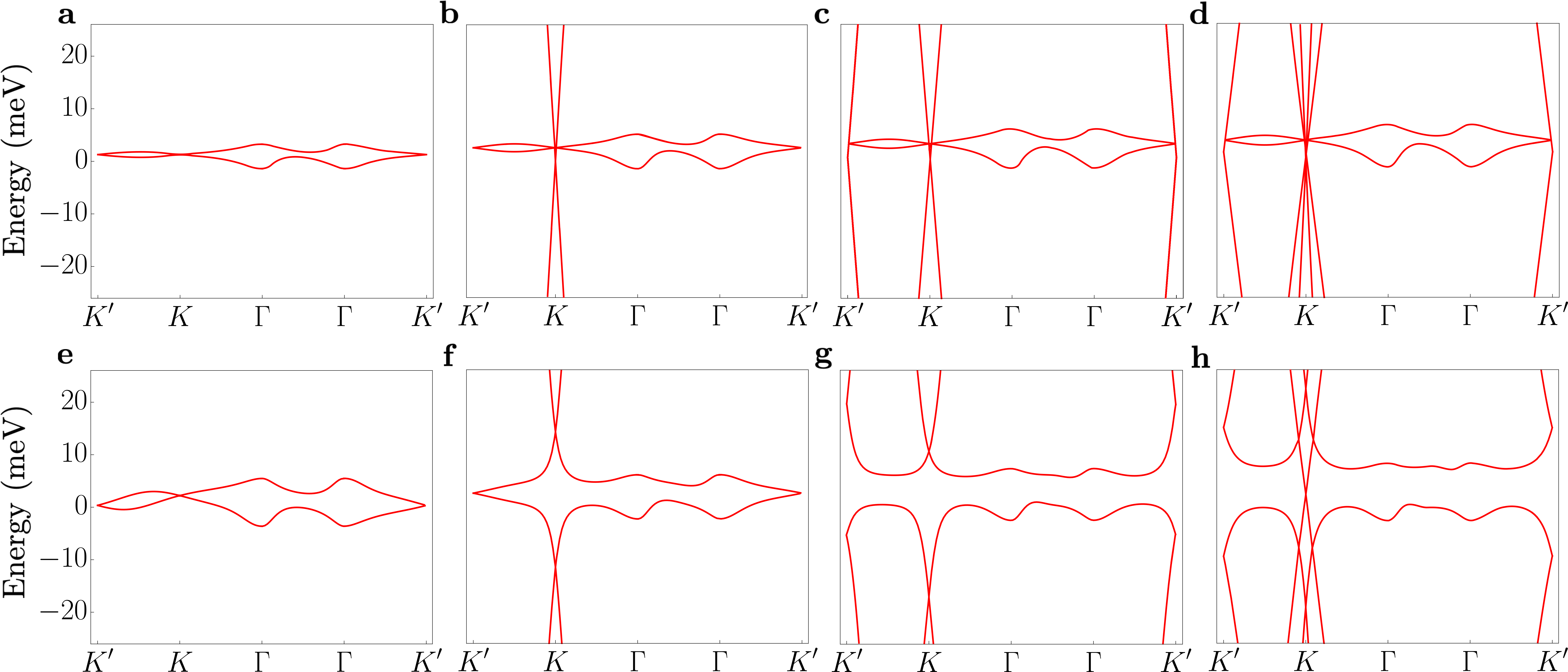}
\end{center}
\caption{
ATMG band structures with and without displacement field: Panels \textbf{a}-\textbf{d} have no displacement field and correspond to $n = 2,3,4,5$ respectively. Panels \textbf{e}-\textbf{h} have $V=50$ meV and also correspond to $n = 2,3,4,5$ respectively. We used the parameters $\kappa = 0.58$ and $\theta = \lambda_1 1.05^\circ$ for $\lambda_1 = 1, \sqrt{2}, \varphi, \sqrt{3}$ for $n=2,3,4,5$ respectively. Note, $n=2$ which corresponds to ordinary MATBG shows little change with displacement field ( \textbf{a} vs. \textbf{e}). In contrast, a gap opens at $n=3, 4, 5$ on applying moderate $V$ ( \textbf{b, c, d} vs. \textbf{f, g, h}).
}
\label{fig:bands}
\end{figure*}

For four layers we have
\begin{widetext}
\beq
    H^{\rm Full}_{\rm dec} =  \left(\begin{array}{cccc}
            -i v_F \bsigma_+ \cdot \nabla - r_1 \Gamma_{+}  & \varphi T(\br) & \frac{2}{\sqrt{5}} \Gamma_{+} & 0 \\
\varphi T^\dagger(\br) & -i v_F \bsigma_- \cdot \nabla + r_1 \Gamma_- & 0 & \frac{2}{\sqrt{5}} \Gamma_- \\
\frac{2}{\sqrt{5}} \Gamma_+ & 0 & -i v_F \bsigma_+ \cdot \nabla - r_2 \Gamma_+ & \frac{1}{\varphi} T(\br) \\
0 & \frac{2}{\sqrt{5}} \Gamma_- & \frac{1}{\varphi} T^\dagger(\br) & -i v_F \bsigma_- \cdot \nabla + r_2 \Gamma_-
\end{array} \right) 
\eeq
\end{widetext}
where $\varphi$ is the golden ratio $\frac{1 + \sqrt{5}}{2}$ and $r_{1,2} = \frac{2\sqrt{5} \mp 5}{10}$.
The tetralayer system is the first system in which there is more than one TBG subsystem and therefore more than one magic angle (again, here we do not consider the higher magic angles of a given TBG subsystem). The first subsystem has a magic angle at $\varphi \theta_{\rm TBG}$ and the second at $\varphi^{-1} \theta_{\rm TBG}$. 

Like trilayer graphene, the external fields lead to a coupling between the two TBGs. However, they also introduce an effective displacement field and in-plane magnetic field in each of the two TBGs that are rescaled by $\frac{1}{3} r_{1,2}$ and $r_{1,2}$, respectively, relative to the corresponding fields applied for the four layer system. Note that $r_1 \approx -0.05$ is ten times smaller in magnitude than in ordinary MATBG. As shown in supplemental material, this smallness arises due to the very rapid decay of $r_1$ with the number of layers for even $n$ with the large $n$ behaviour given by $r_1 \approx - \frac{\pi^2}{2 n^3}$.

The final case we would like to explicitly consider is that of five layers, $n = 5$, whose Hamiltonian is mapped to
\begin{widetext}
\beq
    H^{\rm Full}_{\rm dec} =  \left(\begin{array}{ccccc}
    -i v_F \bsigma_+ \cdot \nabla   & \sqrt{3} T(\br) & \frac{2}{\sqrt{3}} \Gamma_+ & 0 & 0\\
\sqrt{3} T^\dagger(\br) & -i v_F \bsigma_- \cdot \nabla & 0 & \Gamma_- & 0 \\
\frac{2}{\sqrt{3}} \Gamma_+ & 0 & -i v_F \bsigma_+ \cdot \nabla &  T(\br) & \frac{4}{\sqrt{6}} \Gamma_+ \\
0 & \Gamma_- &  T^\dagger(\br) & -i v_F \bsigma_- \cdot \nabla & 0 \\
0 & 0 & \frac{4}{\sqrt{6}} \Gamma_+ & 0 & -i v_F \bsigma_+ \cdot \nabla
\end{array} \right) 
\eeq
\end{widetext}
We see that the first magic angle is scaled by a factor of $\sqrt{3}$ compared to TBG whereas the second magic angle is the {\em same} as TBG magic angle. Note that here, similar to the trilayer case, the displacement and in-plane fields only introduce a coupling between the different TBG and graphene sectors without introducing an effective displacement field or in-plane orbital magnetic field in each sector which is a consequence of mirror symmetry.

\section{Lattice Relaxation Results}
\label{sec:relax}

\begin{figure*}
    \centering
    \includegraphics[width=0.75\textwidth]{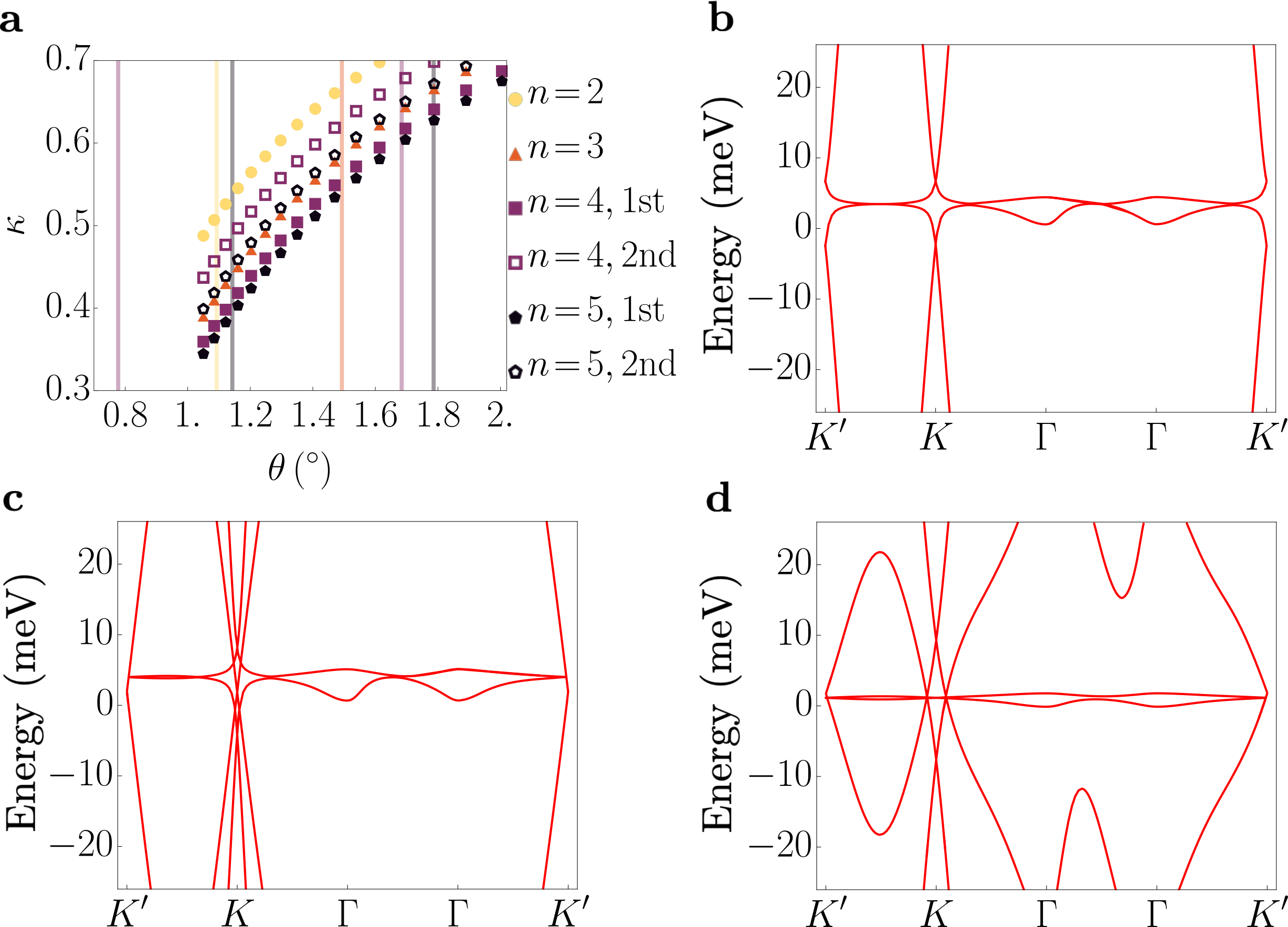}
    \caption{Lattice Relaxation: \textbf{a},  Effective $\kappa$ values for $n=2,3$ and both subsystems for $n=4,5$. Note that the second magic angle for $n=4$ is at $\theta < 1^\circ$ and is beyond the scope of the relaxation calculation. The second magic angle for $n=5$ is within reach and has the minimal value for $\kappa$. \textbf{b-d}, Band structures for the first $n=4$ magic angle and the first and second $n=5$ magic angles respectively. Mixing between the sectors comes from different values of $\kappa$ for the inner and outer interfaces. Note that the second $n=5$ subsystem does not mix with the other sectors between sectors because it is the only subsystem with $M_z = -1$.}
    \label{fig:relax}
\end{figure*}

In TBG, consideration of atomic relaxation at the magic angle has an important and well-known role on the interlayer tunneling \cite{Nam2017,Carr2019}.
We include this effect for the $n$-layer alternating structures with a continuous plate elasticity model \cite{Carr2018relax,Nam2017}, with strain moduli and stacking fault energy parametrized from first principle calculations of bilayer graphene (see SM).
After updating the atomic positions and the inter-layer electronic hopping terms in a twisted supercell  \cite{Fang2016,Carr2018pressure}, the effective $w_0$ and $w_1$ can be obtained from inner products of the tight-binding Hamiltonian with appropriate orbital-projected wavefunctions taken at the $K$ valleys of the monolayers \cite{Fang2019}.

We note that the parameters $w_i$ are sensitive to the parametrization of the atomic potentials and electronic tight-binding models.
See for example Fig. 14 of Ref. \onlinecite{Guinea2019}, where the $\kappa$ in TBG varies between $0.9$ and $0.4$ when considering different inter-atomic potentials and tight-binding parametrizations (because of e.g. electronic screening or corrugation reduction from hBN substrates).
The parametrization of the atomic and electronic functions used here are directly derived from first principles calculations~\cite{Fang2016,Carr2018pressure}, and the bands derived from these functionals show remarkably good agreement with full-scale density functional theory (DFT) calculations (Fig. 5 of Ref.~\onlinecite{Carr2020review}), but even DFT potentials in bilayer systems are known to be highly sensitive on the chosen density functional~\cite{Zhou2015}.
Therefore, the exact values of $\kappa$ presented here should be taken with a grain of salt, but the $\textit{relative}$ changes in $\kappa$ should be universal, and one can take the $\kappa$ value of TBG from our model as a scalable reference point.

In all the multi layer graphene geometries modeled here, the atoms tend to relax in-plane to maximize the amount of AB stacking, and relax out-of-plane to increase in the interlayer distance at AA stacking.
This causes a slight increase in the AB tunneling term $w_1$ accompanied by a significant reduction in the AA tunneling term $w_0$ with decreasing angle (see supplemental material for details). This effect becomes stronger as we increase the number of layers at a fixed angle. In addition, there is a distinction between tunneling at inner interfaces and outer interfaces once $n \geq 4$; the former experiences stronger relaxation effects.

One of the implications of our relaxation results is that the mapping to decoupled TBGs + Dirac is altered due to the different value of $w_0$ and $w_1$ for the inner and outer interfaces. If we start by ignoring $w_0$ (taking the chiral limit) or, more generally, assuming the chiral ratio $\kappa$ is constant between the layers, the decoupling mapping can still be performed even though the value of $w_1$ differs between the inner and outer interfaces. This was already discussed in Ref.~\cite{Khalaf2019} and is expanded upon in the supplemental material. The main consequence of the different value of $w_1$ between inner and outer interfaces for $n = 4,5$ is a slight decrease in the value of the first magic angle and a slight increase in the value of the second second magic angle with the modified magic angle values given in Table \ref{tab:relax}. 

For $n = 4$, the second magic angle is given by $\theta \approx 1.05^\circ/\varphi \approx 0.64^\circ$ which lies outside the range of validity of our relaxation calculations. In addition, for angles smaller than $1^\circ$, the effect of higher harmonics in the BM tunneling potential cannot be neglected \cite{MarginalTBG}. In contrast, the second magic angle for $n = 5$ is the same as TBG first magic angle, $\theta \approx 1.05^o$, in the ideal limit and is slightly increased to $\theta \approx 1.14^o$ with relaxation. For this value, the effect of higher harmonics can be neglected and our relaxation calculation can be trusted.

Introducing the different values of $\kappa = w_0/w_1$ between the inner and outer layer leads to a small coupling between the different sectors. When the number of layers is odd, such coupling only takes place within the same mirror sector since mirror symmetry is still preserved. The effect of such term can be seen in Fig.~\ref{fig:relax} for the first magic angle for $n = 4$ and for the two magic angles for $n = 5$. We notice that the effect of the coupling is qualitatively similar to applying a small displacement field which couples the different TBG sectors for $n = 4$ and couples the TBG sector in the $M_z = +1$ sector with the single Dirac cone for $n = 5$. The value of the coupling between the sectors is a few meV (see supplemental material for details).

Using the mapping to TBGs + Dirac and the relaxation data, we can calculate the effective value of $\kappa$ for each of the two TBG sectors for $n = 4,5$ as shown in Fig.~\ref{fig:relax}a. We note the following. First, the decrease in $\kappa$ due to increased relaxation with increasing the number of layers for $n=2,3,4,5$ is roughly compensated by the increase in the value of the first magic angle leading to an overall slight increase in the value of $\kappa$ for the first TBG block at the first magic angle (see Table \ref{tab:relax}). On the other hand, this implies that the chiral ratio is significantly reduced at the \emph{second} magic angle for $n = 5$.

One remarkable consequence of the previous discussion is that the flat band at the second magic angle for $n = 5$ represents the closest realization for the chiral model at the first magic angle since (i) it lives in the $M_z = -1$ mirror sector and is thus completely decoupled from the rest of the system and (ii) has a relatively small value of $\kappa \approx 0.45$. This makes this system a very promising candidate to search for fractional Chern insulator at zero magnetic field \cite{LedwithFCI, AbouelkomsanFCI, RepellinFCI, LauchliFCI}.

\begin{table}[t]
    \centering
    \begin{tabular}{cccc}
        \hline \hline
        Number of layers $n$ & Magic angle & $\theta$ & $\kappa$  \\
        \hline
        2 & 1st & 1.09${}^o$ & 0.51 \\
        3 & 1st & 1.49${}^o$ & 0.585 \\
        4 & 1st & 1.68${}^o$ & 0.614 \\
        5 & 1st & 1.79${}^o$ & 0.627 \\
        5 & 2nd & 1.14${}^o$ & 0.450 \\
        \hline \hline
    \end{tabular}
    \caption{Magic angles and effective $\kappa = w_0/w_1$ values for TBG subsystems of relaxed ATMG. Here we define the magic angle as the angle for which the bands are exactly flat in the chiral limit \cite{Tarnopolsky}.}
    \label{tab:relax}
\end{table}

\section{ Interacting Multilayers without external fields: magic TBG and Dirac cones decouple modulo screening}
\label{sec:screen}

\subsection{Graphene electrons screen the Coulomb interaction for TBG electrons}

  In this section we will argue that ATMG behaves much like MATBG together with extra decoupled layers of the non-magic subsystems even at the interacting level. We will first integrate out the non-magic electrons and obtain an effective model for just MATBG electrons. The usual strong coupling approach can then be applied to the magic sector.

The interacting Hamiltonian is
\begin{equation}
    \begin{aligned}
        \H & =  \sum_{\bk \in \rm BZ} c^\dag_\bk (h_\bk-\mu) c_\bk  + \sum_{\bk \in \rm BZ} f^\dag_\bk (h_{f \bk} - \mu) f_\bk \\
        %   & + \sum_{\abs{\bk} < \Lambda/v} \sum_{\eta = \pm} f_{\vb k \eta}^\dag  v(\eta \bk_\eta \cdot \bgamma-\mu) f_{\vb k \eta} \\
           & +  \frac{1}{2A} \sum_\bq V_\bq  :\rho_{\vb q} \rho_{\vb -\bq}:
\end{aligned}
\label{intham}
\end{equation}
Here $c_{\vb k}$ and $f_\bk$ are vectors of annihilation operators for the MATBG bands.
The total density has the form
\begin{equation}
    \begin{aligned}
        \rho_\bq & = \rho_{c \bq} + \rho_{f \bq} \\ 
        \rho_{c \bq} & = \sum_{\bk \in \rm BZ} c^\dag_k \Lambda_{c\bq}(\bk)c_{\bk + \bq}, \\
        % \rho_{f \bq} & = \sum_{\abs{\bp} < \Lambda/v} f^\dag_\bp f_{\bp + \bq} \\
        \rho_{f \bq} & = \sum_{\bk \in \rm BZ} f^\dag_\bk \Lambda_{\bq}(\bk) f_{\bk + \bq}\\
        \Lambda_{t\bq}(\bk)_{\alpha \beta} & = \braket{u_{t\alpha \bk}}{u_{t\beta \bk+\bq}},
     \end{aligned}
     \label{densitydecomp}
\end{equation}
where $t = c$ for the magic sector electrons and $t=f$ otherwise. We have written the full form factor of multilayer graphene in the block diagonal form $\tilde{\Lambda}_{\bq}(\bk) = \diag (\Lambda_{c\bq}(\bk), \Lambda_{f \bq}(\bk) )$. The indices $\alpha,\beta$ vary over the various bands, spins and valleys. The form factor $\Lambda_f$ is additionally block diagonal in the non-magic subsystems.

We note that the Hamiltonian \eqref{intham} has a separate internal symmetry for every TBG or graphene subsystem. For example, each TBG subsystem has a separate $\U(4)$ symmetry in the chiral limit. Most importantly, \eqref{intham} commutes with charge rotations of opposite sign in the magic and non-magic sectors. The origin of this symmetry is that the Coulomb interaction to a good approximation only sees the total density summed over all layers which separates into MATBG and non-magic densities by symmetry. This layer summed density-density form of the interaction also implies that for an odd number of layers it does not matter that the extra graphene Dirac cone is technically at the moir\'{e} K point: only differences in BZ momenta matter for the density. As a result we may use $\P$ symmetry without needing to worry about the subtleties discussed in the previous section. There are small corrections to the above picture stemming from the fact that the interaction depends on the out of plane distance between the layers; such corrections are suppressed by the square ratio of the interlayer separation to the moir\'{e} scale $d_\perp^2/a_M^2 \sim \theta^2 \ll 1$.

\begin{figure}
\begin{center}
\includegraphics[width=0.45\textwidth]{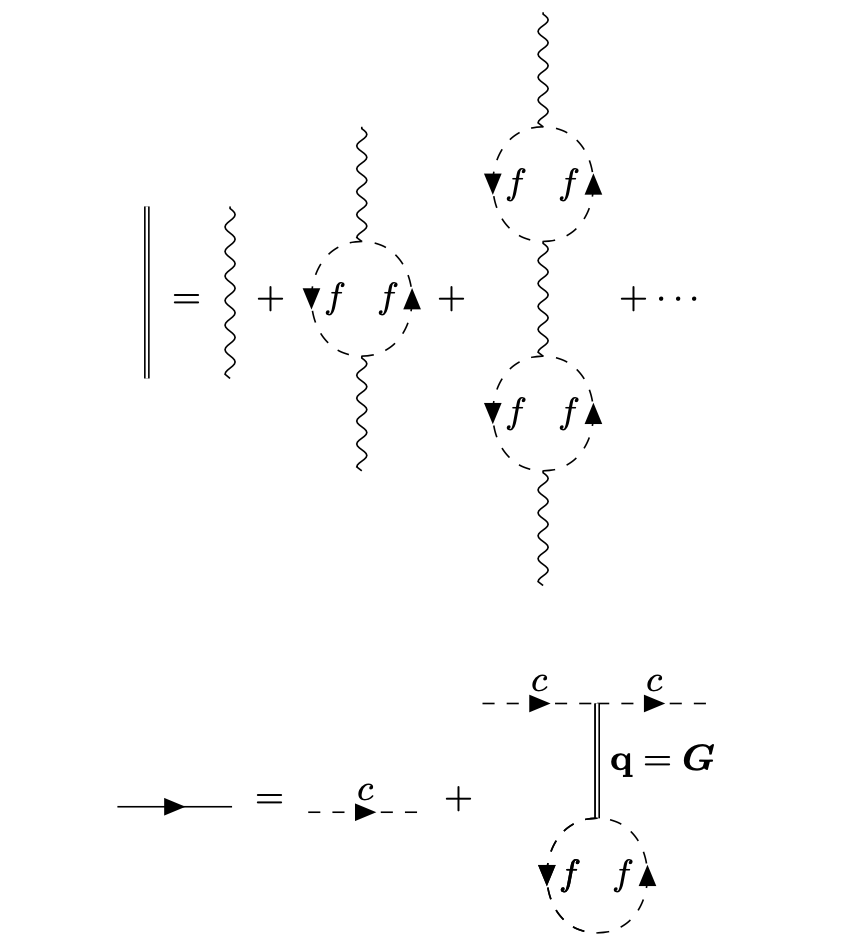}
\end{center}
\caption{Diagrams for integrating out non-magic fermions.  The double line denotes the effective Coulomb interaction $V_{\rm eff}(\bq)$ which is obtained by summing the geometric series of RPA diagrams with non-magic fermion bubbles.  The solid line with an arrow is the effective TBG propagator which differs from the bare propagator by a Hartree-like correction. }
\label{fig:diagrams}
\end{figure}

We now proceed to integrate out the non-magic electrons. Because the non-magic electrons only couple to the TBG density through the Coulomb interaction, they only renormalize the Coulomb interaction and Hartree potential (for $\mu \neq 0$), see Figure \ref{fig:diagrams}.  Additionally, we take the static screening limit as it is a very good approximation given how much faster the non-magic electrons are. We therefore obtain the effective Hamiltonian for MATBG electrons
\begin{equation}
    \begin{aligned}
        \H_{\rm eff} & =  \sum_{\bk \in \rm BZ} c^\dag_\bk (\tilde{h}(\bk) - \mu) c_\bk \\
                     & +  \frac{1}{2A} \sum_\bq V_{\rm eff}(\bq) : \rho_{\vb q} \rho_{\vb -\bq} : .
    \end{aligned}
    \label{manybandtbg}
\end{equation}
where $\tilde{h}$ includes the Hartree term from \ref{fig:diagrams}.

To obtain a calculable form of $V_{\rm eff}(\bq)$ we use the RPA to compute the effective
interaction and approximate the non-magic subsystems with the appropriate number of Dirac cones. We take $V(\bq) = \frac{e^2}{2\epsilon \epsilon_0 q} \tanh(qd_s)$, where $d_s$ is the screening length due to metallic gates.
The effective interaction is then given by
\begin{equation}
    \begin{aligned}
    V_{\rm eff} (\bq) & = \frac{e^2}{2 \epsilon_{\rm eff}(\bq) \epsilon_0 q} \tanh(qd_s)\\
    \epsilon_{\rm eff}(\bq) & = \epsilon + g \tanh(qd_s)\frac{e^2}{8 \epsilon_0 v} f(q/2k_F)
    \end{aligned}
    \label{RPA_interaction}
\end{equation}
where $f(x) = 1$ for $x \gtrsim 1$ and $f(x) = 4/\pi x$ for $x \lesssim 1$ \cite{Hwang2007}.  The number of fermion flavors is nominally $g = 4(n-2)$, however in practice we take $g=4n$ to also include screening from remote MATBG bands. For $qd \gg 1$ the effective dielectric constant is $\epsilon_{\rm eff} = \epsilon + \frac{g\pi \alpha}{8}$, where $\alpha \approx 2.2$ is
the fine structure constant in graphene. For nonzero $\mu$ the Dirac electrons will fill a Fermi sea with momentum $k_F = \mu/v$. The main effect of this is to shorten the effective screening length: by calculating $V_{\rm eff}(\bq = 0)$ we get $d_{\rm eff} = d/(1+g\alpha k_F d_s/\epsilon)$. 

\subsection{Insulating ground states for magic sector}

We may now proceed as in Ref. \cite{KIVCpaper} to determine the MATBG ground states at integer fillings: we integrate out the remote MATBG bands to obtain an effective model just for the eight flat bands and use approximate symmetries to determine the quantum Hall ferromagnetic ground states. 

The flat band projected Hamiltonian is
\begin{equation}
    \begin{aligned}
        \H_{\rm eff} & =  \sum_{\bk \in \rm BZ} c^\dag_\bk (h_c(\bk) - \mu  - h_\mu(\bk)) c_\bk \\
                     & +  \frac{1}{2A} \sum_\bq V_{\rm eff}(\bq) \delta \rho_{c \vb q} \delta \rho_{c \vb -\bq},
    \end{aligned}
    \label{flatbandprojHam}
\end{equation}
where the density operators 
\begin{equation}
    \begin{aligned}
        \delta \rho_{c\bq} & = \rho_{c\bq} - \ov \rho_\bq \\
        \ov \rho_\bq & = \frac{1}{2}\sum_{\bk,\bG} \delta_{\bG,\bq}\tr \Lambda_{\bG}(\bk)
    \end{aligned}
\end{equation}
give the density measured with respect to charge neutrality, $\delta n_f$ is the average density of the non-magic electrons measured from charge neutrality, and $h_\mu(\bk) \propto \mu$ is a Hartree term generated from integrating out non-magic electrons. It is expected to be small because the non-magic electrons do not fill much past charge neutrality. The Hartree term from filled magic sector electrons is contained in interaction term which now only contains magic sector densities measured from charge neutrality. This form \cite{KIVCpaper} of the flat band projected Hamiltonian is convenient because the density operators are exactly odd under particle hole symmetry and the dispersion term is particle hole symmetric if $\mu = 0$.

We now discuss the Chern basis for these flat bands and the approximate $\U(4) \times \U(4)$ symmetry.
The flat bands of twisted bilayer graphene may be viewed in a sublattice basis where the dispersion is off diagonal but the bands have a Chern number $C = \sigma_z \tau_z$ \cite{KIVCpaper,Khalaf2020}. Here we use the Chern and valley basis
\begin{equation}
    \begin{aligned}
        \gamma_{x,y,z} & = (\sigma_x,\sigma_y\tau_z,\sigma_z\tau_z) \\
        \eta_{x,y,z} & = (\sigma_x\tau_x, \sigma_x\tau_y,\tau_z).
        \label{chernbasis}
    \end{aligned}
\end{equation}
In this basis, The dispersion $h_c$ is mostly proportional to $\gamma_{x,y}\eta_0$ whereas the Hartree dispersion is mostly diagonal in the flat band flavors \cite{KIVCpaper,Khalaf2020, KhalafSoftMode}. 
Including spin, we have four bands with $C = 1$ and four bands with $C=-1$. The $\U(4) \times \U(4)$ symmetry in the chiral dispersionless limit corresponds to independent rotations in each Chern sector.  

The ground state may be understood by starting in the chiral dispersionless limit and adding perturbations. The ground states are described by \cite{KIVCpaper,Khalaf2020,KhalafSoftMode}
    \begin{equation}
        Q_{\alpha \beta} = 2 \langle c^\dag_\beta c_\alpha \rangle - \delta_{\alpha \beta}
    \end{equation}
    where $Q$ is an $8 \times 8$ matrix that is $\bk$-independent with $\tr Q = 2\nu$ and $Q^2 = \mathbbm{1}$. The $\pm 1$ eigenvalues of $Q$ label whether a particular band is filled or empty. We also must have $[Q,\gamma_z] = 0$, such that there is a definite Chern number $\tr Q \gamma_z = 2C$. All such $Q$ are ground states in the chiral dispersionless limit at charge neutrality because such states are annihilated by $\delta \rho_\bq$ for all $\bq$. Away from charge neutrality we generate a Hartree terms from the extra filled bands since $\delta \rho_\bG$ no longer annihilates the ground state. The Hartree terms do not split the ferromagnetic states but can instead favor metals or charge density waves \cite{PierceCDW}.
    
    We now discuss symmetry breaking terms. The dispersion $\hat h_c$ may be regarded as a tunneling between Chern sectors and breaks $\U(4) \times \U(4)$ to its diagonal subgroup by favoring $Q \propto \gamma_z$. Moving away from the chiral limit results in incomplete sublattice polarization and less symmetric form factors; this perturbation favors states with $[Q, \gamma_x \eta_z] = 0$. We may summarize these perturbations with the energy function
    \begin{equation}
        E(Q) =  \frac{J}{4} \tr (Q \gamma_x)^2 - \frac{\lambda}{4} \tr (Q \gamma_x \eta_z)^2.
        \label{Qenergy_mt}
    \end{equation}
    for some parameters $J, \lambda > 0$ that may be computed in perturbation theory and in Hartree Fock \cite{KIVCpaper,Khalaf2020, KhalafSoftMode}. The ground state that minimizes both perturbations is the Kramers' intervalley coherent state (KIVC) with $Q \propto \eta_{x,y} \gamma_z$. A valley Hall (VH) state with $Q \propto \eta_z \gamma_z = \sigma_z$ minimizes the $J$ term and is penalized by the $\lambda$ term; it is degenerate with the KIVC state in the chiral limit and is favored if the hBN substrate is aligned. Other states include the valley polarized (VP) state, with $Q = \eta_z$, which is degenerate with the KIVC state if there is no dispersion, and also the TIVC state $Q = \eta_{x,y}$ which is disfavored by both $J$ and $\lambda$.
    \begin{table}[t]
    \centering
    \begin{tabular}{ccc}
        \hline \hline
        State & $Q_c$ & Broken Symmetry   \\
        \hline
        KIVC & $\eta_{x,y}\gamma_z$ &  $U(1)_V, C_2, \T$ \\
        TIVC & $\eta_{x,y}$ &  $U(1)_V$ \\
        VH & $\eta_z \gamma_z$ & $C_2, C_2 \T$ \\
        VP & $\eta_z$ & $C_2, \T$ \\
        NSM & $\gamma_{x,y}$ & $C_3$ \\
        MSM & $\gamma_{x,y} \eta_z$ & $C_3, C_2, \T$ \\
        BM-SM & $\gamma_{x,y}$ & None \\
        \hline \hline
    \end{tabular}
    \caption{State names and order parameters. We drop the spin structure of the order parameters for brevity. The KIVC insulator breaks $\T = \eta_x K$ but preserves the Kramers' time reversal $\T' = \T \eta_z = \eta_y K$. The BM-SM and NSM are distinguished by $C_3$; the BM-SM $Q_c$ has vortices at $K_M$ and $K'_M$ whereas the NSM $Q_c$ has two vortices at $\Gamma$. }
    \label{tab:statenames}
\end{table}
    In practice there are also low energy nematic semimetal (NSM) states with $Q \propto \gamma_{x,y}$ \cite{ShangNematic}. These semimetals can in principle have any valley structure, but we will highlight an ordinary time reversal symmetric NSM state with $Q \propto \gamma_{x,y} \eta_0$, favored by strain, and a magnetic semimetal (MSM) that breaks $C_2$ and $\T$ with $Q \propto \gamma_{x,y} \eta_z$; we will show that the latter is favored by an in-plane magnetic field. Topological considerations obstruct this $Q$ from being $\bk$-independent and so it is nominally not in the low energy manifold of quantum Hall ferromagnetic ground states \cite{ShangNematic, KhalafSoftMode}. However, if the Berry curvature is highly concentrated, as it is for large $\kappa$ at the $\Gamma$ point, the state can be nearly $\bk$-independent with the exception of two vortices localized in a small vicinity of the $\Gamma$ point \cite{ShangNematic, KhalafSoftMode}. A useful limit to consider is one in which the Berry curvature is solenoidal and may be gauged away with a singular gauge transformation \cite{KhalafSoftMode}. With this transformation, the bands lose their topology, Chern sector becomes another flavor, and the nematic semimetal is realized as a low energy quantum Hall ferromagnet. The model also has an approximate $\U(8)$ symmetry with associated nematic soft modes \cite{KhalafSoftMode}. We review this perspective in the supplemental material in section \ref{Suppsubsec:intraTBGnis2}.
    
    If one works far enough away from the magic angle the strong coupling picture breaks down and a completely symmetric ``Bistritzer-Macdonald'' semimetal (BM-SM) becomes the ground state.

\subsection{Non-magic electrons remain semimetallic and decoupled} 

We now ask what happens to the non-magic electrons.  We will argue that they exist essentially independently from the MATBG electrons provided the MATBG electrons form a quantum hall ferromagnetic ground state. Such a result is in stark contrast to doped electrons in twisted bilayer graphene on top of the KIVC ground state which interact with order parameter fluctuations and may condense in pairs to form a
superconductor \cite{kozii2020}.

The essential reason that there is little interaction that the two subsystems are only coupled through the Coulomb interaction. A charge $\delta \rho_{c\bq}$ from order parameter fluctuations must be present to interact with non-magic electrons. In a quantum hall ferromagnetic ground state this charge is the skyrmion density,
\begin{equation}
        \delta \rho_{\rm top}  = \frac{i}{16\pi}\varepsilon_{ij} \tr Q \partial_i Q \partial_j Q,
        \label{topcharge}
\end{equation}
which contains two derivatives and two powers of a soft mode $\phi = Q-Q_0$. This interaction is expected to be very weak at low energy. Furthermore, truly gapless fluctuations within the KIVC manifold of states carry zero topological charge as may be seen by noting that the above trace vanishes when $[Q,\eta_z] = 0$.

At integer fillings we expect a Chern diagonal $Q$ matrix in the magic sector together with metallic or semimetallic non-magic electrons.  For small doping away from integers, the non-magic subsystems will gain charge, and only after the chemical potential reaches some critical level will charge begin to populate the magic sector.  However, as in TBG, small amounts of strain in ATMG may favor a metallic state of the magic sector instead at some integers. 

\section{Effect of external fields on magical insulators and Dirac cones}
\label{sec:externalfields}

We now move on to adding external fields to ATMG and for now focus on their influence at integer fillings. We begin with $n = 2,3$ and then discuss $n=4$ which combines the methods in $n=2,3$. We then make some comments on patterns for larger $n$. We will use two $Q$ matrices $Q_c$ and $Q_f$ to label the filled bands of the magic and non-magic electrons respectively in the zero-field states that we perturb.

\subsection{$n=2$ MATBG}
\label{subsec:extfields_tbg}

For $n=2$ MATBG we have a single $Q$-matrix $Q=Q_c$ that describes the magic electron occupation. The external fields introduce terms in $E(Q)$ that favor some states over others. Here we describe these terms to second order in $V,\bB$; a full calculation is given in the supplemental material in section \ref{Suppsubsec:intraTBGnis2}.

\begin{figure}
    \centering
    \includegraphics[width=0.5\textwidth]{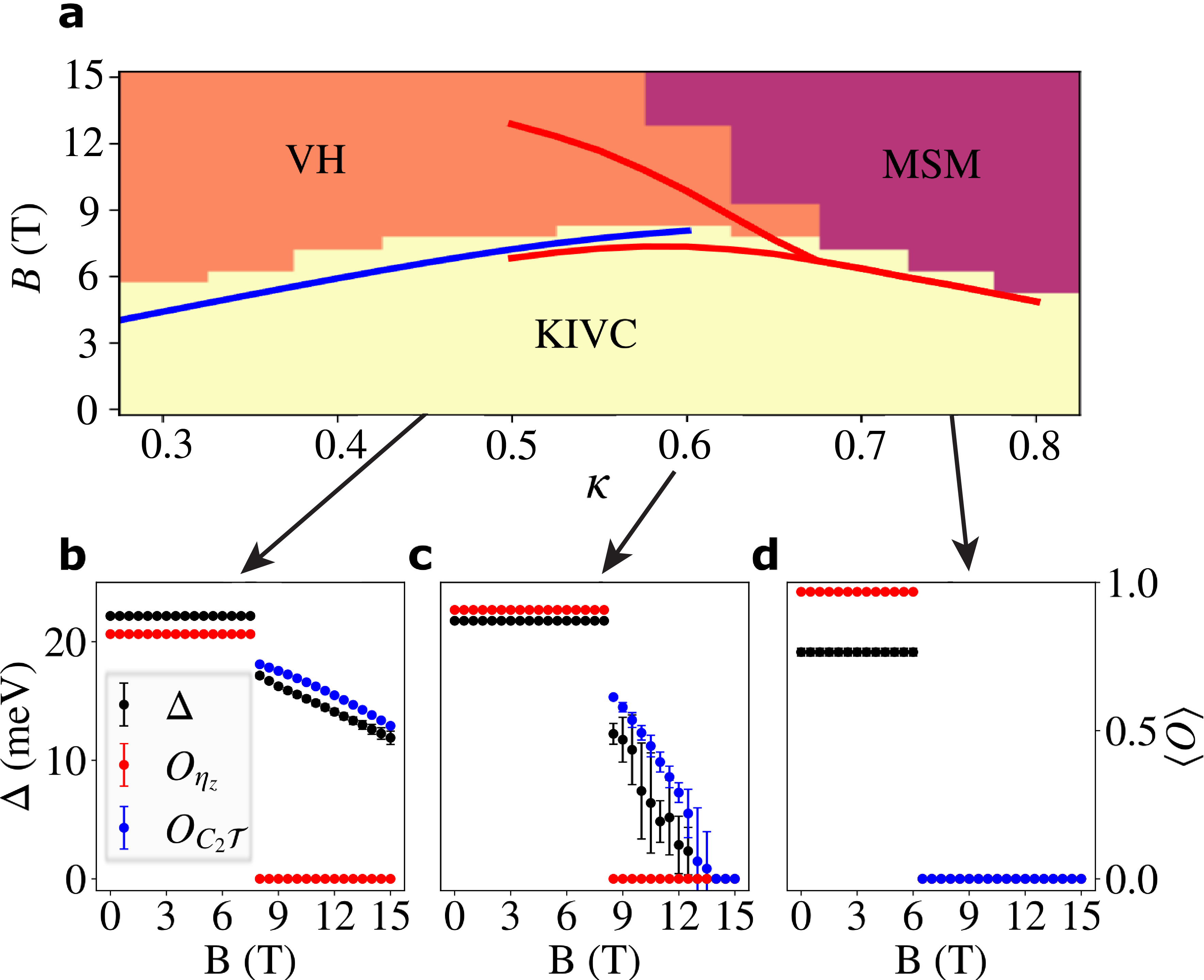}
    \caption{\textbf{a}, Hartree Fock phase diagram of MATBG at charge neutrality in an in-plane magnetic field $B$.  We used the parameters $\theta = 1.05^\circ$, $d_s = 20nm$, the hBN screening $\epsilon = 6.6$, and the screened interaction \eqref{RPA_interaction}.  The blue line and red lines are analytic predictions for the critical fields for small $\kappa$ and large $\kappa$ respectively. The lower red line is for the first order KIVC\textrightarrow VH transition whereas the upper red line is for the second order VH\textrightarrow MSM transition. 
    \textbf{b-d}, Fixed $\kappa$ phase diagram cuts. Here we plot the Hartree Fock band gap $\Delta$ and the order parameters $O_{\eta_z}$ and $O_{C_2\T}$, see Eq. \eqref{OrderParams}, which are nonzero in the KIVC and VH states respectively; the MSM does not break either symmetry. Note, in the  KIVC phase the $C_2 \T$ order parameter is not shown since it is sensitive to the spontaneous choice of IVC phase.
    }
    \label{fig:tbg_phase}
\end{figure}

The electric field has little effect in TBG. The electric field projected to the flat bands induces a dispersion $h_V(\bk)$ that is odd under $\bk \mapsto -\bk$ and is proportional to $\eta_z$. This dispersion has no effect at first order but at second order it leads to a term $\frac{\tilde{\lambda}_V}{4} \tr (Q \eta_z)^2$. In practice this term is very small, $\lambda_V \lesssim 10^{-5}(\rm meV)^{-1}\frac{V^2}{4} $, because the flat bands have very little layer polarization and the dispersion $h_V(\bk)$ is proportional to the layer polarization. 

The in-plane magnetic field has a much stronger effect. Projecting the magnetic field to the flat bands induces a dispersion $h_B(\bk)$ that is proportional to $\eta_z$ and off diagonal in Chern sector. As a result, within perturbation theory, we obtain the term
\begin{equation}
    \frac{\lambda_B}{4} \tr (Q \eta_z \gamma_x)^2.
\end{equation}
The sign of this term is opposite to that of the zero field $\lambda$ term that arises due to imperfect sublattice polarization. Thus, as $\bB$ increases we obtain a transition to a valley Hall state once $\lambda_B>\lambda$. 

The coefficient $\lambda_B$ has larger magnitude than naive dimensional analysis suggests because $h_B$ has angular momentum under $C_3$ and can therefore couple to nematic soft modes. In particular, we obtain $\lambda_B \approx c_B \frac{g_{\rm orb}^2 \mu_B^2 B^2}{4}$ where $c_B$ increases with $\kappa$. In the range $\kappa = 0-0.8$ we have $c_B = 0.1-0.3 (\rm meV)^{-1}$ which corresponds to an effective gap size much less than the interaction scale of $\approx 15 \rm{meV}$. This perturbation theory may be used to find the critical field for the KIVC\textrightarrow VH transition at small $\kappa$; here the zero field gap is small and the Berry curvature is spread out so the nematic modes have a larger gap. At larger $\kappa$ however perturbation theory breaks down for fields well below the critical field.

Instead, at larger $\kappa$ we may view the in plane field as a Zeeman field for the MSM state with $Q \propto \gamma_{x,y} \eta_z$. This state anticommutes with $C_2$ and $\T$ but preserves $C_2 \T$. At zero magnetic field, the MSM state is separated from the KIVC ground state by a gap of around $0.5$ meV per electron which shrinks as $\kappa$ increases. The energy of this state decreases linearly when the magnetic field is added. Because $\{Q_{\rm VH}, Q_{\rm MSM}\} = 0$, the VH state can cant into the MSM state to reduce its energy as well
\begin{equation}
    Q_{\rm VH}(B) = \cos \theta_B Q_{\rm VH} + \sin \theta_B Q_{\rm MSM}.
\end{equation}
The above perturbative superexchange corresponds to this canting at first order in $\theta_B$, while also allowing $\theta_B$ to be $\bk$-dependent. Here instead we take $\theta_B$ to be $\bk$-independent. We do so because at large $\kappa$ we may regard $Q_{\rm MSM}$ to be a generalized quantum Hall ferromagnet, in the sense that it is $\bk$-independent after a singular gauge transformation at the $\Gamma$ point where the Berry curvature is concentrated \cite{KhalafSoftMode}. For a review of this perspective and a detailed calculation of the critical fields see the supplemental material section \ref{Suppsubsec:intraTBGnis2}. This calculation becomes unreliable at small $\kappa$ where the Berry curvature is more spread out - there we rely on the perturbative calculation described above. 

At the largest values $\kappa \gtrsim 0.7$ we obtain a direct first order transition KIVC\textrightarrow MSM. For $\kappa \lesssim 0.7$ we first have a first order transition KIVC\textrightarrow VH, then a second order transition VH\textrightarrow MSM. 

We have performed numerical Hartree Fock simulations for spinless TBG at charge neutrality for comparison with our analytic predictions. We identified the phase of the ground state using the order parameters
\begin{equation}
    \langle O_S\rangle = \frac{A_M}{4A}\sum_\bk \tr \left(\frac{1}{2}(Q_\bk-S^\dag Q_\bk S)\right)^2
    \label{OrderParams}
\end{equation}
for $S = \eta_z, C_2 \T$. The order parameters are normalized such that they have a maximal value of $1$ when $\{Q,S\} = 0$. The KIVC state has $\langle O_{\eta_z}\rangle > 0$ whereas the VH state has $\langle O_{\eta_z}\rangle = 0$ but $\langle O_{C_2\T} \rangle > 0$. The MSM state on the other hand has $O_{\eta_z} = O_{C_2 \T} = 0$.

The Hartree Fock phase diagram is shown in Fig. \ref{fig:tbg_phase}a. The small and large $\kappa$ analytic predictions for the phase boundaries are also plotted, corresponding to perturbative superexchange and non-perturbative canting respectively. We also plot the Hartree Fock band gap and the order parameters we used for phase identification along fixed $\kappa$ cuts in Fig. \ref{fig:tbg_phase}b-d. In particular, the canting of the VH order into the MSM is visible in \ref{fig:tbg_phase}c. The transition to the MSM is accompanied with a gap closing that may correspond to the gap closing of the $\nu = 2$ correlated insulator in Ref. \cite{Dean-Young}. Note also that a transition to the VH or MSM state would destroy skyrmion superconductivity; the former corresponds to an easy-axis anisotropy that strongly disfavors skyrmions and the latter is outside the skyrmion manifold of insulators.

\subsection{$n=3$ sandwiches: Mirror Superexchange and Graphene mass generation}
\label{subsec:extfields_ttg}

\begin{figure*}[h!tb]
    \centering
    \includegraphics[width=\textwidth]{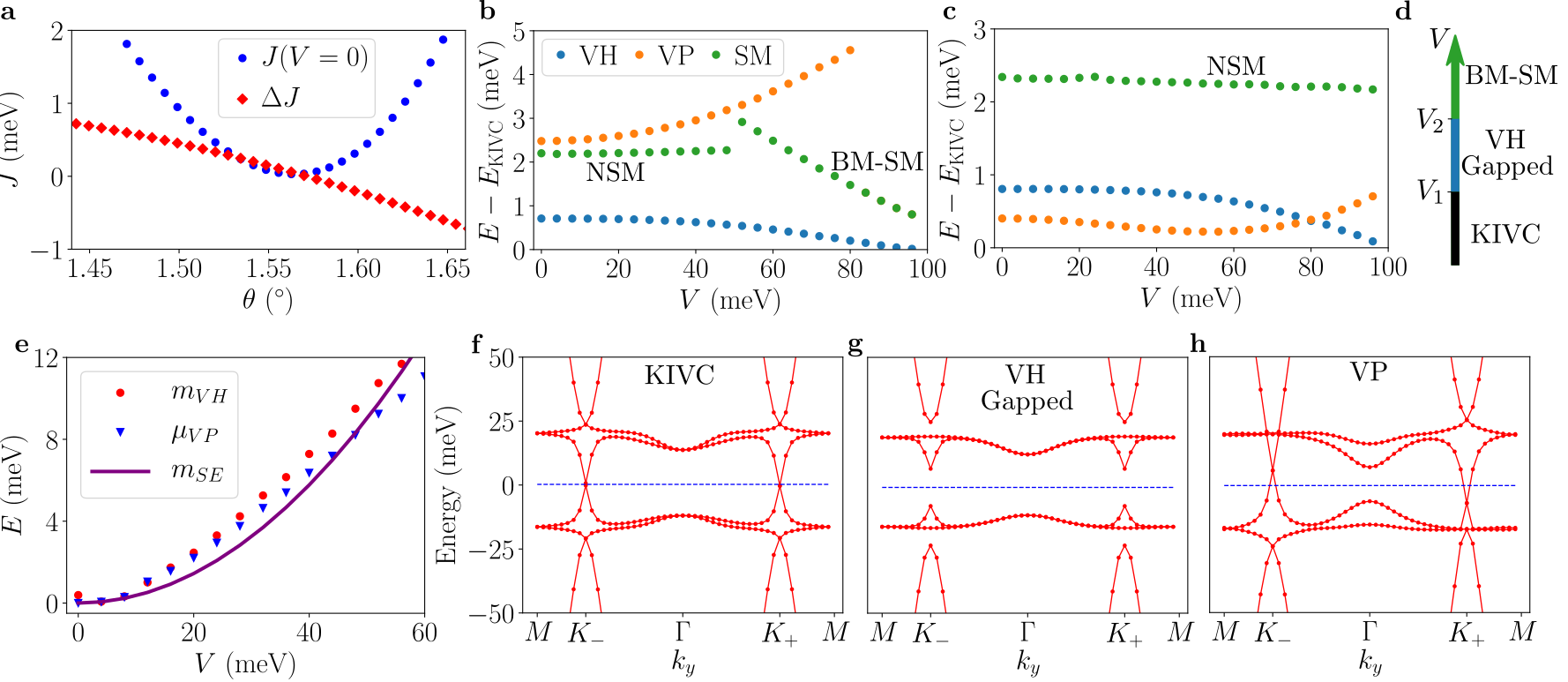}
    \caption{
    Analytic peturbation theory and Hartree Fock numerics for twisted trilayer graphene at charge neutrality.
    \textbf{a}, Chern superexchange parameter $J$ at zero displacement field and finite field contribution $\Delta J(\theta) = J(V = 50 \text{meV}) - J(0)$ as a function of angle. When $\theta$ passes through the magic angle the sign of the dispersion $h_c$ flips leading to a sign change in $\Delta J(\theta)$ as a function of angle. When $\theta$ passes through the magic angle the sign of the dispersion $h_c$ flips leading to a sign change in  $\Delta J$ . 
    \textbf{b, c}, Energies of the valley Hall (VH), semimetal (SM), and valley polarized (VP) states per unit cell relative to the Kramers' intervalley coherent (KIVC) ground state as a function of displacement field for $\theta = 1.47^\circ, 1.61^\circ$. For both angles there is a transition from  $\rm{KIVC} \to \rm{VH}$ at $V \approx 100$ meV. At the smaller angle the gap to the VP state increases monotonically because the displacement field generated dispersion has the same sign as $h_c$ and $E_{\text{VP}} - E_{\text{KIVC}} = 2J$. At the larger angle the VP gap slightly decreases for small $V$ before increasing at larger $V$. For the smaller angle there is also a transition from the nematic SM to the symmetric Bistritzer-Macdonald SM (BM-SM) which decreases in energy and eventually becomes the ground state. For the larger angle there is no transition to the BM-SM for the displacement fields shown. \textbf{d}, Schematic displacement field dependence of the ground state. $V_1 \approx V_1(\kappa)$ is the transition to the VH state. It increases with increasing $\kappa$ because the zero field gap to the VH state does. The field $V_2 \approx V_2(\theta)$ is the transition to the BM-SM. It increases with increasing $\theta$. \textbf{e}, Mass and valley doping of the graphene Dirac cones in the VH and VP states respectively compared with the perturbative superexchange value \eqref{maintext_superex}, \eqref{mse}.
    \textbf{f, g, h}, Hartree Fock band structure for the KIVC, VH, and VP states at $V = 40$ meV. The bands are plotted along a line of constant $k_x$ that passes through all high symmetry points.
    }
    \label{fig:trilayerHF}
\end{figure*}

In this section we analytically discuss effects of $M_z$ breaking external field perturbations in trilayer graphene \eqref{efieldbfield} and then compare to Hartree Fock numerics. While nonzero $V,\bB$ strongly reconstructs the noninteracting band structure near $K_M$ and $K_M'$ in the respective graphene valleys, the effect on an interacting TBG ground state at integer filling is expected to be smaller since the TBG sector will have a gap to single particle excitations at least at $\bk = \bK$. Due to the presence of this gap\footnote{Note that a gap in the graphene sector is \emph{not} required: the external field terms always change the charge at $\bK$ in the TBG system and the TBG system is assumed to have a charge gap there}, we may perform a well defined perturbation theory that gives a "mirror-superexchange" energy at second order in $V,\bB$ and, depending on the ground state in the TBG sector, modifies the graphene dispersion. 

We start with an unperturbed low energy Hamiltonian that consists of the flat band projected TBG sector \eqref{flatbandprojHam} together with the low energy part of the graphene Dirac cone. The external fields contribute the following terms to the effective low energy Hamiltonian
\begin{equation}
\begin{aligned}
    \H_\xi & = \sum_{\bp} c^\dag_\bk \xi_\bp f_\bp + f^\dag_\bp \xi^\dag_\bp c_\bk , \\
    (\xi_{\bp \eta})_{\gamma \gamma'} & = \left(\frac{V}{2}\delta_{\gamma \gamma'} + \eta g_{\rm orb} \mu_B \bB \times \bgamma_{\gamma \gamma'}\right) t_{\bp \eta \gamma}, \\
    t_{\bp \eta \gamma} & = \int_{\rm Cell} d^2 \br\, u^*_{c\bk \eta \gamma +}(\br) e^{i \bG \cdot \br}.
    \label{eBfieldproj}
    \end{aligned}
\end{equation}
where $\bp = \bk + \bG$ where $\bk$ is in the first BZ and $u_{c \bk \eta \gamma \mu}(\br)$ is the MATBG moir\'{e} periodic wavefunction in graphene valley $\eta$ and Chern sector $\gamma$ evaluated in TBG ``layer'' $\mu$ and position $\br$. We have chosen to use the continuous translation symmetry of the graphene subsystem to label states with $\bp$. These terms are the low energy projection of \eqref{efieldbfield}; they take this form because \eqref{efieldbfield} anticommutes with $M_z$. 

We now discuss the effect of \eqref{eBfieldproj} in perturbation theory. The tunneling between mirror sectors has zero effect at first order since the expectation value of the term vanishes by $M_z$ symmetry. However, within second order perturbation theory \eqref{eBfieldproj} is expected to have an effect through a ``superexchange": electrons gain energy by virtually tunneling between mirror sectors. We compute this energy in the supplemental material in section \ref{Suppsubsec:trilayersuperexchange}. It is simplest to understand the result at charge neutrality; in the supplement we show that, qualitatively, the results are general. To a good approximation we obtain
\begin{equation}
\begin{aligned}
    E(Q_c, Q_f) & = \frac{1}{2}\sum_{\bp \eta} \frac{1}{v\abs{\bp_\eta} + U_\bk} \tr(Q_c \xi_{\bp \eta} Q_{f \eta \bp} \xi^\dag_{\bp \eta}).
    \end{aligned}
    \label{maintext_superex}
\end{equation}
The matrix $Q_{f\alpha \beta} = 2 \langle f^\dag_\beta f_\alpha \rangle - 1$ describes the occupation of the graphene Dirac cones and $Q_{f \eta}$ is its projection into valley $\eta$. As one would expect, the energy scales with the square of the external fields and is suppressed by a combination of the Dirac dispersion and the TBG-sector exchange energy $U_{\bk \approx \bK_{\pm}} \approx 18$ meV. The trace measures to what extent the superexchange is Pauli blocked in valley $\eta$ at momentum $\bk$.

We find that the energy vanishes for the ``unperturbed" states $Q_c \sim \gamma_{0,z}$ and $Q_f \sim \gamma_{x,y}$; the $V^2$ and $B^2$ terms vanish by the trace over Chern sector and the $VB$ terms vanish by $C_3$. The energy also vanishes if the TBG sector is in a NSM state as well; in this case the $V^2$ and $B^2$ terms vanish by $C_3$ and the $VB$ terms vanish by the trace over Chern sector. To take advantage of this term, $Q_c$ must develop a $C_3$-symmetric $\gamma_{x,y}$ component or $Q_f$ must develop a Chern diagonal or nematic component depending on $Q_c$. These two options both happen simultaneously and may be considered separately in the next order of perturbation theory which we now describe. We will focus on the displacement field; the magnetic field may be understood analogously but has very little effect at this order in perturbation theory due to its small scale. 

Let us first fix $Q_{f \eta}(\bp)$ to be unperturbed. Then we may understand \eqref{maintext_superex} as a  dispersion term $\frac{1}{2} \tr Q_c h_V$ with 
\begin{equation}
    (h_{V}(\bk))_{\gamma \eta \gamma' \eta'} = \frac{V^2}{4}\sum_{\bp \eta}  \frac{\delta_{[\bp],\bk} \delta_{\eta \eta'}t_{\gamma \eta \bp} t_{\gamma' \eta \bp}^*}{v \abs{\bp_\eta} + U_\bk } \widehat{\bp_\eta} \cdot \bgamma_{\gamma \gamma'} 
\end{equation}
The valley-even part $h_{V+} \propto \eta_0$, where $h_{V\pm} = \frac{1}{2}\left( h_V \pm \eta_x h_V \eta_x \right)$, combines with $h_c$ to generate the inter-Chern superexchange quantified by the parameter $J = J(V) \propto (h_c + h_{V +} )^2$. More precisely we obtain
\begin{equation}
\begin{aligned}
    J(V) & = 2\frac{A_M}{A} \sum_{\bk \bk'} (z_{c_\bk} + z_{V\bk})^* (R^{-1})_{\bk \bk'} (z_{c\bk'}+ z_{V\bk'}) \\
    & = J(0) + 4\frac{A_M}{A}\sum_{\bk \bk'} \Re z_{c\bk}^* (R^{-1})_{\bk \bk'} z_{V\bk'} + O(V^4) \\
    R_{\bk \bk'} & = \frac{1}{A} \sum_\bq V(\bq) \left( \abs{\Lambda_{c \bq}(\bk)}^2 \delta_{\bk \bk'} - \delta_{[\bk+\bq],\bk'} \Lambda^\dag_\bq(\bk)^2  \right),
    \end{aligned}
\end{equation}
where we write $h_{c,V} = h_{c,V x} \gamma_x + h_{c,V y} \gamma_y$ and $z_{c,V} = h_{c,V x} + i h_{c,V y}$. The zero field $J(0)$ as well as the order $V^2$ contribution to $J$ are plotted in Fig. \ref{fig:trilayerHF}a as a function of twist angle (for simplicity we neglect the $O(V^4)$ term in this discussion and in Fig. \ref{fig:trilayerHF}a). As $\theta$ passes through the magic angle, the dispersion $h_c$ essentially flips sign. Accordingly, we see that $J(0)$ has a minimum at the magic angle where $h_c$ has the smallest magnitude. The order $V^2$ contribution to $J$ is negative for large angles. This is because for large angles the dispersion $h_c$ has the same sign as the Dirac dispersion and $h_V$ has the opposite sign as the Dirac dispersion because it is proportional to $Q_f$. When $\theta$ drops below the magic angle, $h_c$ flips sign and thus $J(V)$ decreases with $V$ for these angles. The specific angle for which the $V^2$ contribution to $J$ switches sign is parameter dependent and also dependent on the Hartree Fock subtraction scheme but is approximately $1.57-1.58^\circ$ with our conventions. To see an appreciable decrease in $J$, one needs larger angles $\theta \gtrsim 1.63^\circ$ because otherwise the positive $O(V^4)$ term will quickly take over.

The valley odd $h_{V-}$ generates the term $\frac{\lambda_V}{4} \tr(Q_c \eta_z \gamma_x)^2$ with $\lambda_V \sim V^4$. This is somewhat similar to the magnetic field dispersion in TBG and acts against the zero field $\lambda$, though because this term is $C_3$-symmetric it does not couple to nematic soft modes or act as a Zeeman field that favors the nematic semimetal. It favors the VH state over the KIVC state.

Next we fix $Q_c$ to be unperturbed and Chern-diagonal. We then consider \eqref{maintext_superex} to be an effective dispersion for the graphene electrons. This only works if $Q_c$ is valley diagonal; the graphene electrons are low energy at opposite $\bK_\eta$-points in each valley. For valley-diagonal states at the $\bK_\eta$ points the graphene electrons obtain
\begin{equation}
    m_{\rm SE} = \frac{\abs{t_{K ++}}^2}{U_K} \left( \frac{V^2}{4}Q_c + g_{\rm orb}^2 \mu_B^2 B^2 \gamma_x Q_c \gamma_x \right).
    \label{mse}
\end{equation}
as an additonal dispersion term.
For $Q_c \propto \gamma_z$ this acts as a mass term gapping the Dirac cones. If the TBG bands are in a Chern insulator state then the Dirac electrons will develop a Chern number generating mass whereas if the TBG bands are in a VH state the Dirac electrons will also pick up a VH generating mass. For the VP state, $Q_c = \gamma_0 \eta_z$, we instead obtain a valley doping. In our Hartree Fock simulations we see these terms in the Hartree Fock band structures and their magnitude is consistent with our analytic perturbation theory, see Fig \ref{fig:trilayerHF}d-g. The mass terms, together with an energy optimizing $Q_f$, lower the energy of states with $Q_c \propto \eta_{0,z}\gamma_z$ by an amount $\propto V^4$; see the supplemental information section \ref{Suppsubsec:trilayersuperexchange} for a detailed calculation. This combines with the valley odd dispersion to favor VH over KIVC by an amount $\propto V^4$.

We have performed Hartree Fock simulations on charge neutral spinless trilayers to test our analytic predictions. We use the Hamiltonian \eqref{intham} but with the interaction \eqref{RPA_interaction} to account for the screening effects that Hartree Fock does not include. 
We use $\epsilon = 6.6$ for hBN screening, the interaction \eqref{RPA_interaction}, $\kappa = 0.58$, and $d_s = 20$ nm.
The Hartree Fock state energies are plotted in Fig. \ref{fig:trilayerHF}b-c and show good qualitative agreement with the analytic results described above. In particular, we see that for small angles, panel b, $E_{\text{VP}} - E_{\rm KIVC} = 2J$ increases with displacement field and for sufficiently large displacement field the flat band dispersion $h_c + h_V$ becomes large enough to favor a $C_3$ symmetric ``BM" semimetal over the nematic semimetal; The ``BM" semimetal is essentially the non-interacting ground state which is favored by dispersion rather than interactions.
The BM semimetal decreases in energy with displacement field relative to the IVC state and shows no signs of $C_3$ breaking, whereas the energy of the nematic semimetal is nearly independent of displacement field and its band structure strongly breaks $C_3$. For larger angles, panel c, $J$ first decreases then increases after the order $V^4$ terms take over. There is no transition from the nematic semimetal to the BM semimetal. For both large and small angles, the VH state eventually becomes the ground state at large displacement fields due to $O(V^4)$ terms such as the one generated by $h_{V-}$.

 This also lowers the VH energy relative to KIVC as seen in the Hartree Fock state energies Fig \ref{fig:trilayerHF}b,c. The valley polarized doping also lowers the energy of the valley polarized state but only by an amount $\propto V^6$.
 
 We note that the parameter $J$ is responsible for both the pairing and effective mass of skyrmions; the latter calculation leads to $T_c \propto J$ for the skyrmion superconductor \cite{Khalaf2020}. We therefore conclude that for smaller angles superconductivity should be enhanced at small displacement fields before eventually vanishing due to a transition out of the strong coupling regime or to a valley Hall parent state. This matches the behavior in Refs.~\onlinecite{Park2021TTG,Hao2021TTG} where superconductivity is first enhanced and then destroyed by the displacement field; note also that displacement-field-induced resistance peaks are consistent with the appearance of a gapped VH ground state as discussed in Ref.~\onlinecite{ChristosTTG}.

\subsection{$n=4$ and beyond}
\label{subsec:extfields_tqgandbeyond}

For $n=4$ ATMG the external fields contribute both as intra-TBG external fields as well as tunneling terms between the two TBG subsystems. Each effect may be handled as in $n=2,3$ respectively. We label the occupation of the magic TBG subsystem with $Q_c$ and use $Q_f$ for the non-magic subsystem.

We first discuss the intra-TBG effect: it leads to the same term
$\frac{\lambda_B}{2} \tr (Q_c \eta_z \gamma_x)^2$ but with a different coefficient $\lambda_B = c_B r_k^2 g_{\text{orb}}^2 B^2$. 
The coefficient $r_1 \approx 0.05$ strongly suppresses this effect and multiplies the critical field for valley Hall order by around ten.

The inter-TBG superexchange may be computed in a very similar way to the trilayer graphene superexchange; the main difference is that there are Dirac points at \emph{both} $\bK_\zeta$ points in each valley in non-magic TBG and so the final result is more symmetric. For four layers we have
\begin{equation}
\begin{aligned}
    E(Q_c, Q_f) & = \frac{1}{2}\sum_{\bk} \frac{1}{\abs{h_f(\bk)} + U_\bk} \tr(Q_c \xi_{\bk \eta} Q_{f \eta \bk} \xi^\dag_{\bk \eta}), \\
    (\xi_{\bk \eta})_{\gamma \gamma'} & = \frac{2}{\sqrt{5}} \left(\frac{V}{3}\delta_{\gamma \gamma'} + \eta g_{\rm orb} \mu_B \bB \times \bgamma_{\gamma \gamma'}\right)t_{\bk \eta \gamma}, \\
    t_{\bk \eta \gamma} & = \bra{u_{c\bk \eta \gamma}}\ket{u_{f \bk \eta \gamma}} .
    \end{aligned}
    \label{maintext_superex_4lay}
\end{equation}
Note that here the non-magic subsystem develops a gap when $Q_c$ describes an IVC state as well. If the TBG subsystem develops a gap at charge neutrality, as some two layer devices do \cite{Efetov, EfetovScreening}, then we expect that the displacement field will quickly induce a gap in the non-magic sector as well and the entire four layer device would be gapped at charge neutrality for $V \neq 0$. The displacement field does not favor valley diagonal states for four layers because all correlated states can transfer their order to the non-magic TBG. Indeed, the result \eqref{maintext_superex_4lay} is invariant under rotations of the valley pseudospin $\eta$. Hence it cannot favor valley-diagonal states over IVC states. Note, however, that the displacement field couples to the non-magic subsystem directly with a not-small coefficient. Its projection is proportional to $\eta_z$ and is odd under $\bk \to -\bk$. In practice the mass generation wins out at large enough displacement fields, but the IVC masses are slightly favored because they anticommute with this dispersion and can therefore always open up a gap. We do not expect any transition to the VH state for four layer structures. 

We may interpret \eqref{maintext_superex_4lay} as an effective dispersion for the magic TBG subsystem. We find that the correction to $J$ is around fifty percent larger for four layers. There is no $O(V^4)$ valley-odd superexchange for four layers.

For larger $n$ the intra-MATBG superexchange may largely be ignored because the coefficient $r_1 \approx -\frac{-\pi^2}{2n^3}$ decays very quickly. The inter-subsystem superexchange remains however, and the mass generation is in general more complex for higher $n$ because the non-magic subsystems begin to tunnel into each other as well. This is discussed explicitly in the supplemental material for $n=5$ \ref{Suppsubsec:higher_n_superexchange}. The dispersion generation for the magic sector is qualitatively similar.

\section{Weak coupling theory of pair-breaking}
\label{sec:weakcoupling}
In this section, we switch perspective and consider a weak coupling approach to superconductivity. While the strong coupling approach discussed so far is potentially relevant in the limit of small doping (relative to $\nu = \pm 2$), for sufficiently large doping we adopt 
a phenomenological weak coupling perspective to understand superconductivity as a Fermi surface instability. It turns out this approach makes the effect of in-plane magnetic field very transparent.

\subsection{$n=2$}
\label{subsec:tbgweakcoupling}
We will start by discussing the case of TBG ($n = 2$) before considering the case of general $n$. We start by assuming that the $\nu = \pm 2$ is spin-polarized. Flavor polarization is suggested by Hall data measurements \cite{PabloMott, PomeranchukYoung} as well as the observation of Cascade transitions \cite{CascadeShahal, CascadeYazdani}. In fact, the ground state in the strong coupling limit is a spin-polarized version of the KIVC as shown in Ref.~\cite{KIVCpaper}. Without including the intervalley Hund's coupling, the spin polarization can be chosen independently in the two valleys. However, once intervalley Hund's coupling $J_H$ is included, it selects a spin ferromagnet where the spin direction in the two valleys is aligned if $J_H < 0$ and a spin-valley locked state where the spin direction in the two valleys is anti-aligned if $J_H > 0$ \cite{KIVCpaper}. Once this state is doped, the doped electrons only fill one spin species in each of the valleys and pairing takes place between these doped species. This implies that, regardless of the pairing mechanism, we expect spin-triplet pairing for $J_H < 0$ and a mixture of singlet and triplet pairing for $J_H > 0$ \cite{Symmetry4e, LakeSenthil}. Only the latter exhibits a finite temperature BKT transitions making it the more likely scenario experimentally \cite{Symmetry4e, LakeSenthil}. This scenario is also more compatible with the observed decrease of the $\nu = 2$ insulating gap with in-plane magnetic field \cite{Dean-Young}.

In either case, we can think of pairing that takes place essentially between spinless electrons. Since the KIVC state only preserves a Kramers time-reversal symmetry $\T'$ \cite{KIVCpaper} in addition to $C_2$ and $C_3$, the effective KIVC Hamiltonian close to $\nu = \pm 2$ (see supplemental material for details) is invariant under the spinful representations of wallpaper group $p6$ with valley playing the role of spin:
\begin{gather}
    \T' = \tau_y \K, \qquad C'_2 = i \sigma_x \tau_x, \quad C'_3 = - e^{\frac{2\pi i}{3}\sigma_z \tau_z} \\
    {\T'}^2 = {C'_2}^2 = {C'_3}^3 = -1, \quad [\T',C'_2] = [\T',C'_3] = [C'_2,C'_3] = 0
\end{gather}
Here, $\T'$ is the Kramers time-reversal symmetry which combines spinless time-reversal $\T = \tau_x \K$ with $\pi$ $\U(1)$ vally rotation which remains a symmetry of the KIVC state \cite{KIVCpaper}. This symmetry ensures the degeneracy of the KIVC bands at the $\Gamma$ point but the bands are in general non-degenerate away from this point. Thus, upon doping, we obtain two $\T'$-symmetric Fermi surfaces (FS) 
(see Ref.~\cite{kozii2020}). For the intermediate doping regime of interest, the two FSs are sufficiently far apart in momentum which justifies the assumption that pairing takes place independently in each FS.  

Thus, we can focus on a single FS. The pairing amplitude between $\T'$ related electrons is defined as
\begin{equation}
    \Delta(\bk) = \langle c_\bk (\T' c_\bk \T'^{-1}) \rangle
\end{equation}
An important subtlety to note here is that, when expressed in the basis of the KIVC band, the operator $\T'$ has to be momentum-dependent to satisfy the relation $\T'^2 = -1$ (this is similar to what happens when considering the implementation of spinful time-reversal at the surface of a topological insulator). The same consideration also applies for $C'_2$ such that
\begin{equation}
    \T' c_\bk {\T'}^{-1} = e^{i \chi_\bk} c_{-\bk}, \quad C'_2 c_\bk {C'_2}^{-1} = e^{i \phi_\bk} c_{-\bk}, 
\end{equation}
where $\chi_\bk - \chi_{-\bk} = \phi_\bk + \phi_{-\bk} = \pi$. This yields the pairing function
\begin{equation}
    \Delta(\bk) = e^{i \chi_\bk} \langle c_\bk c_{-\bk} \rangle.
\end{equation}
which satisfies $\Delta(-\bk) = \Delta(\bk)$. This pairing function has even parity under $C'_2$ since
\begin{align}
    C'_2 \Delta(\bk) {C'_2}^{-1} &= e^{i\chi_\bk} \langle (C'_2 c_\bk {C'_2}^{-1}) (C'_2 c_{-\bk} {C'_2}^{-1}) \rangle \nonumber \\
    &= e^{i \chi_\bk} e^{i (\phi_\bk + \phi_{-\bk})} \langle c_{-\bk} c_\bk \rangle = \Delta(\bk)
\end{align}
where the minus sign from exchanging $c_\bk$ and $c_{-\bk}$ cancels against the minus sign from 
$e^{i (\phi_\bk + \phi_{-\bk})}$. Note that we could have defined $\Delta(\bk)$ without the extra phase factor in which case we would get $\Delta(-\bk) = -\Delta(\bk)$ but the action of $C'_2$ will be unchanged. This means that $\Delta(\bk)$ can be decomposed into even angular momentum channels and since we only have $C_6$ symmetry (rather than contineous rotation), we are left with the channels corresponding to angular momenta $l = 0$ ($s$-wave) and $l = \pm 2$ ($d$-wave).

In weak coupling, $T_c$ is determined from the strength of the pairing interaction in the different channels $V_l$, $l = 0, \pm 2$ with time-reversal symmetry enforcing $V_2 = V_{-2}$. Without specifying the origin of pairing, we cannot determine the values of $V_{0,2}$ and which one is larger. Assuming at least one of $V_0$ and $V_2$ is attractive and denoting the larger of $V_0$ and $V_2$ by $V_{\rm max}$, the BCS formula gives $T_c = \Lambda e^{-1/N(0) V_{\rm max}}$ where $\Lambda$ is some cutoff and $N(0)$ is the density of states at the FS.

\begin{figure}
    \centering
    \includegraphics[width = 0.38 \textwidth]{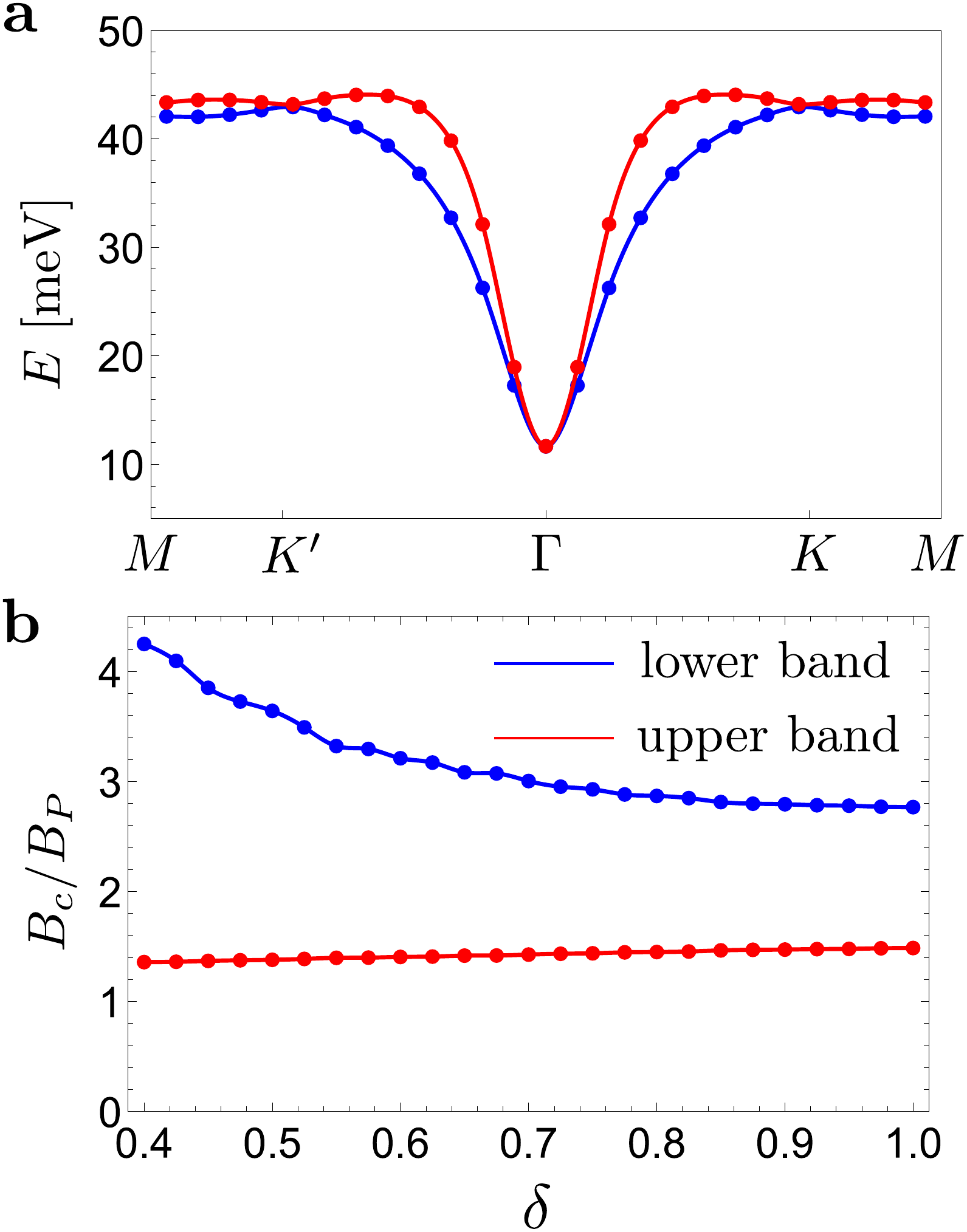}
    \caption{(a) The top two bands for the IVC bands relevant for electron doping the KIVC state close to half-filling $\nu = 2 + \delta$ at $\kappa = 0.7$. (b) The ratio of the critical field due to orbital in plane field for intraband superconductivity in the upper/lower KIVC band, $B_c$, and the Pauli field, $B_P$, as a function of doping away from half-filling $\delta$.}
    \label{fig:Bc}
\end{figure}

We would now like to understand the effect of in-plane orbital field on such superconductor. In the following, we will ignore Zeeman field which for the effectively spinless superconductor considered here does not act by pair breaking \footnote{More precisely, it has no effect for the ferromagnetic case and it leads to canting of the spins in the spin-valley locked case without affecting the pairing.}. The effect of the in-plane field is obtained by projecting the term $\frac{g_{\text{orb}}}{2} \mu_B \mu_z \hat z \times \bB \cdot \bsigma$ onto the KIVC bands:
\beq
H_\bB = \mu_B \bg_{n,\bk} \cdot \bB, \qquad \bg_{n,\bk} = \frac{g_{\text{orb}}}{2} \langle u_{n,\bk}| \mu_z (\sigma_y, -\sigma_x \tau_z) | u_{n, \bk} \rangle
\label{gnk}
\eeq
where $n$ labels to the two KIVC bands. 
$\T'$ implies that $\bg_{n,-\bk} = -\bg_{n,\bk}$ which means that this term acts as pair breaking leading to the loss of superconductivity at $\mu_B B_c \sim T_c$ which is of the same order as the Pauli field. Remarkably, the critical in-plane orbital field turns out to be independent of the strength of the interaction as long as $V_0 \neq V_2$ and given by the simple expression (see supplemental material for details)
\begin{equation}
    \frac{B_c}{B_p} = 3.5 
    \frac{1}{\sqrt{\gamma_0}}, \qquad
    \gamma_0 = \int_{\rm FS} \frac{d\phi_\bk}{2\pi} (g_{x,\bk}^2 + g_{y,\bk}^2) 
\end{equation}
The value of $B_c/B_p$ is shown in Fig.~\ref{fig:Bc} for the two FSs. We see that it only depends on the doping and the FS where pairing takes place. In addition, it is always a number of order 1 and is extremely close to 1 when pairing takes place in the second (smaller) FS. Thus as far as response to in-plane field is concerned, this superconductor will behave very similarly to a spin-singlet Pauli limited superconductor.

\subsection{$n=3$}
\label{subsec:ttgweakcoupling}
The situation is very different for the trilayer case. Here, we note the the magnetic field is odd under both mirror symmetry $M_z$ and Kramers time-reversal $\T'$ \cite{MacdonaldInPlane} which means that the combination of the two $M_z \T'$ remains unbroken in the presence of in-plane magnetic field. This symmetry ensures that in the $M_z = +1$ sector, where the KIVC order resides, Kramers time-reversal symmetry remains unbroken and pairing between $\T'$-related electrons is unaffected. Our argument here is similar to the one of Ref.~\cite{MacdonaldInPlane} with the main difference being that we consider pairing on top of KIVC order rather than intervalley pairing with unbroken valley $\U(1)$ symmetry.

The absence of pair-breaking effect of in-plane field relies on $M_z$ symmetry. This symmetry is broken in the presence of vertical displacement field which couples the two mirror sectors. However, we note here that we expect this pair breaking effect to remain small as long as the displacement field is small. The reason is that any pair breaking term should arise from a process where an electron tunnels from the KIVC band to the Dirac band and back to the KIVC band. For a given point on the KIVC Fermi surface, such term will be suppressed by $\frac{V}{\hbar v_F k_{\rm min}}$ where $k_{\rm min}$ is the minimum distance to the Dirac Fermi surface. The two Fermi surfaces may overlap at a few points but for most of the KIVC fermi surface, the distance $k_{\rm min}$ is of the order of the size of the Moire BZ which yields a denominator $\hbar v_F k_{\rm min} \sim 100$ meV. Thus, for displacement fields smaller than this value, we do not expect a significant pair breaking effect due to in-plane field. We note that our analysis here is also compatible with that of Ref.~\cite{MacdonaldInPlane} which only observed pair-breaking due to in-plane field for relatively large values of vertical displacement field.

In conclusion, if we assume the pairing has the same structure in TBG ($n=2$) and trilayer ($n=3$), we find that superconductivity in the latter will survive to much higher in-plane magnetic field than the former. The reason is that, while both are expected to survive the Zeeman part of the field, TBG will experience relatively strong orbital component which destroys superconductivity at values close to the Pauli limit. This result is fully consistent with the experimentally observed behavior where superconductivity is destroyed by an in-plane field of the order of the Pauli field in TBG \cite{PabloTBGNematicity} whereas it persists to much higher fields in TTG \cite{PabloTrilayerInPlane}.

\subsection{General $n$}
\label{subsec:any_n_weakcoupling}
For even $n$, we expect an orbital effect of the magnetic field for each TBG subsystem similar to that case of $n = 2$. However, note that for the first TBG block (the one with the largest magic angle), the in-plane field enters the Hamiltonian multiplied by a very small prefactor. For example, as discussed in Sec.~\ref{sec:BandStructure}, the effect of in-plane magnetic field on the first TBG block in tetralayer graphene ($n = 4$) is scaled by a factor $|r_1| \approx 0.05$ which will give rise to a pair breaking field that is ten times larger than that for TBG. Thus, although symmetry-wise the case of even $n$ is similar to TBG rather than trilayer, practically, we expect superconductivity in the first TBG block (at the first magic angle) to be very robust to in-plane field similar to the trilayer case.

For odd $n$, we can apply the same reasoning as in the trilayer case to deduce that the in-plane field will not have a pair-breaking effect at zero displacement field and only a weak pair-breaking effect at finite but small field.

\section{Conclusions}

In this manuscript we have developed a quantitative and controlled analysis of the ATMG systems that enable us to compare and contrast these multilayer systems with $n=2$ MATBG. While the low energy behavior of ATMG is dominated by the magic sector, the different couplings of external fields and different levels of lattice relaxation give ATMG much more tunability than twisted bilayer graphene. Such increased tunability will be vital for stress-testing theories of correlated physics. 

Let us conclude by highlighting the predictions made by our theory.  First, we predict that strain-free four layer systems can be gapped at charge neutrality with small displacement fields. Second, ATMG samples with $n>2$ and larger (smaller) than-magic angles can see superconductivity weakened (strengthened) by small displacement fields. To see an appreciable suppression of $J$ and $T_c$, one may need $\theta \gtrsim 1.63^\circ$ for TTG. Third, in a parallel magnetic field superconductivity can exceed the Pauli limit for $n>2$, when tuned to their first magic angle. Fourth, we predict a novel semimetallic state for $n=2$ MATBG, the magnetic semimetal (MSM - related to but distinct from the nematic semimetal), which is stabilized by an in-plane magnetic field. 
Finally, our lattice relaxation calculation additionally points to $n=5$ ATMG tuned to their second magic angle, $\theta = 1.14^o$(which happens to be close to the {\em first} magic angle of MATBG) as a promising host for zero-field FCIs.

\section{Acknowledgements}
We thank Tomohiro Soejima, Nick Bultinck, Michael Zaletel, and Daniel Parker for insightful discussions, collaborations on related projects, and development of the Hartree Fock codebase.
We additionally thank Tomohiro Soejima for comprehensive and patient help in performing Hartree Fock simulations.
Patrick Ledwith was partly supported by the Department of Defense (DoD) through the National Defense Science and Engineering Graduate Fellowship (NDSEG) Program. 
Eslam Khalaf was supported by the German National Academy of Sciences Leopoldina through grant LPDS 2018-02 Leopoldina fellowship. 
Ziyan Zhu is supported by the STC Center for Integrated Quantum Materials, NSF Grant No. DMR-1231319, ARO MURI Grant No. W911NF14-0247, and NSF DMREF Grant No. 1922165. 
Stephen Carr was supported by the National Science Foundation under grant No. OIA-1921199.
Ashvin Vishwanath was supported by a Simons Investigator award and by the Simons Collaboration on Ultra-Quantum Matter, which is a grant from the Simons Foundation (651440,  AV).
\bibliography{sources}

\pagebreak
\widetext
\begin{center}

\textbf{\large Supplemental material}
\end{center}
%%%%%%%%%% Merge with supplemental materials %%%%%%%%%%
%%%%%%%%%% Prefix a "S" to all equations, figures, tables and reset the counter %%%%%%%%%%
\setcounter{equation}{0}
\setcounter{figure}{0}
\setcounter{table}{0}
\setcounter{page}{1}
\makeatletter
\renewcommand{\theequation}{S\arabic{equation}}
\renewcommand{\thefigure}{S\arabic{figure}}
\renewcommand{\thesection}{S\Roman{section}}
\renewcommand{\bibnumfmt}[1]{[S#1]}
\setcounter{section}{0}

\section{Review of the mapping twisted multilayer graphene to TBG}
\label{SuppSec:mappingreview}

The alternating twist multilayer graphene (ATMG) is defined by stacking $n$ graphene layers such that the relative twist angle between layers have equal magnitude by alternating signs i.e. $\theta_l = (-1)^l \theta/2$, $l = 1,\dots, n$. In general, this generates $n-1$ Moir\'e patterns which have the same wavelength but may be displaced relative to each other. We restrict ourselves to the case where such relative displacements vanish causing these Moir\'e patterns to align. This configuration was shown in Ref.~\cite{Carr2020} to be the energetically most stable configuration for $n=3$ and this conclusion is likely to hold for general $n$. In the following, we will always assume the case where the Moir\'e patterns are all aligned.

The Hamiltonian for ATMG with $n$ layers for a single flavor (single spin-valley)  is given by
\beq
H(\br) = \left(\begin{array}{ccccc}
-i v_F \bsigma_+ \cdot \nabla  & T(\br) & 0 & 0 & \dots \\
T^\dagger(\br) & -i v_F \bsigma_- \cdot \nabla & T^\dagger(\br) & 0 & \dots \\
0 & T(\br) & -i v_F \bsigma_+ \cdot \nabla  & T(\br) \\
0 & 0 & T^\dagger(\br) & -i v_F \bsigma_- \cdot \nabla & \dots \\
\dots & \dots & \dots & \dots & \dots
\end{array} \right)
\label{HATMGsupp}
\eeq
where $\bsigma_\pm = e^{\pm \frac{i}{4} \theta \sigma_Z} \bsigma e^{\mp \frac{i}{4} \theta \sigma_z}$ and $T(\br)$ is given by
\begin{gather}
    T(\br) = \left(\begin{array}{cc}
w_0 U_0(\br) & w_1 U_1(\br) \\
w_1 U^*_1(-\br) & w_0 U_0(\br)
\end{array} \right), \qquad U_m(\br) = \sum_{n=1}^3 e^{\frac{2 \pi i}{3} m (n - 1)} e^{-i \bq_n \cdot \br}, \\ \bq_n = 2 k_D \sin \left(\frac{\theta}{2}\right) R_{\frac{2 \pi (n-1)}{3}}(0,-1)
\end{gather}
where $R_\phi$ denotes counter clockwise rotation by $\phi$ and  $k_D = \frac{4\pi}{3 \sqrt{3} a_{\rm CC}}$. We would also like to consider the effect of vertical displacement field and in-plane magnetic field given by
\beq
H^U_{l,l'} = U \delta_{l,l'} \left[\frac{l-1}{n-1} - \frac{1}{2}\right], \qquad H^B_{l,l'} = g_{\text{orb}} \mu_B \delta_{l,l'} \bsigma_{(-1)^{l-1}} \cdot (\hat z \times \bB_{\parallel}) \left[l - 1 - \frac{n-1}{2}\right] 
\eeq
In the following, we will ignore the small $\theta$ rotation 
where $g_{\text{orb}} = \frac{e v_F d}{\mu_B}$ where $d \approx 3.5 A^o$ is the interlayer separation. For $v_F \approx 10^6$ m/s, $g_{\text{orb}} = 6.14$. We notice that the external fields only enter the Hamiltonian in the combinations
\beq
\Gamma_\pm = \frac{U}{n-1} + g_{\text{orb}} \mu_B \bsigma_\pm \cdot (\hat z \times \bB_{\parallel})
\eeq
Note that in the limit of small angles, we can ignore the rotation in $\bsigma_\pm$ and take $\bsigma_\pm \approx \bsigma$ which leads to $\Gamma_\pm = \Gamma = \frac{U}{n-1} + g_{\text{orb}} \bsigma \cdot (\hat z \times \bB_{\parallel})$. We will however prefer to keep this rotation for the sake of generality. We can then write the effect of displacement field and in-plane magnetic field as
\beq
H^{U+B}_{ll'} = \delta_{ll'} \Gamma_{(-1)^{l-1}} \left[l - 1 - \frac{n-1}{2}\right].
\eeq

A remarkable result proved in Ref.~\cite{Khalaf2019} is that the Hamiltonian (\ref{HATMGsupp}) maps to $m$ copies of TBG with rescaled tunneling matrices $T$ for $n = 2m$ and to $m$ copies of TBG (also with rescaled tunneling matrices $T$) in addition to a single Dirac cone for $n = 2m + 1$. To understand this mapping, we follow Ref.~\cite{Khalaf2019} and perform the transformation $\tilde H = P^T H P$ with $P$ defined as
\beq
P = \sum_{k=1}^{n_e} \delta_{2k, n_o + k} + \sum_{k=1}^{n_o} \delta_{2k-1, k}
\label{Pk}
\eeq
where $n_{e/o}$ simply denote the number of layers with even/odd index (assuming layers are number from 1 to $n$) which are given by
\beq
n_e = \lfloor n/2 \rfloor, \qquad n_o = \lceil n/2 \rceil
\eeq
This transformation simply relabels the layers such that the Hamiltonian has a block matrix structure in the even-odd layer index
\beq
\tilde H = P^T H P = \left( \begin{array}{cc} -i v_F \bsigma_+ \cdot \nabla & T(\br) \otimes W \\
T^\dagger(\br) \otimes W^\dagger & -i v_F \bsigma_- \cdot \nabla \end{array} \right), 
\eeq
with $W$ being an $n_o \times n_e$ describing the structure of the tunneling between layers. Since tunneling only takes place between nearest neighboring layers, $W$ has the simple form $W_{ij} = \delta_{i,j} + \delta_{i,j+1}$. The mapping can be established by writing the singular value decomposition for $W = A \Lambda B^T$ which implies that
\beq
H' = \left( \begin{array}{cc} A^T & 0 \\ 0 & B^T \end{array} \right) \tilde H \left( \begin{array}{cc} A & 0 \\ 0 & B \end{array} \right) = \left( \begin{array}{cc} -i v_F \bsigma_+ \cdot \nabla & T(\br) \otimes \Lambda \\
T^\dagger(\br) \otimes \Lambda^\dagger & -i v_F \bsigma_- \cdot \nabla \end{array} \right)
\label{Mapping}
\eeq
Here, $\Lambda$ is an $n_o \times n_e$ matrix with non-vanishing entries only along the diagonal which we denote by $\lambda_k$ where $k = 1, \dots, n_e$. For even $n$, $n_e = n_o = n/2$ and this mapping means that the Hamiltonian $H'$ decomposes into a sum of $n/2$ TBG Hamiltonians with the tunneling $T(\br)$ rescaled by $\lambda_k$:
\begin{equation}
    H^{\rm even}_{\rm dec} = P H'_{\rm even} P^T = \left(\begin{array}{ccccc}
     -i v_F \bsigma_+ \cdot \nabla & \lambda_1 T(\br) & 0 & 0 & \dots \\
    \lambda_1 T^\dagger(\br) & -i v_F \bsigma_- \cdot \nabla & 0 & 0 & \dots \\
    0 & 0 & -i v_F \bsigma_+ \cdot \nabla & \lambda_2 T(\br) & \dots \\
    0 & 0 & \lambda_2 T^\dagger(\br) & -i v_F \bsigma_- \cdot \nabla & \dots \\
    \dots & \dots & \dots & \dots & \dots  
    \end{array}\right)
\end{equation}
For odd $n$, the Hamiltonian $H'$ reduces to $(n-1)/2$ TBG Hamiltonians with the tunneling $T(\br)$ rescaled by $\lambda_k$ in addition to a single Dirac cone that is decoupled from the rest of the system:
\begin{equation}
    H^{\rm odd}_{\rm dec} = P H'_{\rm odd} P^T = \left(\begin{array}{cccccc}
     -i v_F \bsigma_+ \cdot \nabla & \lambda_1 T(\br) & 0 & 0 & \dots & \dots \\
    \lambda_1 T^\dagger(\br) & -i v_F \bsigma_- \cdot \nabla & 0 & 0 & \dots & \dots \\
    0 & 0 & -i v_F \bsigma_+ \cdot \nabla & \lambda_2 T(\br) & \dots & \dots \\
    0 & 0 & \lambda_2 T^\dagger(\br) & -i v_F \bsigma_- \cdot \nabla & \dots & \dots \\
    \dots & \dots & \dots & \dots & \dots & \dots \\
    \dots & \dots & \dots & \dots & \dots & -i v_F \bsigma_+ \cdot \nabla
    \end{array}\right)
\end{equation}
The general unitary matrix which maps $H$ to a decoupled sum of TBGs possibly in addition to a Dirac cone has the explicit form 
\beq
V = P \left(\begin{array}{cc}
A & 0 \\ 0 & B \end{array} \right) P^T, \quad \implies \quad H_{\rm dec} = V^T H V
\eeq
The eigenvalues $\lambda_k$ as well as the matrices $A$ and $B$ can be obtained by finding the eigenvalues and eigenvectors of $W^T W$ and $W W^T$
\begin{gather}
W^T W u_k = \lambda_k u_k, \qquad k=1, \dots, n_e \nonumber \\
W W^T v_k = \lambda_k v_k \qquad k=1, \dots, n_o
\end{gather}
It is straightforward to verify that
\begin{gather}
    \lambda_k = 2 \cos \frac{\pi k}{n+1},\\ u_{k,m} = \sin \left(\frac{\pi}{n+1} m (2k - \delta_{n \, {\rm mod} \, 2, 0})\right) , \qquad m = 1, \dots, n_e \\
    v_{k,m} = \sin \left(\frac{\pi}{n+1} k (2m - \delta_{n \, {\rm mod} \, 2, 1})\right) \qquad m = 1, \dots, n_o
\end{gather}
The normalization of the wavefunctions is given by
\beq
\beta_k = \left(\sum_{m=1}^{n_e} |u_{k,m}|^2 \right)^{-1/2} = \frac{2}{\sqrt{n+1}}, \qquad \chi_k = \left(\sum_{m=1}^{n_o} |v_{k,m}|^2 \right)^{-1/2} = \frac{2}{\sqrt{n+1}} \begin{cases}
\frac{1}{\sqrt{2}} &: k = n_e + 1 \\
1 &: \text{otherwise}
\end{cases}
\eeq
This leads to the explicit expressions for $A$ and $B$ as
\beq
A_{km} = \chi_k v_{k,m}, \qquad B_{km} = \beta_k u_{k,m}
\eeq

The effect of vertical displacement field and in-plane magnetic field in the decoupled picture can be simply obtained via
\beq
H^{U+B}_{\rm dec} = V^T H^{U+B} V
\eeq
In the following, let us consider some explicit cases.

\subsection{Trilayer $n=3$}
The simplest example of ATMG corresponds to the trilayer case with $n = 3$. The Hamiltonian for this case is given explicitly by
\beq
H(\br) = \left(\begin{array}{ccc}
-i v_F \bsigma_+ \cdot \nabla  & T(\br) & 0 \\
T^\dagger(\br) & -i v_F \bsigma_- \cdot \nabla & T^\dagger(\br) \\
0 & T(\br) & -i v_F \bsigma_+ \cdot \nabla \end{array} \right)
% \label{HK}
\eeq
The decoupling matrix $V$ can be obtained easily following the general procedure explains above. Its explicit form is given by
\beq
    V = P \left(\begin{array}{cc}
A & 0 \\ 0 & B \end{array} \right) P^T = \left(\begin{array}{ccc}
\frac{1}{\sqrt{2}} & 0 & -\frac{1}{\sqrt{2}}  \\
0 & 1 & 0 \\
\frac{1}{\sqrt{2}} & 0 & \frac{1}{\sqrt{2}} \end{array} \right)
\eeq
The full Hamiltonian in the decoupled basis has the form
\beq
 H^{\rm Full}_{\rm dec} =  \left(\begin{array}{ccc}
-i v_F \bsigma_+ \cdot \nabla  & \sqrt{2}  T(\br) & \Gamma_+ \\
\sqrt{2} T^\dagger(\br) & -i v_F \bsigma_- \cdot \nabla & 0 \\ 
\Gamma_+ & 0 & -i v_F \bsigma_+ \cdot \nabla \end{array} \right)
\eeq
We see that the Hamiltonian maps to a direct sum of a TBG Hamiltonian with the interlayer tunneling scaled by $\sqrt{2}$ and a single Dirac cone. The two are coupled through the displacement field $U$ and the in-plane field $\bB$.

\subsection{Tetralayer $n=4$}
We can similarly work out the Hamiltonian for the tetralayer case $n = 4$. The Hamiltonian for this case is given explicitly by
\beq
H(\br) = \left(\begin{array}{cccc}
-i v_F \bsigma_+ \cdot \nabla  & T(\br) & 0 & 0\\
T^\dagger(\br) & -i v_F \bsigma_- \cdot \nabla & T^\dagger(\br) & 0 \\
0 & T(\br) & -i v_F \bsigma_+ \cdot \nabla & T(\br) \\
0 & 0 & T^\dagger(\br) & -i v_F \bsigma_- \cdot \nabla 
\end{array} \right)
% \label{HK}
\eeq
The decoupling matrix $V$ is
\beq
    V = \frac{2}{\sqrt{5}} \left(\begin{array}{cccc}
\sin \frac{\pi}{5} & 0 & -\sin \frac{2\pi}{5} & 0 \\
0 & \sin \frac{2\pi}{5} & 0 & -\sin \frac{\pi}{5}\\
\sin \frac{2\pi}{5} & 0 & \sin \frac{\pi}{5} & 0 \\
0 & \sin \frac{\pi}{5} & 0 & \sin \frac{2\pi}{5}\end{array} \right)
\eeq
leading to the Hamiltonian
\beq
    H^{\rm Full}_{\rm dec} =  \left(\begin{array}{cccc}
    -i v_F \bsigma_+ \cdot \nabla - r_1 \Gamma_+  & \varphi T(\br) & \frac{2}{\sqrt{5}} \Gamma_+ & 0 \\
\varphi T^\dagger(\br) & -i v_F \bsigma_- \cdot \nabla + r_1 \Gamma_- & 0 & \frac{2}{\sqrt{5}} \Gamma_- \\
\frac{2}{\sqrt{5}} \Gamma_+ & 0 & -i v_F \bsigma_+ \cdot \nabla - r_2 \Gamma_+ & \frac{1}{\varphi} T(\br) \\
0 & \frac{2}{\sqrt{5}} \Gamma_- & \frac{1}{\varphi} T^\dagger(\br) & -i v_F \bsigma_- \cdot \nabla + r_2 \Gamma_-
\end{array} \right) 
\eeq
where $\varphi$ is the golden ratio $\frac{1 + \sqrt{5}}{2}$ and $r_{1,2} = \frac{2\sqrt{5} \mp 5}{10}$.
Thus, the tetralayer system maps to a pair of TBGs whose interlayer tunneling is scaled by $\varphi$ and $1/\varphi$. This yields two sequences of magic angles obtained from TBG magic angles either by multiplying or dividing by the golden ratio $\varphi$ \cite{Khalaf2019}. This mapping can also be understood in terms of bonding and antibonding orbitals as follows. The first TBG (with $T$ rescaled by $\varphi$) is built using the bonding orbitals between the first and third (layer 1) and second and fourth (layer 2) whereas the second TBG (with $T$ rescaled by $1/\varphi$) is built using the antibonding orbitals between the first and third (as layer 1) and second and fourth (as layer 2). Including vertical displacement field and/or in-plane magnetic field leads to a coupling between the two TBGs. In addition, it also introduces an effective displacement field and in-plane magnetic field in each of the two TBGs that are rescaled by $\frac{1}{3} r_{1,2}$ and $r_{1,2}$, respectively, relative to the corresponding fields applied for the four layer system.

We notice that the scaling ratio for the external fields relevant to the first TBG block is very small $r_1 \approx -0.05$. This smallness can be explained by explicitly computing the parameter $r_k$ for $n = 2p$ which is given by
\begin{eqnarray}
r_k = -\frac{4}{2p + 1} \sum_{m=1}^p (2m - 1/2 - p) \sin^2 \frac{2 \pi k m}{2p+1} = - \frac{3 + \cos \frac{4 k \pi}{2p+1} - 4 \cos \frac{4 k \pi (p+1)}{(2p + 1)}}{4 (2p + 1) \sin^2 \frac{2 \pi k}{2p + 1} } \approx -\frac{\pi^2 k^2}{16 p^3} + O(p^{-4})
\end{eqnarray}
For $k \ll p$ and $p \gg 1$, this implies that $r_k$ is very small. In particular, this means that $r_1$, relevant for the first TBG block where the first magic angle is realized, is always small for any even number of layers $n = 2p$ with $p > 1$ due to the very rapid decrease of $r_1$ with $p$. 

\subsection{$n=5$}
Finally, let us consider the case $n = 5$. The Hamiltonian for this case is given explicitly by
\beq
H(\br) = \left(\begin{array}{ccccc}
-i v_F \bsigma_+ \cdot \nabla  & T(\br) & 0 & 0 & 0\\
T^\dagger(\br) & -i v_F \bsigma_- \cdot \nabla & T^\dagger(\br) & 0 & 0 \\
0 & T(\br) & -i v_F \bsigma_+ \cdot \nabla & T(\br) & 0\\
0 & 0 & T^\dagger(\br) & -i v_F \bsigma_- \cdot \nabla & T^\dagger(\br) \\
0 & 0 & 0 & T(\br) & -i v_F \bsigma_+ \cdot \nabla 
\end{array} \right)
% \label{HK}
\eeq
The decoupling matrix $V$ is
\beq
    V = \left(\begin{array}{ccccc}
\frac{1}{\sqrt{6}} & 0 & -\frac{1}{\sqrt{2}} & 0 & \frac{1}{\sqrt{3}} \\
0 & \frac{1}{\sqrt{2}} & 0 & -\frac{1}{\sqrt{2}} & 0 \\
\frac{\sqrt{2}}{\sqrt{3}} & 0 & 0 & 0 & - \frac{1}{\sqrt{3}} \\
0 & \frac{1}{\sqrt{2}} & 0 & \frac{1}{\sqrt{2}} & 0 \\
\frac{1}{\sqrt{6}} & 0 & \frac{1}{\sqrt{2}} & 0 & \frac{1}{\sqrt{3}} \end{array} \right)
\eeq
leading to the Hamiltonian
\beq
    H^{\rm Full}_{\rm dec} =  \left(\begin{array}{ccccc}
    -i v_F \bsigma_+ \cdot \nabla   & \sqrt{3} T(\br) & \frac{2}{\sqrt{3}} \Gamma_+ & 0 & 0\\
\sqrt{3} T^\dagger(\br) & -i v_F \bsigma_- \cdot \nabla & 0 & \Gamma_- & 0 \\
\frac{2}{\sqrt{3}} \Gamma_+ & 0 & -i v_F \bsigma_+ \cdot \nabla &  T(\br) & \frac{4}{\sqrt{6}} \Gamma_+ \\
0 & \Gamma_- &  T^\dagger(\br) & -i v_F \bsigma_- \cdot \nabla & 0 \\
0 & 0 & \frac{4}{\sqrt{6}} \Gamma_+ & 0 & -i v_F \bsigma_+ \cdot \nabla
\end{array} \right) 
\eeq
Note that here, similar to the trilayer case, the displacement and in-plane fields only introduce a coupling between the different TBG/Dirac sectors without introducing an effective displacement field or magnetic field in each sector.

\section{Lattice relaxation}
\label{Suppsec:LatticeRelax}
\subsection{Summary of relaxation calculation}
\label{Suppsubsec:summaryofrelaxcalc}
Here, we provide details of the lattice relaxation model. 
We employ a continuum model based on local configuration to obtain the relaxation pattern of the multi-layered heterostructures~\cite{Carr2018relax}. 
Every real space position $\vec{r}$ can be uniquely parametrized by a local shift vector $\vec{b}$ that describes its local environment with respect to its neighboring layers. 
Letting $A_1$, $A_2$ be the matrices whose columns are the lattice vectors of layers 1 and 2 respectively and defining $A_{\delta1} = I - A_1 A_2^{-1}$ and $A_{\delta2} = I - A_2 A_1^{-1}$, the local configuration vectors are given as $\vec{b}^{(\ell)} = A_{\delta\ell} \vec{r}$ for $\ell=1,2$. Note that $\vec{b}^{(1)} = -\vec{b}^{(2)}$, and only $\vec{b}^{(1)}$ and $\vec{b}^{(2)}$ are needed because there is only a single twist angle and every other layer is aligned. 
The relaxation vectors $\vec{u}^{(\ell)}$ have a symmetry dictated by the symmetry of the system depending on whether the number of layer $n$ is even or odd. 
If $n=2k$, where $k \in \mathbb{Z}$, $\vec{u}^{(\ell)} = -\vec{u}^{(n+1-\ell)}$. If $n=2k+1$, $\vec{u}^{(\ell)} = \vec{u}^{(n+1-\ell)}$ $\forall \ell \neq \frac{n+1}{2}$. 

We minimize the total energy as a function of the relaxation displacement vectors in local configuration space, and the total energy is consisted of intralayer and interlayer terms. The intralayer term can be obtained from linear elasticity theory:
\begin{eqnarray}
E^\mathrm{intra} (\vec{u}^{(\ell)}) 
= \sum_{\ell=1}^n \int \frac{G}{2} (\partial_x u^{(\ell)}_x + \partial_y u^{(\ell)}_y)^2  
+ \frac{K}{2} 
\left \{ (\partial_x u^{(\ell)}_x - \partial_y u^{(\ell)}_y)^2 + (\partial _x u^{(\ell)}_y + \partial_y u^{(\ell)}_x)^2
\right \} 
d^2 \vec{b},
\end{eqnarray}
where $G$ and $K$ are shear and bulk modulus of a monolayer graphene, which we take to be $G = 47352 \, \mathrm{meV/unit \ cell}$, $K = 69518 \, \mathrm{meV/unit \ cell}$~\cite{Carr2018relax}. 

The interlayer energy accounts for the energy cost of the layer misalignment, which is described by generalized stacking fault energy (GSFE) ~\cite{Zhou2015}, obtained using first principles Density Functional Theory (DFT) with the Vienna Ab initio Simulation Package (VASP)~\cite{Kresse1996,Kresse1999}. The GSFE is the total energy of an untwisted bilayer per unit cell as a function of the local stacking between the adjacent layers, and with the lowest energy stacking identified as zero energy. For bilayer graphene, the GSFE is maximized at the AA stacking and minimized at the AB/BA stacking. Letting $\vec{b} = (b_x, b_y)$ be the relative stacking between two layers, we define the following unitless vector $\vec{v}= (v, w) \in [0, 2 \pi ]^2$ to describe all possible configurations:
\begin{equation}
	\begin{pmatrix} 
		v \\ w
	\end{pmatrix}
		= \frac{2 \pi}{a_0} \mqty[\sqrt{3}/2 & -1/2 \\ \sqrt{3}/2 & 1/2]
	\begin{pmatrix}
		b_x \\ b_y
	\end{pmatrix},
\end{equation}
where $a_0 = 2.4595$~\AA is the graphene lattice constant. 
We then parametrize the GSFE as follows, 
\begin{align}
V^\mathrm{GSFE} = c_0 + & c_1(\cos v + \cos w + \cos (v + w) ) \nonumber \\
+ & c_2 (\cos (v + 2w) + \cos(v-w) + \cos(2 v + w)) \nonumber \\
+ & c_3 (\cos(2 v ) + \cos (2 w) + \cos(2 v + 2 w)),\label{eqn:vgsfe}
\end{align}
where we take $c_0 = 6.832\, \mathrm{meV/cell}$, $c_1 = 4.064 \, \mathrm{meV/cell}$, $c_2 = -0.374 \, \mathrm{meV/cell}$, $c_3 = -0.0095\, \mathrm{meV/cell}$~\cite{Carr2018relax}.
These energies were derived from DFT calculations with a van der Waals force implemented through the vdW-DFT method and the SCAN+rVV10 functional~\cite{dion2004van, vydrov2010nonlocal, klimevs2011van,peng2016versatile}. In terms of $V^\mathrm{GSFE}$, the total interlayer energy can be expressed as follows:
\begin{align}
	E^\mathrm{inter} (\vec{u}^{(\ell)}) &= \frac{1}{2} \sum_{\ell=1}^{n-1} \int V^\mathrm{GSFE} (\vec{B}^{\ell \rightarrow \ell+1}) \,d^2 \vec{b},
\end{align}
where $\vec{B}^{\ell \rightarrow \ell+1} = \vec{b}^{(\ell+1)} + \bu^{(\ell+1)} - \bu^{(\ell)}$ is the relaxation modified local shift vector.
Note that we only consider the interaction energy between adjacent pairs of layers. 
The total energy is obtained by summing over uniformly sampled configuration space, justified by the ergodicity of incommensurate lattices~\cite{cazeaux2016, massatt2016}.

\begin{figure}[ht!]
    \centering
    \includegraphics[width = 0.5\textwidth]{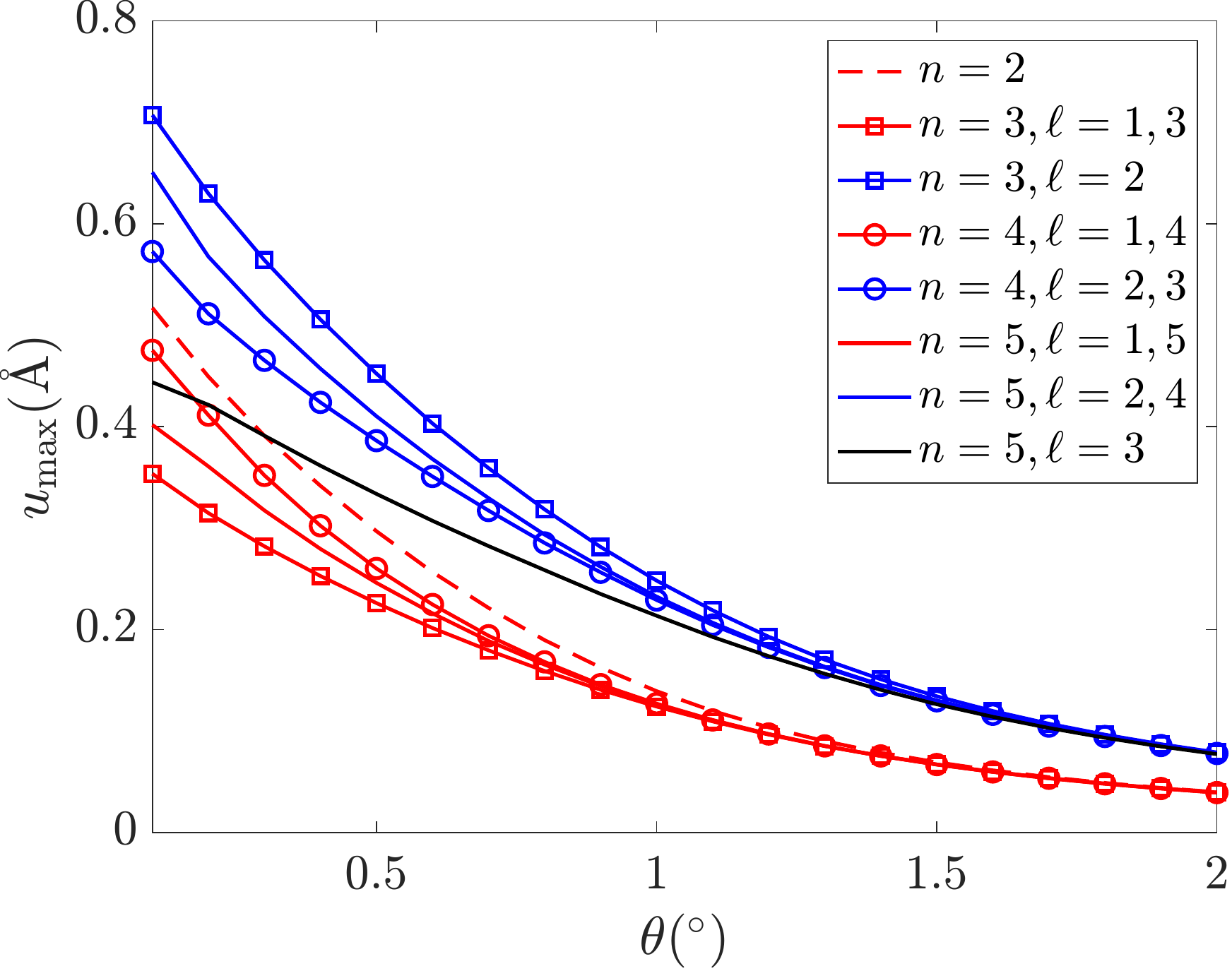}
    \caption{Maximum magnitude of the relaxation displacement vectors in layer $l$ as a function of $\theta$ in ATMG for $n=2,3,4,5$. Results for layers with one neighboring layer are plotted in red, while those with two neighbors are plotted in blue (the middle layer of the $n=5$ ATMG is plotted in black).}
    \label{fig:relaxation}
\end{figure}

Figure~\ref{fig:relaxation} shows the magnitude of relaxation as a function of $\theta$ for $n = 2,3,4,5$. For a generic $n > 2$, the relaxation magnitude has two asymptotic values, depending on the number of neighboring layers. For example, when $n=3$, layers 1 and 3 have only one adjacent layer, and layer 2 is sandwiched between the top and bottom layers. As a result, when $\theta \gtrsim 1^\circ$, layers 1 and 2 have the same magnitude as that of the bilayer graphene, and layer 2 has roughly doubled magnitude due to the doubled interlayer interaction. This can be generalized to an arbitrary $n > 2$ -- the relaxation magnitude of the middle layers is twice as that of the top and bottom layers. 
The magnitude at a smaller angle depends on the number of layers, but for the magic angles considered in this work ($\theta > 1^\circ$), this effect is negligible. 

\subsection{Effect of lattice relaxation on magic angles and the multilayer mapping}
\label{effectofrelax}
We now discuss the effects of relaxation on the magic angle values and the multilayer mapping. We start by writing a generalization for Eq.~\ref{HATMGsupp} which allows for different interlayer coupling \cite{Khalaf2019}:
\beq
H(\br) = \left(\begin{array}{ccccc}
-i v_F \bsigma_+ \cdot \nabla  & T_{12}(\br) & 0 & 0 & \dots \\
T_{12}^\dagger(\br) & -i v_F \bsigma_- \cdot \nabla & T_{32}^\dagger(\br) & 0 & \dots \\
0 & T_{32}(\br) & -i v_F \bsigma_+ \cdot \nabla  & T_{34}(\br) \\
0 & 0 & T_{34}^\dagger(\br) & -i v_F \bsigma_- \cdot \nabla & \dots \\
\dots & \dots & \dots & \dots & \dots
\end{array} \right), \qquad T_{ij}(\br) = \left(\begin{array}{cc}
w^{ij}_0(\theta) U_0(\br) & w^{ij}_1(\theta) U_1(\br) \\
w^{ij}_1(\theta) U^*_1(-\br) & w^{ij}_0(\theta) U_0(\br)
\end{array} \right)
\label{HATMG_diffCouplings}
\eeq
To define the magic angle in this case we go to the chiral limit by setting $w_0^{ij}$ to zero; away from the chiral limit, there is no unique way to define the magic angle \cite{Tarnopolsky}, so we always define the magic angle to be the one at the chiral limit. We can follow the same steps as before using the matrix $P$ from Eq.~\ref{Pk} to write the Hamiltonian in the even-odd layer basis as
\begin{equation}
     T(\br) = w_1 \left(\begin{array}{cc}
0 & U_1(\br) \\
 U^*_1(-\br) & 0
\end{array} \right), \quad W(\theta) = \frac{1}{w_1} \left(\begin{array}{cccc}
w^{12}_1(\theta) & 0 & 0 & \dots\\
w^{23}_1(\theta) & w^{34}_1(\theta) & 0 & \dots\\
0 & w^{45}_1(\theta) & w^{56}_1(\theta) & \dots \\
\dots & \dots & \dots & \dots 
\end{array} \right)
\end{equation}
where $w_1 \approx 110$ meV is the relaxation-independent value which we introduced to ensure notational consistency with the case of equal interlayer coupling considered earlier. As in the uniform case, we can introduce the singular value decomposition of $W$ and use it to map the multilayer Hamiltonian to a set of decoupled TBG Hamiltonians (plus a graphene Hamiltonian for $n$ odd). Denoting the eigenvalues of $W(\theta)$ by $\lambda_k(\theta)$ as before, we can write the magic angles for the decoupled TBG as
\begin{equation}
    \frac{\theta}{\lambda_k(\theta)} = \theta^{\rm TBG}_{\rm magic} 
\end{equation}
where $\theta^{\rm TBG}_{\rm magic} $ is the TBG magic angle. Due to the $\theta$ dependence of $\lambda_k$, this equation can only be solved numerically leading to the value in Table \ref{tab:relax} in the main text. All magic angles of interest (first magic angle for $n = 3, 4$ and second magic angle for $n = 5$) follow from taking $\theta^{\rm TBG}_{\rm magic}$ to be the first magic angle of TBG which is equal to 1.06${}^o$ assuming the $\theta$-independent value $w_1 = 110$ meV.

\begin{figure}
    \centering
    \includegraphics[width = 0.9\textwidth]{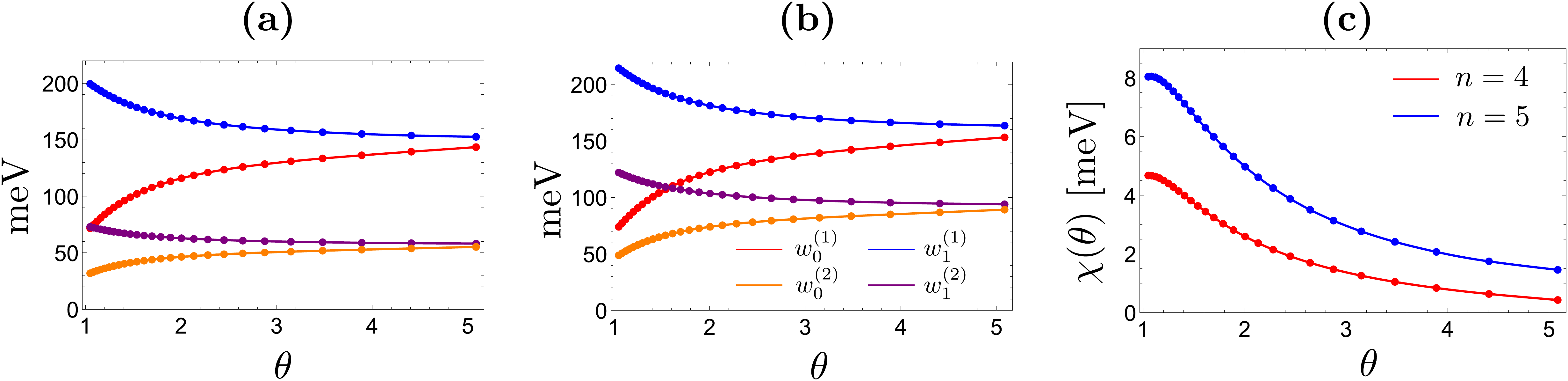}
    \caption{{\bf Mapping the relaxation parameters for $n = 4$ and $n = 5$.} (a), (b) plot the parameters $w_{0,1}^{(l)}$ defined in (\ref{Ttchi}) as a function of $\theta$ for $n = 4$ and $n = 5$, respectively. (c) plots the value of the coupling $\chi(\theta)$ between the different sectors (cf.~Eqs.~\ref{Ttchi}) as a function $\theta$ for $n = 4$ and $n = 5$. }
    \label{fig:relaxMapping}
\end{figure}

Restoring the layer and angle dependent intra-sublattice tunneling $w^{ij}_0(\theta)$, and apply the mapping in Eq.~\ref{Mapping}, we generally get Hamiltonian of TBGs sectors (plus a Dirac cone for odd $n$) with small coupling between the sectors. For $n = 4$, we can write this Hamiltonian explicitly as
\begin{gather}
    H^4_{\rm dec} = V H V^T = \left(\begin{array}{cccc}
     -i v_F \bsigma_+ \cdot \nabla & T_\theta^{(1)}(\br) & 0 & t_\theta(\br)  \\
    {[T_\theta^{(1)}(\br)]}^\dagger & -i v_F \bsigma_- \cdot \nabla & -t^\dagger_\theta(\br) & 0 \\
    0 & -t_\theta(\br) & -i v_F \bsigma_+ \cdot \nabla & T_\theta^{(2)}(\br)  \\
    t^\dagger_\theta(\br) & 0 & {[T_\theta^{(2)}(\br)]}^\dagger & -i v_F \bsigma_- \cdot \nabla 
    \end{array}\right) \\
    T_\theta^{(l)}(\br) = \left(\begin{array}{cc}
    w^{(l)}_0(\theta) U_0(\br) & w^{(l)}_1(\theta) U_1(\br) \\
    w^{(l)}_1(\theta) U^*_1(-\br) & w^{(l)}_0(\theta) U_0(\br)
    \end{array}\right), \qquad t_\theta(\br) = \left( \begin{array}{cc}
    \chi(\theta) U_0(\br) & 0 \\
    0 & \chi(\theta) U_0(\br) 
    \end{array} \right)
    \label{Ttchi}
\end{gather}
The values of $w_{0,1}^{(l)}(\theta)$ for two sectors $l = 1,2$ as well as the coupling $\chi(\theta)$ as shown in Fig.~\ref{fig:relaxMapping}. In addition, the value of the chiral ratio in each sector $\kappa^{(l)}(\theta) = \frac{w_0^{(l)}(\theta)}{w_1^{(l)}(\theta)}$ is plotted in Fig.~\ref{fig:relax} in the main text. As we can see the coupling $\chi(\theta)$ is much smaller than $w_{0,1}^{(l)}(\theta)$ and is not expected to lead to any significant qualitative differences. This coupling may also be treated perturbatively as a tunneling between sectors in a similar way to the external fields, see section \ref{Suppsubsec:higher_n_superexchange}. 

For $n = 5$, we have
\begin{gather}
    H^5_{\rm dec} = V H V^T = \left(\begin{array}{ccccc}
     -i v_F \bsigma_+ \cdot \nabla & T_\theta^{(1)}(\br) & 0 & 0 & 0  \\
    {[T_\theta^{(1)}(\br)]}^\dagger & -i v_F \bsigma_- \cdot \nabla & 0 & 0 & -t^\dagger_\theta(\br) \\
    0 & 0 & -i v_F \bsigma_+ \cdot \nabla & T_\theta^{(2)}(\br)  \\
    0 & 0 & {[T_\theta^{(2)}(\br)]}^\dagger & -i v_F \bsigma_- \cdot \nabla & 0 \\
    0 & -t_\theta(\br) & 0 & 0 & -i v_F \bsigma_+ \cdot \nabla
    \end{array}\right) 
\end{gather}
with $T_\theta^{(l)}(\br)$ and $t_\theta(\br)$ have the same definition as in \ref{Ttchi}. The values of $w_{0,1}^{(l)}(\theta)$ for two sectors $l = 1,2$ as well as the coupling $\chi(\theta)$ as shown in Fig.~\ref{fig:relaxMapping}.

\section{Intra-TBG effects of external fields}
\label{Suppsec:intraTBG}

For an even number of layers, there is no $M_z$ symmetry and so the electric and magnetic fields act within the TBG sectors, not only as tunnelings between them. To understand the intra-TBG effects of the external fields in perturbation theory it is sufficient to consider $n=2$ MATBG and then straightforwardly generalize the results to $n>2$ even.

\subsection{External fields in $n=2$ MATBG}
\label{Suppsubsec:intraTBGnis2}

We begin with the unperturbed interacting Hamiltonian after integrating out remote MATBG bands
\begin{equation}
  \begin{aligned}
    \H & = \sum_\bq V_{\rm eff}(\bq) \delta \rho_\bq \delta \rho_{-\bq}, \qquad  \delta \rho_\bq = \rho_\bq - \ov{\rho}_\bq \\
    \rho_\bq & = \sum_\bk c^\dag_\bk \Lambda_\bq(\bk) c_{\bk + \bq}, \qquad \ov{\rho}_\bq = \frac{1}{2} \sum_{\bG, \bk} \delta_{\bG,\bq} \tr \Lambda_{\bG}(\bk),
\end{aligned}
  \label{unpert_tbg}
\end{equation}
where, in this section, $\Lambda_\bq(\bk)$ is chirally symmetric form factor for MATBG \cite{KIVCpaper}:
\begin{equation}
  \Lambda_\bq(\bk) = \Lambda^S_{c \bq}(\bk) = F_\bq(\bk) e^{i \phi_\bq(\bk) \gamma_z}.
  \label{symm_formfactor}
\end{equation}
and $C_2$ symmetry implies that 
\begin{equation}
    F_{-\bq}(-\bk) = F_\bq(\bk) \qquad \phi_{-\bq}(-\bk) = \phi_\bq(\bk).
    \label{symm_formfactor_C2}
\end{equation}
In other subsections we will use the subscript $c$ to clarify that it refers to the form factors of the magic sector. The superscript $S$ which we will usually omit differentiates this symmetric part of the form factor from the much smaller chirally-asymmetric part $\Lambda^A_\bq(\bk) = -\gamma_z \Lambda^A_\bq(\bk) \gamma_z$. The presence of $\Lambda^A$ gives rise to the anisotropy $\lambda = \frac{A_M}{2A^2} \sum_{\bq \bk} V(\bq) \Lambda_\bq(\bk)^2$, whereas the chern off-diagonal dispersion $h_c$ gives the superexchange $J\propto h^2/U$ \cite{Khalaf2020}. 

The Hamiltonian \eqref{unpert_tbg} has a manifold of quantum Hall ferromagnetic ground states with a $\bk$-independent $Q$ matrix with definite Chern number: $[Q,\gamma_z] = 0$. It also has a $\U(2n) \times \U(2n)$ symmetry, where $n$ is the number of spin components, consisting of arbitrary unitary rotations in each Chern sector. For large $\kappa$ the model \eqref{unpert_tbg} has an approximate $\U(4n)$ symmetry due to concentration of Berry curvature \cite{KhalafSoftMode}; this emergent symmetry will be important when we consider MSM states at the end of this section. We briefly review this now. Consider the limit where the Berry curvature is perfectly solenoidal at the $\Gamma$ point such that the Berry connection $\bA_\bk = \gamma_z(-k_y,k_x)/\abs{\bk}^2$. Then the Berry connection vanishes after the singular gauge transformation 
\begin{equation}
  c_\bk \to e^{i \theta_\bk \gamma_z} c_\bk, \qquad \theta_\bk = \arg(k_x + i k_y). 
  \label{gaugetransf_U8}
\end{equation}
After this transformation all $4n$ Chern bands have the form factor $\Lambda'_\bq(\bk) \approx  F_\bq(\bk)$ and may be rotated into each other. Ground states consist of $\U(4n)$ ferromagnets with $Q' = e^{i \theta_\bk \gamma_z} Q e^{-i \theta_\bk \gamma_z}$. This transformation does nothing to Chern diagonal states but turns nematic semimetals into quantum Hall ferromagnets with $Q' \propto \gamma_{x,y}$ and $\bk$-independent.

We will use the hatted notation for second quantized operators above for the rest of this section, together with the following identities for expectation values \cite{Khalaf2020}:
\begin{equation}
        \hat A = \sum_{\alpha \beta} c^\dag_\alpha A_{\alpha \beta} c_\beta, 
        \qquad \langle \hat A \rangle  = \Tr \tilde{P} A,  \qquad
    \langle \hat A \hat B \rangle = \Tr \tilde{P} AB - \Tr A \tilde{P} B \tilde{P} + \Tr\tilde{P} A \Tr \tilde{P} B,
\label{identity_expval}
\end{equation}
The upper case trace denotes a sum over momentum and orbital indices whereas a lower case trace denotes a sum over orbital indices only. This notation is useful in part because ``hatting" preserves commutators:
\begin{equation}
    [c^\dag_\alpha c_\beta, c^\dag_\gamma c_\delta] = \delta_{\beta \gamma} c^\dag_{\alpha} c_{\delta} - \delta_{\alpha \delta} c^\dag_\gamma c_\beta \quad  \implies \quad [\hat A, \hat B] = \widehat{[A,B]} 
\end{equation}
In this notation, the density operator is
\begin{equation}
    \rho_\bq = \hat{\Lambda}_\bq, \qquad (\Lambda_\bq)_{\bk \bk'} = \Lambda_\bq(\bk) \delta_{\bk + \bq, \bk'}.
\end{equation}

We project external field terms
\begin{equation}
  \frac{1}{2}\left(V + g_{\text{orb}} \mu_B (\hat{\bz} \times \bB) \cdot \bsigma \right) \mu_z,
\end{equation}
to the Chern/valley flat band basis. The operator $\mu_z = \pm$ labels the two TBG layers. The electric field term is diagonal in Chern sector and valley. It is symmetric under $C_2 \T$ and is therefore proportional to $\gamma_0$ after projection. The factor of $\mu_z$ means that it is odd under (unitary) $\P \T = i\sigma_x \tau_x \mu_y$ symmetry. On the flat bands $\P \T$ acts as $\eta_x \gamma_z$, and so the electric field term must be proportional to $\eta_z$ after projection. The magnetic field is off diagonal in Chern sector and even under both $\P \T$ and $C_2 \T$. This fixes its form to be $\gamma_{x,y} \eta_z$. After projection we obtain 
\begin{equation}
  \begin{aligned}
  \hat{h}_{V,B}&  = \sum_\bk c^\dag_\bk h_{V,B}(\bk) c_\bk,  \\
  \qquad h_V(\bk)_{\gamma \eta \gamma' \eta'} & =  \bra{ u_{\bk \gamma \eta}}\frac{V}{2}\mu_z \ket{u_{\bk \gamma' \eta'}} = h_0(\bk) (\eta_z)_{\eta \eta'} \delta_{\gamma \gamma'} , \\
  \qquad h_B(\bk)_{\gamma \eta \gamma' \eta'} & =  \bra{ u_{\bk \gamma \eta}}\frac{1}{2}g_{\rm orb} \mu_B \hat{\bz} \times \bB \cdot \bsigma \mu_z \ket{u_{\bk \gamma' \eta'}} = (\eta_z)_{\eta \eta'} \bb_\bk \cdot (\bgamma)_{\gamma \gamma'} ,
\end{aligned}
  \end{equation}
  where $h_0(\bk) = -h_0(-\bk)$ by $C_2$ symmetry. The influence of these terms at first order is zero for Chern diagonal states: the magnetic field term is because it is off diagonal in Chern sector and $Q$ is diagonal, whereas the electric field term is odd under $\bk \mapsto -\bk$ and $Q$ is $\bk$ independent. They both contribute at second order. 

  To obtain the second order terms we parameterize rotations outside the ground state subspace in the usual way
  \begin{equation}
      \ket{\Psi_M} = e^{-i \hat{M}} \ket{\Psi_Q}, \qquad \{M,Q\} = 0.
  \end{equation}
  For the electric field superexchange we want $[M, \gamma_z] = 0$ whereas for the magnetic field term $\{M ,\gamma_z \} = 0$. The interaction energy is
  \begin{equation}
  E_{\rm int} = -\frac{1}{4A} \sum_\bq V(\bq) \langle (\ad^2_{\hat M} \rho_\bq)\rho_{-\bq} + \rho_{\bq} \ad^2_{\hat M} \rho_{-\bq} + 2 \ad_{\hat M} \rho_\bq \ad_{\hat M} \rho_{-\bq} \rangle 
      + \frac{1}{2A} \sum_\bG V(\bG) \langle (\ad^2_{\hat M} \rho_\bG)\ov \rho_{-\bG} \rangle.
      \label{TBGinteractionterms}
  \end{equation}
  The first and second terms entirely cancel against the background terms because $Q$ is $\bk$-independent and Chern-diagonal, and thus $[Q,\Lambda_\bq] =0$. We then evaluate
  \begin{equation}
  \begin{aligned}
  2 \langle \ad_{\hat M} \rho_\bq \ad_{\hat M} \rho_{-\bq} \rangle & = 2\langle \hat{[M,\Lambda_\bq]} \hat{[M,\Lambda_{-\bq}]} \rangle = 2 \Tr P [M,\Lambda_\bq] [M,\Lambda_{-\bq}] = \Tr [M,\Lambda_\bq] [M,\Lambda_{-\bq}] \\
  & = -2\sum_\bk \tr\left(F_{\bq}(\bk)^2 M_\bk^2 - M_\bk \Lambda_{\bq}(\bk) M_\bk \Lambda^\dag_\bq(\bk) \right)
  \end{aligned}
      \label{TBGadad}
  \end{equation}
  where we used $[Q,M]=[Q,\Lambda_\bq] = 0$ to drop the $\Tr P [M,\Lambda_\bq]P[M,\Lambda_{-\bq}]$ term and the $Q$-dependent part of $\Tr P [M,\Lambda_\bq][M,\Lambda_{-\bq}]$.

  \subsubsection{Electric field}

  We now specialize to the case of an electric field and handle a magnetic field after. In this case $[M_\bk, \gamma_z] = [M_\bk, \Lambda_\bq(\bk)] = 0$. We separate the different Chern sectors and expand $M$ via 
  \begin{equation}
      M = \sum_{\gamma \mu} m_{\gamma \mu} r^\mu \gamma_\gamma, \qquad Q = \sum_\gamma Q_\gamma \gamma_\gamma,
  \end{equation}
  where $r^\mu$ form a basis for $2n \times 2n$ Hermitian matrices and $m_{\gamma \mu}$ are real because $M^\dag = M$. We then have
  \begin{equation}
      E_{\rm int} = \frac{1}{2A}\sum_\bq V(\bq) F_\bq(\bk)^2 \tr (M_\bk^2 - M_\bk M_{[\bk + \bq]})
      = \frac{1}{2}\sum_{\bk \mu \gamma} m_{\gamma \mu \bk} R_{\bk \bk'} m_{\gamma \mu \bk'}.
  \end{equation}
  where
  \begin{equation}
      R_{\bk \bk'} = \frac{1}{A}\sum_\bq  V(\bq) F_\bq(\bk)^2(\delta_{\bk \bk'} - \delta_{[\bk +\bq], \bk'}),
  \end{equation}
  and $[\bk + \bq]$ denotes the first Brillouin zone component of $\bk + \bq$.
  The first order term coupling $M$ to the displacement field is
  \begin{equation}
      E_V = -i \langle [\hat M, \hat h_V] \rangle = -\frac{i}{2} \Tr Q[M,h_V] = i \sum_{\bk \mu \gamma} m_{\mu\gamma \bk} \tr r^\mu Q_\gamma h_{V\bk},
  \end{equation}
  and so we obtain the full energy
  \begin{equation}
  E_{\rm int} + E_V = \frac{1}{2}\sum_{\bk \mu \gamma} m_{\gamma \mu \bk} R_{\bk \bk'} m_{\gamma \mu \bk'} + i \sum_{\bk \mu \gamma} m_{\mu\gamma \bk} \tr Q_\gamma[r^\mu,h_{V\bk}].
  \end{equation}
  Minimizing with respect to $m$ then gives
  \begin{equation}
  \begin{aligned}
  E_{\rm SE} & = -\frac{1}{2}\sum_{\bk \mu \gamma} \left( i \tr  r^\mu Q_\gamma h_{V \bk} \eta_z  \right) R^{-1}_{\bk \bk'} \left( i \tr  r^\mu Q_\gamma h_{V \bk'} \eta_z \right) \\
  & = -\frac{1}{2} \sum_{\bk \mu \gamma} h_{V\bk} R^{-1}_{\bk \bk'} h_{V\bk'} \tr r^\mu \eta_z Q_\gamma \tr r^\mu  Q_\gamma \eta_z.
  \end{aligned}
  \label{TBG_SE_preFierze}
  \end{equation}
  Technically, $R$ is not invertible here because it has a zero mode with $M_\bk$ independent of $\bk$. This is the goldstone mode for $\U(4) \times \U(4)$ symmetry breaking. However, $h_V(\bk)$ does not couple to this goldstone mode because it is odd under $\bk \mapsto - \bk$. In practice, we may add a term $C\tr(M_\bk + M_{-\bk})^2$ to the interaction energy to gap the goldstone mode but not influence the superexchange mediated by $h_V$. 

  We now apply the Fierze identity for $r^\mu$,
  \begin{equation}
      \tr A r^\mu \tr B r^\mu = \frac{1}{2}\left( \tr AB - \tr QAQB \right),
  \end{equation}
  to \eqref{TBG_SE_preFierze}. We have, up to an unimportant constant from the first term,
  \begin{equation}
      E_{\rm SE} = \frac{\tilde{\lambda}_V}{4} \tr Q \eta_z Q \eta_z, \qquad \tilde{\lambda} =  \sum_{\bk \mu \gamma} h_{V\bk} R^{-1}_{\bk \bk'} h_{V\bk'}.
  \end{equation}
  We find $\tilde{\lambda}_V = \frac{1}{4} c_V V^2$, where $c_V$ is highly $\kappa$ dependent but always very small $c_V \approx 10^{-5} (\rm{meV})^{-1}$. This is because the flat bands have almost no layer polarization, and so the electric field term projected to the flat bands is very small.

  \subsubsection{Magnetic field}

  If we instead have an in-plane magnetic field then $\{M, \gamma_z \} = 0$ and so it may be written in a block off-diagonal form in Chern sector,
  \begin{equation}
     M = \begin{pmatrix} 0 & m \\ m^\dag & 0 \end{pmatrix}_{\gamma}, \qquad m = m_\mu r^\mu, \qquad r^\mu = - Q_+ r^\mu Q_-.
  \end{equation}
  The interaction energy is
  \begin{equation}
      E_{\rm int}= \frac{1}{2}\sum_{\bk \mu \gamma} m^*_{\mu \bk} R_{\bk \bk'} m_{\gamma \mu \bk'}, \qquad
      R_{\bk \bk'} = \frac{1}{A}\sum_\bq  V(\bq) F_\bq(\bk)^2(\delta_{\bk \bk'} - \delta_{[\bk +\bq], \bk'} e^{-2i\phi_\bq(\bk)}),
  \end{equation}
  where the extra factor of $e^{-2i\phi_\bq(\bk)}$ arises because $M$ anticommutes with the part of the form factor that is proportional to $\gamma_z$. Its presence ensures that there are no gapless inter-Chern modes.

  The first order term is
  \begin{equation}
      E_B = -i \langle \widehat{[M, h_B]} \rangle = -\frac{i}{2} \Tr Q [M,h_B] = -i \Tr QMh_B = -i \sum_{\bk \mu} \left( m_\mu z^*_{b \bk} \tr r^\mu \eta_z Q_+  - m^*_\mu z_b \tr Q_+ \eta_z r^{\mu \dag}  \right),
  \end{equation}
  where $z_{b \bk} = b_{x \bk} + i b_{y \bk}$.
  Integrating out $m$ we obtain the energy
  \begin{equation}
      E_{\rm SE} = -2 \sum_{\bk \bk'} z^*_{b \bk} R^{-1}_{\bk \bk'} z_{b \bk'} \sum_\mu \tr r^\mu \eta_z Q_+ \tr r^{\mu \dag} Q_+ \eta_z
  \end{equation}
  If we use the Fierz identity,
  \begin{equation}
      \tr r^\mu A \tr r^{\mu \dag} B = \frac{1}{2}\left(\tr AB - \tr Q_+ A Q_- B  \right),
  \end{equation}
  we have up to an unimportant constant
  \begin{equation}
    E_{\rm SE}\frac{A_M}{A} = \frac{1}{2}\lambda_B \tr Q_+ \eta_z Q_- \eta_z, \qquad \lambda_B =  \frac{2A_M}{A}\sum_{\bk \bk'} z^*_{b \bk} R^{-1}_{\bk \bk'}z_{b \bk}.
  \end{equation}
  Explicitly we find $\lambda_B = \frac{1}{4}c_B g_{\text{orb}}^2 B^2$, where the coefficient $c_B$ increases with $\kappa$. As $\kappa$ ranges from $0$ to $0.8$ we have $c_B=0.1-0.3 (\rm{meV})^{-1}$. The reason for the large coefficient $c_B \gg U_\bk^{-1}$ and the strong $\kappa$ dependence is that the magnetic field couples to a nemmatic near-soft mode. The nematic soft mode becomes softer as $\kappa$ increases because the Berry curvature becomes more concentrated \cite{KhalafSoftMode}. For this reason perturbation theory breaks down before we reach the critical field once $\kappa > 0.6$ or so. Before this value of $\kappa$ we may use the above expression to predict the critical field for the KIVC$\to$VH transition.

  As discussed in the main text for large $\kappa$ it is better to think of the magnetic field as an effective ``Zeeman'' field for a low lying MSM state. We may model the MSM and VH states as $\U(4n)$ quantum Hall ferromagnets with $Q'= \bn \cdot \bgamma \eta_z$ for a three dimensional real unit vector $\bn$. For $\bn$ in the $x-y$ plane we obtain a MSM state whereas for $\bn$ along the $z$ axis we obtain a VH state. The perturbative superexchange calculation above corresponds to a small $\bk$-dependent canting away from the VH state, whereas here we can consider a non-perturbative canting that is $\bk$-independent within the manifold of $\U(4n)$ quantum Hall ferromagnets. The Zeeman energy is given by
  \begin{equation}
    E_{B} = \frac{1}{2} \Tr Q h_B = \frac{1}{2} \sum_\bk \tr \left(\bb_\bk \cdot \bgamma \eta_z e^{2 i \theta_\bk \gamma_z} \bn \cdot \bgamma \eta_z \right) = 2nA \bb' \cdot \bn, \qquad b'_x + b'_y = \frac{1}{A} \sum_\bk e^{2i\theta_\bk} (b_x + i b_y)
    \label{MSMzeeman}
  \end{equation}
  where $\bb'$ is nonzero because $\bb_\bk$ winds twice around the $\Gamma$ point. The vector $\bb'$ may be thought of as the projection of the magnetic field $\bB$ onto the space of $\U(4n)$-related ground states; it rotates once when the magnetic field does as one would expect. This is the Zeeman energy we use to predict the transitions in the main text for large $\kappa$.

\subsection{Even n>2}
\label{Suppsubsec:intraTBGnisbiggerthan2}

We now discuss the effects of the external fields for even $n>2$. For odd $n$ there is no such external field contribution because the external fields only connect opposite $M_z$ sectors. We again assume that one TBG is magic and the other TBGs are weakly coupled. As for the odd $n$ case there are superexchange terms arising from field-induced tunneling between different TBGs. However, the most important effect for the magic $n=2$ TBG is the intra-TBG superexchange. This may be computed in the same way as for $n=2$, but with a coefficient that depends on how the external fields couple to the TBG in question.

For example, consider $n=4$ with the BM Hamiltonian copied here for convenience,
\beq
    H^{\rm Full}_{\rm dec} =  \left(\begin{array}{cccc}
    -i v_F \bsigma_+ \cdot \nabla - r_1 \Gamma_+  & \varphi T(\br) & \frac{2}{\sqrt{5}} \Gamma_+ & 0 \\
\varphi T^\dagger(\br) & -i v_F \bsigma_- \cdot \nabla + r_1 \Gamma_- & 0 & \frac{2}{\sqrt{5}} \Gamma_- \\
\frac{2}{\sqrt{5}} \Gamma_+ & 0 & -i v_F \bsigma_+ \cdot \nabla - r_2 \Gamma_+ & \frac{1}{\varphi} T(\br) \\
0 & \frac{2}{\sqrt{5}} \Gamma_- & \frac{1}{\varphi} T^\dagger(\br) & -i v_F \bsigma_- \cdot \nabla + r_2 \Gamma_-
\end{array} \right).
\eeq
The strength of the intra-TBG superexchange is proportional to the couplings $r_{1,2} = \frac{2\sqrt{5} \mp 5}{10}$. It is important to note that $r_1 \approx 1/20$, and therefore critical fields could be as much as ten times larger for four-layer graphene for this TBG subsystem compared to $n=2$ TBG. As discussed in \ref{SuppSec:mappingreview}, this is generic for $n>2$ even; we have $r_1 \approx \frac{\pi^2}{2n^3}$.

\section{Field mediated superexchange for $n>2$}
\label{Suppsec:superexchange}
In this section we compute the energy associated with superexchange between TBG or graphene sectors mediated by the displacement field. We follow a very similar methodology to \cite{Khalaf2020} where the $J$ and $\lambda$ terms in \eqref{Qenergy_mt} were calculated, but the discussion here will be self contained. We focus on the $n=3$ trilayer case first and then describe the modifications associated with TBG-TBG superexchange in four and higher layers.

\subsection{n=3 MATTG mirror superexchange}
\label{Suppsubsec:trilayersuperexchange}
For $n=3$ we start with the Hamiltonian
\begin{equation}
    \begin{aligned}
        \H_{\rm eff} & = \H_{D} + \H_{\rm int} \\
    \H_{\rm D} & = \sum_{\bp} \sum_\eta f_{\bp}^\dag  \left( v\eta\bp_\eta \cdot \bgamma - \mu \right) f_{\bp}, \\ 
    \H_{\rm int} & = \frac{1}{2A} \sum_\bq V_{\rm eff}(\bq) \delta \rho_{ \bq} \delta \rho_{-\bq},  \\
    \delta \rho & = \delta \rho^S + \delta\rho_f
    \end{aligned},
                     \label{unperturbed}
\end{equation}
where $\bp_\eta = \bp - \bK_\eta$ and $\bp = \bk + \bG$ where $\bk$ is restricted to the first Brillouin zone. While the TBG flat bands are natural to describe in the first moir\'{e} Brillouin zone, we will opt to label the graphene electrons with an unfolded $\bp$. The density operator and form factor break into TBG and graphene parts:
\begin{equation}
    \rho_\bq = \hat{\tilde{\Lambda}}_\bq, \qquad (\tilde{\Lambda}_\bq)_{\bk \bk'} = \tilde{\Lambda}_\bq(\bk) \delta_{\bk + \bq, \bk'},\qquad \tilde{\Lambda}_\bq(\bk) = \begin{pmatrix} \Lambda_{c\bq}(\bk) & 0 \\ 0 & \Lambda_{f\bq}(\bk) \end{pmatrix}.
\end{equation}
The mean field ground state of this Hamiltonian is a Slater determinant state specified by 
\begin{equation}
    \tilde{P} = \begin{pmatrix}P_c & 0 \\ 0 & P_f \end{pmatrix} = \frac{1}{2}(1 + \tilde{Q}), \qquad \tilde{Q} = \begin{pmatrix} Q_c & 0 \\ 0 & Q_f \end{pmatrix},
    \qquad P_{c \alpha \beta} = \langle c^\dag_\beta c_\alpha \rangle \qquad P_{f \alpha \beta} = \langle f^\dag_\beta f_\alpha \rangle,
\end{equation}
where $Q_c$ is $\bk$-independent, at least approximately near the $\bK_\eta$ points where the superexchange is active, and
\begin{equation}
    Q_f(\bp) = \sum_{\eta = \pm} Q_{f \eta}(\bp)\eta_\eta, \qquad Q_{f \eta}(\bp) = -\eta \frac{\bp_\eta}{\abs{\bp_\eta}} \cdot \bgamma \Theta(v\abs{\bp_\eta} - \abs{\mu}) + \sgn{\mu}\,\Theta(\abs{\mu} - v\abs{\bp_\eta})
    \label{unperturbedQ}
\end{equation}
describes occupation of graphene electrons. We used the notation 
\begin{equation}
    \eta_\pm = \frac{1}{2}(1 \pm \eta_z) \qquad \gamma_\pm = \frac{1}{2}( 1 \pm \gamma_z).
\end{equation}

It is important to note that small graphene masses and dopings are arbitrarily low energy excited states and may be favored by small displacement fields or in plane magnetic fields, depending on the TBG ground state. We will allow for this possibility and therefore only use \eqref{unperturbedQ} in the computation of the energy gap to excited states. Similarly, we will allow the magic sector to gain a piece $\propto \gamma_{x,y}$ and therefore will only assume $Q_c \propto \gamma_z$ in the computation of the energy gap. Note that we have ignored terms proportional to $\rho^A$ and also the dispersion in the TBG sector.  The effects of these terms are captured within perturbation theory in the $J$ and $\lambda$ terms. 

We will similarly compute the order $V^2, \, B^2$ terms on top of the unperturbed model \eqref{unperturbed}. The perturbations we add are the displacement field and parallel magnetic field terms \eqref{efieldbfield} projected to the flat bands. In this section it will be convenient to label both the flat band and graphene electrons with the unfolded $\bp$ and use a periodic gauge $c_{\bp} = c_{\bk}$ for the flat bands where $\bk = [\bp]$ is the first BZ part of $\bp$. We then obtain
\begin{equation}
    \hat{\Xi} =  \sum_{\bp \in \rm BZ} \tilde c^\dag_\bp \Xi_\bp \tilde c_\bp = \sum_{\bp \in \rm{BZ}} c^\dag_\bp \xi_\bp f_\bp + f^\dag_\bp \xi^\dag_\bp c_\bp, \qquad \tilde{c} = \begin{pmatrix} c \\ f \end{pmatrix}, \qquad \Xi_\bp = \begin{pmatrix} 0 & \xi_\bp \\ \xi^\dag_\bp & 0 \end{pmatrix},
    \label{pertterm}
\end{equation}
where $\xi_\bp = v_\bp + b_\bp$. 

The tunnelings $\xi_\bp$ are given by projecting the perturbations to the BM hamiltonian to the low energy subspace of flat bands and graphene Dirac cones. We use the complete sublattice polarization of the unperturbed TBG flat bands to only keep the Chern diagonal part of $v_\bp$ and the Chern off diagonal part of $b_\bp$. We obtain
\begin{equation}
    \xi_\bp = \sum_{\eta = \pm} \xi_{\bp \eta} \eta_\eta, \qquad (\xi_{\bp \eta})_{\gamma \gamma'} = \frac{1}{2}\left(V\delta_{\gamma \gamma'} + \eta \bA \cdot \bgamma_{\gamma \gamma'}\right) t_{\bp \eta \gamma}
    \label{tbgdiracoverlap}
\end{equation}
with no implied summation over $\eta$ or $\gamma$. The spatial wavefunction overlap is
\begin{equation}
  t_{\bp \eta \gamma} = \int_{\rm Cell} d^2 \br\, u^*_{+ \bp \eta \gamma +}(\br) = \int_{\rm Cell} d^2 \br u^*_{+ \bk \eta \gamma +}(\br) e^{i \bG \cdot \br},
\end{equation}
where $u_{+ \bp \eta \gamma \mu}(\br)$ is mirror $+1$ moir\'{e} periodic wavefunction in graphene valley $\eta$ and Chern sector $\gamma$ evaluated in TBG ``layer'' $\mu$ and position $\br$. As before $\bp = \bk + \bG$.

The operator $\hat{\Xi}$ is zero in the ground state subspace of $Q$ matrices that are mirror diagonal. Its contribution comes from superexchange at second order. Application of $\hat{\Xi}$ generates rotations outside of the ground state manifold; these excited states may be paramterized as
\begin{equation}
    \ket{\Psi_M} = e^{-i \hat M} \ket{\Psi_{\tilde{Q}}},  \qquad \{M, M_z \} = \{ M, \tilde{Q} \} =  0.
\end{equation}
We will calculate the expectation value of the Hamiltonian in the state $\ket{\Psi_M}$ up to second order in $M$. We will then integrate out $M$ to obtain the superexchange contribution to the energy functional. 

First it is convenient to find a simple parameterization of $M$. The anticommutation with $M_z$ and $\tilde{Q}$ imply that we can write $M$ as
\begin{equation}
    M = \begin{pmatrix} 0 & m \\ m^\dag & 0 \end{pmatrix} \qquad Q_c m Q_f = -m.
    \label{Mrep1}
\end{equation}
We also flip momenta in one valley within the graphene sector so that the location of the graphene Dirac point in both valleys can be the same; this recovers some of the pseudospin isotropy. This is carried out by the change of variables
\begin{equation}
    m_\bp = w_{1 \bp} \eta_+ + w_{2-\bp} \eta_- + w_{3\bp}\eta_x \eta_+ + w_{4 -\bp} \eta_x \eta_- , \qquad
    w_\bp = w_{1 \bp} \eta_+ + w_{2\bp} \eta_- + w_{3\bp}\eta_x \eta_+ + w_{4 \bp} \eta_x \eta_- .
\end{equation}
such that $Q_c w Q'_f = - w$ where
\begin{equation}
    Q'_f(\bp) = Q_{f+}(\bp)\eta_+ + Q_{f-}(-\bp) \eta_- = -\frac{\bp_+}{\abs{\bp_+}} \cdot \bgamma \Theta(v\abs{\bp_+} - \abs{\mu}) + \sgn{\mu}\,\Theta(\abs{\mu} - v\abs{\bp_+})
\end{equation}
is now the identity in pseudospin space.

Finally, for the purpose of eventually integrating out $w$, we expand
\begin{equation}
    w_\bp = \sum_{\mu \nu} C^{\mu \nu}_\bp w_{\bp \nu} t_\mu \qquad t_\mu C^{\mu \nu}_\bp = \frac{1}{2}\left(t_\nu - Q_c t_\nu Q_{f \bp}' \right)
    \label{projected_expansion}
\end{equation}
where $t^\mu$ are a basis for $4n \times 4n$ matrices and $n$ is the number of spin components. Important cases are $n=1$ for an effective spinless model on top of a spin polarized state at $\nu = \pm2$ and $n=2$ for $\nu = 0$. The matrix $C = C^2 = C^\dag$ projects onto the generators that are odd under $A \mapsto Q_c A Q'_{f \bp}$.

    We now evaluate the energy of the two terms in the Hamiltonian \eqref{unperturbed} and that of the perturbation \eqref{pertterm}. To evaluate expectation values we use the expansion
\begin{equation}
    \bra{\Psi_M} \hat A \ket{\Psi_M} = \langle e^{-i \ad_{\hat M}} A \rangle = \langle \hat A \rangle -i \langle \ad_{\hat M} A \rangle -\frac{1}{2} \langle \ad_{\hat M}^2 A \rangle + \cdots
\end{equation}
where $\ad_A B = [A,B]$. We will use that this operation satisfies the product rule $\ad_A (BC) = (\ad_A B)C + B \ad_A C$. The expectation values on the right hand side are in the state $\ket{\Psi_{\tilde{Q}}}$.

\subsubsection{Graphene Dispersion}

We first evaluate the graphene dispersion term. We write it as $\H_D = \sum_\bp f^\dag_\bp h_D f_\bp$ with $h_D = \sum_\eta \eta \eta_\eta v\bp_\eta \cdot \bgamma - \mu$. The zeroth order term is a constant, and the first order term vanishes by $M_z$ symmetry. The second order term is
\begin{equation}
    \begin{aligned}
        E_{D} & = -\frac{1}{2}\left\langle \ad_{\hat M} \ad_{\hat M} \begin{pmatrix} 0 & 0 \\ 0 & \hat h_D \end{pmatrix} \right\rangle = -\frac{1}{2} \left\langle \widehat{\left[M,\left[M,h_D\frac{1}{2}(1-M_z)\right]\right]} \right\rangle = -\frac{1}{2} \Tr \tilde{P} \left[M,\left[M,h_D\frac{1}{2}(1-M_z)\right]\right] \\
              & = \Tr P_c m h_D m^\dag - \frac{1}{2} \Tr P_f \{h_D, m^\dag m \},
\end{aligned}
\end{equation}
where in moving from the first to second line we explicitly evaluated the commutators and split the trace into separate mirror sectors. We now use $Q_c m Q_f = -m$ and its adjoint to show
\begin{equation}
    \Tr P_c m h_D m^\dag = \frac{1}{2} \Tr P_f^\perp h_D m^\dag m + \frac{1}{2}\Tr P_f^\perp m^\dag m h_D,
\end{equation}
where $P_{c,f}^\perp = \frac{1}{2}(1-Q_{c,f})$ and the two terms on the right hand side are equal. We now combine the traces in the two mirror sectors to obtain
\begin{equation}
    E_D = -\frac{1}{2}\Tr Q_f \{h_D , m^\dag m \} = - \frac{1}{2}\Tr \{ Q_f, h_D \} m^\dag m 
    %\approx \sum_{\bk \eta} v\abs{\bk_\eta} \tr m^\dag_\bk m_\bk \eta_\eta,
    \label{diracenergy}
\end{equation}
and with the expansion \eqref{projected_expansion},
\begin{equation}
    E_D =  \frac{1}{2}\sum_\bp \tr \{h'_D,Q'_f\} w^\dag_\bp w_\bp = \frac{1}{2} \sum_{\bp \mu} w^*_{\bp \mu} ( C R^D C )^{\mu \nu}_{\bp \bp'} w_{\bp \nu}, \qquad R^{D \mu \nu}_{\bp \bp'} = \delta_{\bp \bp'}\tr \{h'_D, Q'_f\} t_\mu^\dag t_\nu
    \label{dispenergy_munonzero}
\end{equation}
At charge neutrality the energy takes the simple form 
\begin{equation}
    R^{D \mu \nu}_{\bp \bp'} = 2v\abs{\bp_+} \delta_{\bp \bp'} \delta^{\mu \nu}.
    \label{dispenergy_CN}
\end{equation}

\subsubsection{Interaction Term}

We now evaluate the interaction term. First we rewrite it using $\delta \rho_\bq = \rho_\bq - \ov\rho_\bq$:
\begin{equation}
    \H_{\rm int} = \frac{1}{2A} \sum_\bq V_{\rm eff}(\bq) \rho_\bq \rho_{-\bq} - \frac{1}{A} \sum_\bG V_{\rm eff}(\bG) \rho_\bG \ov \rho_{-\bG}.
\end{equation}
The zeroth order term is part of the constant ground state energy, independent of $M$, and the first order term vanishes by $M_z$ symmetry. The second order terms are
\begin{equation}
    E_{\rm int} = -\frac{1}{4A} \sum_\bq V_{\rm eff}(\bq) \langle (\ad^2_{\hat M} \rho_\bq)\rho_{-\bq} + \rho_{\bq} \ad^2_{\hat M} \rho_{-\bq} + 2 \ad_{\hat M} \rho_\bq \ad_{\hat M} \rho_{-\bq} \rangle 
        + \frac{1}{2A} \sum_\bG V_{\rm eff}(\bG) \langle (\ad^2_{\hat M} \rho_\bG)\ov \rho_{-\bG} \rangle
        \label{interactionterms}
\end{equation}
We begin with the first term 
\begin{equation}
    \begin{aligned}
    \langle (\ad^2_{\hat M} \rho_\bq) \rho_{-\bq} \rangle & = \langle \widehat{[M,[M,\tilde{\Lambda}_\bq]]} \widehat{\tilde{\Lambda}_{-\bq} }\rangle \\
    & = \Tr \tilde{P} [M,[M, \tilde{\Lambda}_{\bq}]] \Lambda_{- \bq}
    - \Tr \tilde{P} [M,[M, \tilde{\Lambda}_{\bq}]] \tilde{P} \tilde{\Lambda}_{- \bq}
    + \Tr \tilde{P} [M,[M,\tilde{\Lambda}_\bq]] \Tr \tilde{P} \tilde{\Lambda}_{-\bq} \\
    & = \Tr \tilde{P} [M,[M, \tilde{\Lambda}_{\bq}]] [\tilde{\Lambda}_{- \bq},\tilde{P}] + \sum_{\bG}\langle [M,[M,\tilde{\Lambda}_\bG]] \rangle \ov\rho_{-\bG} \delta_{\bq,\bG}.
    \end{aligned}
    \label{adsq_term}
\end{equation}
The second term in \eqref{adsq_term}, which is only nonzero when $\bq$ is a reciprocal lattice vector, together with the analogous contribution from the second term in \eqref{interactionterms}, cancels with the background term (final term in \eqref{interactionterms}). We have used $\tilde{P} = \tilde{P}^2$ to write the first two terms in \eqref{adsq_term} in terms of the commutator $[\tilde{\Lambda}_{-\bq},\tilde{P}]$ which is zero in the positive mirror sector but nonzero in the negative mirror sector because $Q_f$ is $\bp$-dependent. An explicit expansion of the commutators yields
\begin{equation}
    \begin{aligned}
        \Tr \tilde{P} [M,[M, \tilde{\Lambda}_{\bq}]] [\tilde{\Lambda}_{- \bq},\tilde{P}] & = \Tr P_f \Lambda_{f \bq} ( m^\dag m \Lambda_{f -\bq} + \Lambda_{f -\bq}m^\dag m - 2 m^\dag \Lambda_{c -\bq} ) (\Lambda_{f -\bq} P_f - \Lambda_{f \bq} P_f ) \\
        & = \Tr (\Lambda_{f \bq} \Lambda_{f -\bq} P_f + \Lambda_{f -\bq} P_f \Lambda_{f \bq} - \Lambda_{f \bq}P_f \Lambda_{f -\bq} P_f - P_f \Lambda_{f -\bq} P_f \Lambda_{f \bq}) m^\dag m\\ 
        & \quad - 2 \Tr [\Lambda_{f -\bq}, P_f] P_f m^\dag \Lambda_{c \bq} m,
    \end{aligned}
    \label{adad1}
\end{equation}
and the analogous contribution from $\langle \rho_\bq \ad_{\hat M}^2 \rho_{-\bq} \rangle$ is
\begin{equation}
    \begin{aligned}
        \Tr \tilde{P} [\tilde{P}, \Lambda_{\bq}]  [M,[M, \tilde{\Lambda}_{-\bq}]] & = 
        \Tr (P_f \Lambda_{f \bq} \Lambda_{f -\bq} + \Lambda_{f -\bq} P_f \Lambda_{f \bq} - P_f \Lambda_{f \bq}P_f \Lambda_{f -\bq} -  \Lambda_{f -\bq} P_f \Lambda_{f \bq} P_f) m^\dag m\\ 
        & \quad - 2 \Tr P_f [P_f, \Lambda_{f -\bq}] m^\dag \Lambda_{c -\bq} m.
    \end{aligned}
    \label{1adad}
\end{equation}
Finally we have
\begin{equation}
    2 \langle \ad_{\hat M} \rho_\bq \ad_{\hat M} \rho_{- \bq} \rangle = 2 \langle \widehat{[M, \tilde{\Lambda}_\bq]} \widehat{[M, \tilde{\Lambda}_{-\bq}]} \rangle = 2\Tr \tilde{P} [M,\tilde{\Lambda}_\bq][M,\tilde{\Lambda}_{-\bq}] - 2 \Tr \tilde{P} [M,\tilde{\Lambda}_\bq] \tilde{P}[M,\tilde{\Lambda}_{-\bq}]
\end{equation}
We evaluate each term by explicitly computing the commutators, splitting the trace into the two mirror sectors, and then eliminating dependence on $Q_c$ by using $m = -Q_c m Q_f$. The first term is 
\begin{equation}
    \begin{aligned}
    2\Tr \tilde{P} [M,\tilde{\Lambda}_\bq][M,\tilde{\Lambda}_{-\bq}] & = 2 \Tr P_c (m \Lambda_{f \bq} - \Lambda_{c \bq}m)(m^\dag \Lambda_{c -\bq} - \Lambda_{f -\bq} m^\dag) \\
     & \quad + 2 \Tr P_f(m^\dag \Lambda_{c \bq} - \Lambda_{f \bq} m^\dag)(m \Lambda_{f -\bq} - \Lambda_{c -\bq}m) \\ 
     & = -2 \Tr m^\dag \Lambda_{c \bq} \Lambda_{c -\bq} m - 2 \Tr (P_f^\perp \Lambda_{f \bq} \Lambda_{f -\bq} + \Lambda_{f-\bq} P_f \Lambda_{f \bq})m^\dag m \\
     & \quad + 2 \Tr (\Lambda_{f -\bq} m^\dag \Lambda_{c \bq} m + \Lambda_{f \bq} m^\dag \Lambda_{c -\bq} m),
    \end{aligned}
    \label{adad_firstterm}
\end{equation}
and the second term is
\begin{equation}
    \begin{aligned}
    -2\Tr \tilde{P} [M,\tilde{\Lambda}_\bq]\tilde{P}[M,\tilde{\Lambda}_{-\bq}] & = -2 \Tr P_c (m \Lambda_{f \bq} - \Lambda_{c \bq}m)P_f(m^\dag \Lambda_{c -\bq} - \Lambda_{f -\bq} m^\dag) \\
    & \quad - 2 \Tr P_f(m^\dag \Lambda_{c \bq} - \Lambda_{f \bq} m^\dag)P_c(m \Lambda_{f -\bq} - \Lambda_{c -\bq}m) \\ 
    & = 2\Tr(P_f^\perp \Lambda_{f \bq} P_f \Lambda_{f -\bq} + \Lambda_{f -\bq} P_f \Lambda_{f \bq} P_f^\perp)m^\dag m.
    \label{adad_secondterm}
    \end{aligned}
\end{equation}

We now collect terms from \eqref{adad1}, \eqref{1adad}, \eqref{adad_firstterm}, and \eqref{adad_secondterm} and insert them back into \eqref{interactionterms}. First there is the term with two factors of $\Lambda_c$ in \eqref{adad_firstterm}; this is the TBG exchange energy
\begin{equation}
\begin{aligned}
     -\frac{1}{4A} \sum_{\bq} V_{\rm eff}(\bq) \left(   -2 \Tr m^\dag \Lambda_{c \bq} \Lambda_{c -\bq} m   \right) = \frac{1}{2A} \sum_{\bp, \bq} V_{\rm eff}(\bq) F_{\bq}(\bp)^2 \tr w^\dag_{\bp} w_\bp & = \frac{1}{2} \sum_\bp 2U_
    \bp  \tr w^\dag_{\bp} w_\bp, \\
    U_\bp & = \frac{1}{2} \sum_{\bp, \bq} V_{\rm eff}(\bq) F_{\bq}(\bp)^2,
\end{aligned}
\end{equation}
where we used \eqref{symm_formfactor}, \eqref{symm_formfactor_C2}.
Next there are the terms with two factors of $\Lambda_f$:
\begin{equation}
\begin{aligned}
    & \Tr (\Lambda_{f \bq} \Lambda_{f -\bq} P_f + \Lambda_{f -\bq} P_f \Lambda_{f \bq} - \Lambda_{f \bq}P_f \Lambda_{f -\bq} P_f - P_f \Lambda_{f -\bq} P_f \Lambda_{f \bq}) m^\dag m \\
    & + \Tr (P_f \Lambda_{f \bq} \Lambda_{f -\bq} + \Lambda_{f -\bq} P_f \Lambda_{f \bq} - P_f \Lambda_{f \bq}P_f \Lambda_{f -\bq} -  \Lambda_{f -\bq} P_f \Lambda_{f \bq} P_f) m^\dag m\\
    & - 2 \Tr (P_f^\perp \Lambda_{f \bq} \Lambda_{f -\bq} + \Lambda_{f-\bq} P_f \Lambda_{f \bq})m^\dag m \\
    & + 2\Tr(P_f^\perp \Lambda_{f \bq} P_f \Lambda_{f -\bq} + \Lambda_{f -\bq} P_f \Lambda_{f \bq} P_f^\perp)m^\dag m\\
    & = \sum_\bp \tr(2P_{f\bp} - 2P_{f \bp} - P_{f\bp + \bq} P_{f\bp} - P_{f\bp} P_{f\bp - \bq} - P_{f\bp}P_{f\bp + \bq} - P_{f\bp - \bq}P_{f\bp} +2P_{f \bp}^\perp P_{f \bp + \bq} + 2P_{f \bp - \bq} P_{f \bp}^\perp ) m^\dag_\bp m_\bp \\
    & =  \sum_\bp \tr \left( -\frac{1}{4}\{ Q_{f\bp}, Q_{f \bp + \bq} + Q_{f \bp - \bq} \} - \frac{1}{2}Q_{f \bp} Q_{f \bp + \bq} - \frac{1}{2} Q_{f\bp - \bq} Q_{f\bp}\right) m^\dag_\bp m_\bp
    \end{aligned}
\end{equation}
which gives the Dirac exchange energy,
\begin{equation}
\begin{aligned}
    & -\frac{1}{4A} \sum_{\bp, \bq} V_{\rm eff}(\bq) \tr \left( -\frac{1}{4}\{ Q_{f\bp}, Q_{f \bp + \bq} + Q_{f \bp - \bq} \} - \frac{1}{2}Q_{f \bp} Q_{f \bp + \bq} - \frac{1}{2} Q_{f\bp - \bq} Q_{f\bp}\right) m^\dag_\bp m_\bp \\
    & = \frac{1}{2A} \sum_{\bp, \bq} V_{\rm eff}(\bq) \tr \frac{1}{2} \{ Q_{f \bp}, Q_{f \bp + \bq} \} m^\dag_\bp m_\bp \\
    & = \frac{1}{2} \sum_{\bp \mu} w^*_{\bp \mu} ( C \delta R^D C )^{\mu \nu}_{\bp \bp'} w_{\bp' \nu}
    \end{aligned}
\end{equation}
where $\delta R^D$ is the renormalization to the dispersion energy $R^D \to R^{D}_{\rm eff} = R^D + \delta R^D$ due to the replacement $h'_D \to h'^{D}_{\rm eff} = h'^D + \delta h'_D$ with 
\begin{equation}
    \delta h'_D(\bp) = \frac{1}{A}\sum_\bq V_{\rm eff}(\bq) Q'_f(\bp + \bq), \qquad \delta R^D = \delta_{\bp \bp'} \tr \{h'_D(\bp), Q'_f(\bp \} t^\dag_\mu t_\nu .
\end{equation}
At charge neutrality this renormalization is that of the Dirac velocity $v \to v_{\rm eff} =  v+\delta v$ where
\begin{equation}
\delta v \abs{\bp_+} = \frac{1}{2A} \sum_\bq V_{\rm eff}(\bq) \cos \left( \theta_{\bp_+}-\theta_{\bp_+ + \bq} \right).
\end{equation}
is the usual velocity renormalization and $\theta_{\ba}$ is the angle of a vector $\ba$. 

Finally, there are the terms with one factor of $\Lambda_c$ and one factor of $\Lambda_f$. They are
\begin{equation}
\begin{aligned}
    & - 2 \Tr [\Lambda_{f -\bq}, P_f] P_f m^\dag \Lambda_{c \bq} m - 2 \Tr P_f [P_f, \Lambda_{f -\bq}] m^\dag \Lambda_{c -\bq} m.\\
    & + 2 \Tr (\Lambda_{f -\bq} m^\dag \Lambda_{c \bq} m + \Lambda_{f \bq} m^\dag \Lambda_{c -\bq} m) \\
    & = 2 \sum_\bp \tr(1 - P_{f \bp} + P_{f \bp + \bq} P_{f \bp})m^\dag_{\bp}F_{\bq}(\bp)e^{i \phi_{\bq}(\bp)} m_{\bp + \bq} + \text{c.c.}\\
    & = 2 \sum_\bp \tr \left(\frac{3}{4} + \frac{1}{4}Q_{f \bp + \bq} Q_{f \bp}\right) m^\dag_{\bp}F_{\bq}(\bp)e^{i \phi_{\bq}(\bp) \gamma_z} m_{\bp + \bq} + \text{c.c.}\\
    & = 2 \sum_\bp \tr Q_{f\bp + \bq} Q_{f\bp}  m^\dag_{\bp}F_{\bq}(\bp)e^{i \phi_{\bq}(\bp)\gamma_z} m_{\bp + \bq} + \text{c.c.}\\
    & = 2 \sum_\bp \tr Q'_{f \bp + \bq} Q'_{f \bp} w^\dag_{\bp}F_{\bq}(\bp)e^{i \phi_{\bq}(\bp) \gamma_z } w_{\bp + \bq} + \text{c.c.}
\end{aligned} 
\end{equation}
where we used $Q_c m_{\bp} Q_{f\bp} = -m_{\bp}$ to obtain the second to last line and the $C_2$ symmetry of the TBG form factors \eqref{symm_formfactor_C2} to obtain the last line.

Putting everything together and using the expansion \eqref{projected_expansion} we have
\begin{equation}
E_D + E_{\rm int}  = \frac{1}{2} \sum_{\bp\bp' \mu\nu} w^*_{\mu \bp} \left(C_\bp R_{\bp \bp'} C_{\bp'}\right)^{\mu \nu} w_{\nu \bp}
\end{equation}
with
\begin{equation}
\begin{aligned}
    R_{\bp \bp'}^{\mu \nu} & = R^{D \mu \nu}_{\text{eff} \bp \bp'} + U_\bp \delta_{\bp \bp'}\delta^{\mu \nu} + \tilde{R}^{\mu \nu}_{\bp \bp'} \\
    \tilde{R}^{\mu \nu}_{\bp \bp'} & = -\frac{1}{A} \sum_\bq  V_{\rm eff}(\bq) \delta_{\bp + \bq, \bp'} \tr Q'_{f \bp'} Q'_{f \bp} t^{\mu\dag} F_{\bq}(\bp)e^{i \phi_{\bq}(\bp) \gamma_z } t^\nu + \text{h.c.}.
    \end{aligned}
\end{equation}

\subsubsection{Displacement and in plane field perturbations}
    Finally we evaluate the perturbation term \eqref{pertterm}. It is nonzero at first order and acts as a source term:
    \begin{equation}
            \begin{aligned}
            E_\Xi & = -i \langle \ad_{\hat{M}} \hat \Xi \rangle = -i \langle \widehat{ [M, \Xi ] } \rangle = -i \tr \tilde{P} [M, \Xi] \\
                & = -i \Tr P_c(m \xi^\dag - \xi m^\dag ) -i \Tr P_f(m^\dag \xi - \xi^\dag m) = -i \Tr Q_c (m \xi^\dag - \xi m^\dag),
        \end{aligned} 
    \end{equation}
    where in the final step we used $Q_c m Q_f = -m$ to relate the traces in each mirror sector. Using \eqref{tbgdiracoverlap} we obtain
    \begin{equation}
        E_\Xi = -i \sum_{\bk \mu } C^{\mu \nu}_\bp w_{\nu \bp} \tr Q_c  t_\mu \xi'^\dag_{\bp}  - w^*_{\nu \bp} C^{\nu \mu} \tr Q_c \xi'_\bp t^{\dag}_\mu.
    \end{equation}
    where $\xi_\bp' = \xi_{\bp,+}\eta_+ + \xi_{-\bp,-}\eta_-$.

\subsubsection{Superexchange energy}
    We now have the effective energy functional
    \begin{equation}
        E[w] = E_D + E_{\rm int} + E_\Xi = \frac{1}{2} \sum_{\bp\bp' \mu\nu} w^*_{\mu \bp} \left(C_\bp R_{\bp \bp'} C_{\bp'}\right)^{\mu \nu} w_{\nu \bp} -i \sum_{\bp \mu } C^{\mu \nu}_\bp w_{\nu \bp} \tr Q_c  t_\mu \xi'^\dag_\bp  - w^*_{\nu \bp} C^{\nu \mu} \tr Q_c \xi'_\bp  t^{\dag}_\mu.
    \end{equation}
    Integrating out $w_{\mu \bp}$ then gives
    \begin{equation}
        E_{\rm SE}(Q_c, Q_f) = -2 \sum_{\bp \mu} \tr \left( Q_c t_\mu \xi'^\dag_\bp \right) \left(C_\bp R^{-1}_{\bp \bp'} C_{\bp'}\right)^{\mu \nu} \tr \left( Q_c \xi_{\bp'} t^\dag_\nu \right).
    \end{equation}
    Above we used that $C = C^\dag = C^2$ is a projector. If we now decompose $t_\nu C^{\nu \mu} = \frac{1}{2}(t^\mu - Q_c t^\mu Q'_{f \bp})$ we obtain
    \begin{equation}
    \begin{aligned}
    E_{\rm SE}(Q_c, Q_f) = -\frac{1}{2}\sum_{\bp \bp' \mu \nu } \biggl( 
    & \tr \left( Q_c t_\mu \xi'^\dag_\bp \right) R^{-1 \mu \nu}_{\bp \bp'}\tr \left( Q_c \xi'_{\bp'}  t^{\dag}_\nu \right)\\
    & -\tr \left(t_\mu Q'_{f \bp} \xi'^\dag_\bp \right) R^{-1 \mu \nu}_{\bp \bp'}\tr \left( Q_c \xi'_{\bp'}  t^{\dag}_\nu \right) \\
    & -\tr \left( Q_c t_\mu \xi'^\dag_\bp \right) R^{-1 \mu \nu}_{\bp \bp'}\tr \left( \xi'_{\bp'}  Q'_{f \bp}t^{\dag}_\nu \right) \\
    & +\tr \left( t_\mu Q'_{f \bp}  \xi'^\dag_\bp \right) R^{-1 \mu \nu}_{\bp \bp'}\tr \left( \xi'_{\bp'}  Q'_{f \bp}t^{\dag}_\nu \right).
    \biggr).
    \label{SE_energy_general}
    \end{aligned}
    \end{equation}
    An analytic evaluation of the above expression without further approximations is difficult because it is hard to invert $R$. However, we may understand its essential features from a symmetry point of view. 
    
    First, we show that the above expression is $Q_c$ independent as long as $[Q_c, \gamma_z] = 0$ and $Q_c$ is approximately $\bk$-independent near the $\bK_\eta$ points. We first consider the first term which is quadratic in $Q_c$. Note that $R$, $\xi'$, and $Q_f'$ have no dependence on pseudospin or real spin which implies $R$ is invariant under $\U(2n)$ basis rotations $t^\mu \mapsto Ut^\mu U^\dag$ where $U \propto \gamma_0$. Also, $[\xi', U] = [Q_f',U] =0$, while this transformation induces $Q_c \mapsto U^\dag Q_c U$. A general ansatz for a quadratic in $Q_c$ term with $U(2n)$ invariance may be written as $\tr (Q_c \gamma_\gamma)A^{\gamma \gamma'} \tr(Q_c \gamma_{\gamma'}) $, for some $2 \times 2$ matrix $A$. Here we used that $Q_c$ is diagonal in Chern sector to restrict Chern sector generators to the diagonal projectors $\gamma_\gamma$. To restrict $A^{\gamma \gamma'}$, we note that $R$ is also invariant under $t^\mu \mapsto \gamma_z t^\mu$. This transformation induces $Q_c \mapsto Q_c \gamma_z$, which flips the sign of the off diagonal terms in $A$. The diagonal terms of $A$ must be equal by $C_2\T$ symmetry, such that $A$ is proportional to the identity matrix the identity matrix and the term is independent of the choice of $Q_c$. For the terms with one factor of $Q_c$ and one factor of $Q'_f$ we also may use invariance under $C_2 \T$ and $\U(2n)$ to show that if $Q'_f$ preserves these symmetries then there may be no dependence on $Q_c$ except for a simple Hartree-like term $\tr Q_c$ when $\mu \neq 0$.
    
    However, mass terms and doping terms are arbitrarily low energy perturbations to the Dirac semimetal that may be generated in perturbation theory. Depending on the symmetry breaking of the TBG ground state, this may occur for the graphene electrons. Additionally, we may consider \eqref{SE_energy_general} itself as a perturbation that acts as a dispersion term for the TBG electrons. We may do this numerically, but at charge neutrality we find very accurate and analytic results are obtainable if we neglect $\tilde{R}$ such that $R^{\mu \nu}_{\bp \bp'} = 2\delta_{\bp \bp'} \delta^{\mu \nu} (v_{\rm eff} \abs{\bp_+} + U_\bp) $. With this approximation we have 
    \begin{equation}
    \begin{aligned}
    E_{\rm SE}(Q_c, Q_f) = -\frac{1}{4}\sum_{\bp \mu \nu } \frac{1}{v\abs{\bp_+} + U_\bp} \biggl( 
    & \tr \left( Q_c t_\mu \xi'^\dag_\bp \right) \tr \left( Q_c \xi_\bp  t^{\dag}_\nu \right)\\
    & -\tr \left(t_\mu Q'_{f \bp} \xi'^\dag_\bp\right) \tr \left( Q_c  \xi'_\bp t^{\dag}_\nu \right) \\
    & -\tr \left( Q_c t_\mu  \xi'^\dag_\bp \right) \tr \left(   \xi'_\bp Q'_{f \bp}t^{\dag}_\nu \right) \\
    & -\tr \left(t_\mu Q'_{f \bp} \xi'^\dag_\bp\right) \tr \left(   \xi'_\bp Q'_{f \bp}t^{\dag}_\nu \right)
    \biggr).
    \end{aligned}
    \end{equation}
    
    We now use the Fierze identity for $t_\mu$,
    \begin{equation}
        \tr A t^\mu \tr B t^{\mu\dag } = \tr AB,
    \end{equation}
Applied to the superexchange energy, we obtain
\begin{equation}
    E_{\rm SE}(Q_c, Q_f) = -\frac{1}{4} \sum_{ \bp } \frac{1}{v\abs{\bp_+} + U_\bp} ( \text{const.} - 2 \tr \xi'^\dag_\bp Q_c \xi'_\bp Q'_{f\bp}).
    \label{generalSEresult}
\end{equation}
which is the result we obtain in the main text after undoing the $\bk \to -\bk$ transformation in valley $\eta = -1$.

As discussed in the main text, the superexchange term also, indirectly, favors valley-diagonal TBG states since the energy can be lowered by developing a Dirac mass or valley dependent doping in this case. 
We show here that this energy is of order $V^4$ or $B^4$ when a Dirac mass is generated, for example the VH state, and of order $V^6,B^6$ when a doping is generated.
Let us consider the VH case with $Q_c = \eta_z \gamma_z$; a Chern insulator may be understood equivalently.
The graphene sector has an effective dispersion
\begin{equation}
  h_D^{\rm{eff}} = h'_D + m_\bp \eta_z \gamma_z, \qquad m_\bp \approx  \frac{m_K}{1 + \frac{v \abs{\bp_+}}{ U_\bp} },
\end{equation}
which is minimized by $Q'_f = -h_D^{\rm eff}/\sqrt{v^2k^2 + m_\bp^2}$. Here $m_K = m_{\rm SE} \sim V^2, B^2$ \eqref{mse}. Without loss of generality we take $m_K>0$. Note that here we have ignored the $\bp$ dependence of the matrix element $t_{\bp++}$ and interaction energy $U_\bk$ but kept the momentum dependence of the energy gap in the denominator; we will see that the mass is still important for momenta with $v\abs{\bp_+} \approx U_K \gg m_K$ but not for momenta with $v \abs{\bp_+} \approx vk_\theta$ which is the scale on which $t$ changes. 

The energy of the VH state per unit cell, relative to an IVC state that does not see any superexchange, is lowered by
\begin{equation}
    \begin{aligned}
        E_{\rm VH} - E_{\rm IVC} & = \frac{A_M}{A}\sum_\bp \frac{1}{2} \tr h_D^{\rm eff} Q'_f - \frac{1}{2}\tr h'_D Q'^{(0)}_f  = -2n \frac{A_M}{A}\sum_\bp \left(\sqrt{v^2\abs{\bp_+}^2 + m_\bp^2} - v \abs{\bp_+} \right)\\
        & = -2nA_M \int \frac{d^2 p}{(2\pi)^2} \sqrt{v^2\abs{\bp_+}^2 + m_\bp^2} - v\abs{\bp}_+
\end{aligned}
\end{equation}
Note that as we claimed above we may not approximate $m_\bp = m_K$; if we do the integral becomes UV divergent. First we split the integral into the two regimes $x = v\abs{\bp_+}/m_K \ll \Lambda$ and $x \gg \Lambda$ with $1 \ll \Lambda \ll x_0 = U_K/m_K$. In the former regime we can take $v\abs{\bp_+} \ll U_K$ and in the latter regime we may take $v\abs{\bp_+} \gg m_K$. 
\begin{equation}
    \begin{aligned}
        E_{\rm SE,\, VH} - E_{\rm SE,\, IVC} & =-\frac{2n A_M m_K^3}{2\pi v^2} \int_0^\Lambda dx (x\sqrt{x^2+1} - x)  + \int_\Lambda^\infty dx\left(x\sqrt{x^2+\frac{1}{(1+\frac{x}{x_0})^2}} - x^2\right) \\
        & \approx -\frac{n A_M m_K^3}{2\pi v^2}\left( \int_0^{\Lambda^2} du (\sqrt{u+1} - \sqrt{u}) + \int_\Lambda^\infty \frac{dx}{\left(1+\frac{x}{x_0}\right)^2} \right)\\
        & \approx -\frac{n A_M m_K^3}{2\pi v^2} \left( \Lambda + \frac{1}{3}x_0 - \Lambda \right) \\
        & = -\frac{n A_M U_K m_K^2}{6 \pi  v^2} = -\frac{8\sqrt{3}n\pi U_K m_K^2}{54  v^2 k_\theta^2}.
    \end{aligned}
\end{equation}
The result is independent of the arbitrary intermediate scale $\Lambda$ we chose and as we claimed it is dominated by $x \approx x_0$. The final result scales as $V^4$ because the energy gain is proportional to the square of the mass and as $(vk_\theta)^{-2}$ because Dirac dispersion restricts the superexchange to a the vicinity of the $K_M$ point.

For a valley doping we have
\begin{equation}
  h_D^{\rm{eff}} = h'_D + \mu_\bp \eta_z.
\end{equation}
We set $\mu_\bp = \mu_K = m_{\rm SE}$ because we will see that only momenta with $v\abs{\bp_+} < \mu_{\bp}$ contribute to the energy and for these momenta $\mu_{\bp} \approx \mu_K$. We also set $\mu_K > 0$ without loss of generality. The energy is minimized by $Q' = -h'_D \Theta(v\abs{\bp_+} - \mu_K) + \sgn \mu_K \Theta(\mu_K - v \abs{\bp_+})$.  We obtain for the energy per unit cell
\begin{equation}
  \begin{aligned}
    E_{\rm SE,\, VP} - E_{\rm SE,\, IVC} & = \frac{A_M}{A}\sum_\bp \frac{1}{2} \tr h_D^{\rm eff} Q'_f - \frac{1}{2}\tr h'_D Q'^{(0)}_f  = -2n \frac{A_M}{A}\sum_\bp (\mu - v\abs{\bp_+}) \Theta(\mu - v\abs{\bp_+}) \\
    & = \frac{-2n A_M}{2 \pi} \int_0^{\mu/v} pdp (\mu - p) = \frac{nA_M\mu^3}{3\pi v^2}.
\end{aligned}
  \label{VPenergy}
\end{equation}
Notably, this energy is order $V^6$ and has little effect in our Hartree Fock simulations compared to the inter-Chern superexchange from the displacement field induced MATBG dispersion.

\subsection{Inter-TBG superexchange for $n=4$ and beyond}
\label{Suppsubsec:higher_n_superexchange}

We now comment on how these results generalize to $n > 3$, in particular to the case when the superexchange occurs between two TBG subsystems. We assume one of the twisted bilayer graphene subsystems is at its magic angle and is strongly coupled near the $\bK_\pm$ points and the others are sufficiently away from their magic angles that they may be approximated by their low energy Dirac cones. This approximation may break down if the screening from the various Dirac cones is strong enough to make the MATBG weakly coupled or if two magic angles are sufficiently close that the weak coupling approximation breaks for one of the nominally non-magic subsystems. 

With this assumption, it is straightforward to generalize the trilayer graphene computation with only slight modifications. In particular, we no longer take $\bk \to -\bk$ in the negative graphene valley. This is because there is a potential for superexchange-induced IVC order in non-magic TBG subsystems since there are Dirac points in \emph{both} valleys for each $\bK_\pm$ point. Furthermore, this doubling of Dirac cones implies that the Dirac dispersion term $R^{D\mu \nu} \propto \delta^{\mu \nu}$ even without taking $\bk \to -\bk$. When the MATBG has tunneling to a non-magic TBG as well as a graphene subsystem, the two superexchanges may be done separately and combined at the end. Alternatively, we may add a spectator graphene Dirac cone that is not tunneled into in order to perform the graphene superexchange computation without the $\bk \to -\bk$ transformation.

We now show the explicit results of this procedure for $n = 4$ and $n=5$. Consider $n=4$ with the BM Hamiltonian copied here for convenience,
\beq
    H^{\rm Full}_{\rm dec} =  \left(\begin{array}{cccc}
    -i v_F \bsigma_+ \cdot \nabla - \lambda_1 \Gamma_+  & \varphi T(\br) & \frac{2}{\sqrt{5}} \Gamma_+ & 0 \\
\varphi T^\dagger(\br) & -i v_F \bsigma_- \cdot \nabla + \lambda_1 \Gamma_- & 0 & \frac{2}{\sqrt{5}} \Gamma_- \\
\frac{2}{\sqrt{5}} \Gamma_+ & 0 & -i v_F \bsigma_+ \cdot \nabla - \lambda_2 \Gamma_+ & \frac{1}{\varphi} T(\br) \\
0 & \frac{2}{\sqrt{5}} \Gamma_- & \frac{1}{\varphi} T^\dagger(\br) & -i v_F \bsigma_- \cdot \nabla + \lambda_2 \Gamma_-
\end{array} \right).
\label{supp_4layerBM}
\eeq
For simplicity we work at charge neutrality, though there is no obstruction to introducing $\mu \neq 0$ in the same way we did for $n=3$. We also neglect the off diagonal $\tilde{R}$ term and restrict the superexchange to the lowest energy BM bands of the non-magic TBG. 
We consider the first TBG to be magic. There is an external field driven superexchange with the second TBG. The second TBG has dispersion $h_f(\bk)$ and occupations determined by $Q_f(\bk)$. Note that instead of working with $\bp = \bk + \bG$ here we work inside the first Brillouin zone since TBG doesn't have continouous translation symmetry like graphene. 
The matrix $R$ for inter-TBG superexchange is essentially the same as the sandwich graphene $R$: the maindifference is that the dispersion part $R^D$ is diagonal in $\mu, \nu$ indices without taking $\bk \to -\bk$ and is proportional to $\abs{h_f(\bk)}$ as opposed to $v \abs{\bp}$. 

The other difference involved in the inter-TBG superexchange is the matrix element $t_{\bk \eta \gamma} = \bra{u_{c \bk \eta \gamma}} \ket{ u_{f \bk \eta \gamma} } $ is $\P \T$ symmetric; it contains no layer structure unlike the trilayer graphene matrix element which only involves the top layer of the magic sector. This symmetry ensures that the mass generated for the non-magic subsystem is the same for all $Q_c \propto \gamma_z$. In particular the IVC mass will be the same as the VH mass, whereas there is no IVC mass in sandwich graphene. 

In addition to the external fields lattice relaxation mixes the two MATBG sectors. Here we compute the superexchange associated with this effect. By redoing the mapping using the relaxed structure parameters, we find that the different values of $\kappa$ on the external interfaces versus the internal interfaces lead to the term
\begin{equation}
  \begin{pmatrix} 0 & -\tilde{w}_0U_0(\br) \\ \tilde{w}_0U_0(-\br) & 0 \end{pmatrix}
  \label{interTBGrelax}
\end{equation}
in the lower left block of \eqref{supp_4layerBM} (and its hermitian conjugate in the upper right block). We find $\tilde{w}_0 = 3.27$ meV. This term is $\P \T$ asymmetric unlike the displacement field and therefore comes with an extra factor of $\eta$ in the projection $\xi_{\eta \bk}$. This doesn't matter much, however, and this term will essentially act as an effective displacement field.

We now focus on five layers. We copy the BM Hamiltonian for five layers here for convenience:
\beq
    H^{\rm Full}_{\rm dec} =  \left(\begin{array}{ccccc}
    -i v_F \bsigma_+ \cdot \nabla   & \sqrt{3} T(\br) & \frac{2}{\sqrt{3}} \Gamma_+ & 0 & 0\\
\sqrt{3} T^\dagger(\br) & -i v_F \bsigma_- \cdot \nabla & 0 & \Gamma_- & 0 \\
\frac{2}{\sqrt{3}} \Gamma_+ & 0 & -i v_F \bsigma_+ \cdot \nabla &  T(\br) & \frac{4}{\sqrt{6}} \Gamma_+ \\
0 & \Gamma_- &  T^\dagger(\br) & -i v_F \bsigma_- \cdot \nabla & 0 \\
0 & 0 & \frac{4}{\sqrt{6}} \Gamma_+ & 0 & -i v_F \bsigma_+ \cdot \nabla
\end{array} \right).
\eeq
First let us consider the first TBG, with tunneling amplitudes scaled by $\sqrt{3}$, to be magic. Let us discuss the inter-TBG superexchange.
The matrix elements
\begin{equation}
  \begin{aligned}
  t_{\bk \eta \gamma } & =  \bra{u^*_{c \bk \eta \gamma}} \begin{pmatrix} \frac{2}{\sqrt{3}} & 0 \\ 0 & 1 \end{pmatrix}_{\rm Layer} \ket{u_{f \bk \eta \gamma'}} 
  \label{5layermelt}
\end{aligned}
\end{equation}
are now neither $\P \T$ symmetric or $\P \T$ asymmetric either. As a result, the Dirac masses for $Q_c \propto \gamma_z \eta_{0,z}$ at the $\bK_\zeta$ points in valley $\eta$ are different for $\zeta = \pm \eta$. The $\P \T$ symmetric and $\P \T$ asymmetric parts of \eqref{5layermelt} give rise to opposite masses for KIVC states that partially cancel out. Note, however, that the displacement field mixes the non-magic TBG with the graphene subsystem. We therefore expect the KIVC mass to not have a large impact. Similarly the valley diagonal masses will mostly be important near $\bK_{-\eta}$ where there is no graphene Dirac cone. 

When the second TBG subsector is at its magic angle external fields couple it to the non-magic TBG subsystem as well as the graphene subsystem, but the two non-magic subsystems do not couple to each other. In this case the masses will have more of an effect since they will not be dominated by mixing between the non-magic sectors. While lattice relaxation mixes the non-magic sectors, this will be overwhelmed once a large enough displacement field is applied.

\section{Weak coupling theory of pair-breaking}
\label{Suppsec:weakcouplingSC}
\begin{figure}
    \centering
    \includegraphics[width = 0.7 \textwidth]{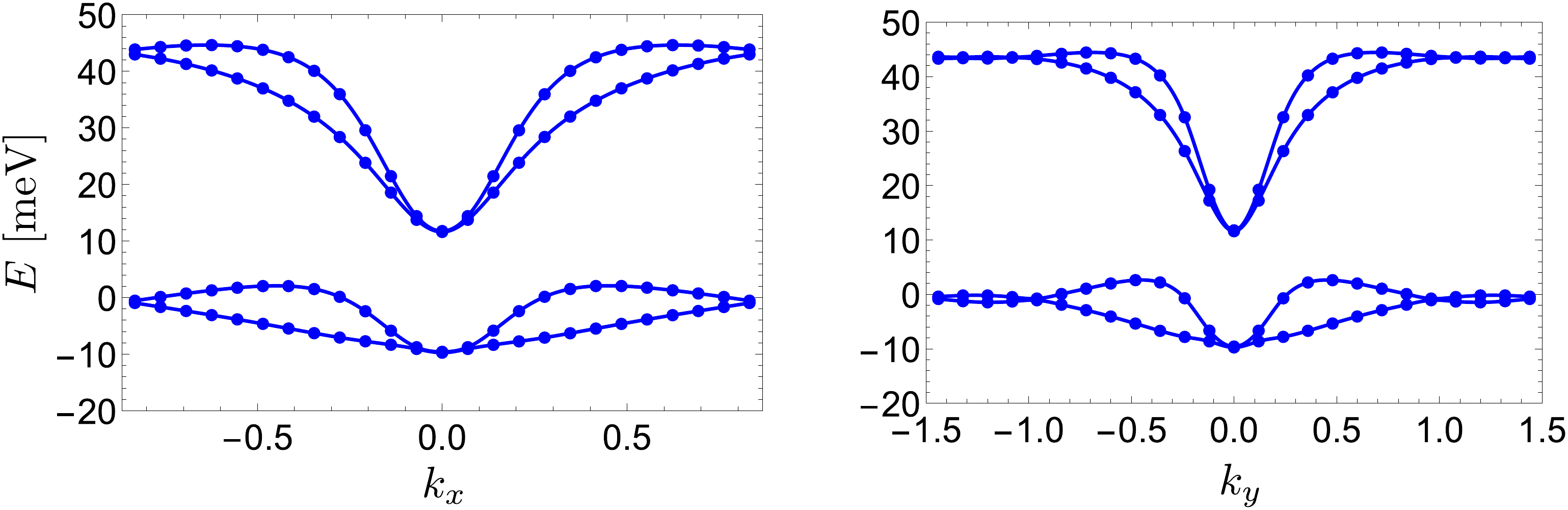}
    \caption{Dispersion of the spin-polarized KIVC bands at $\nu = 2$ given by $Q = P_\uparrow \sigma_0 \tau_0 + P_\downarrow \sigma_y \tau_y$. $k$ is measured in units of the Moir\'e momentum $q_M = \frac{4 \pi}{3\sqrt{3} a_{CC}} \theta$. The non-degenerate upper bands are the KIVC bands for $\downarrow$ whereas the lower bands are three-fold degenerate corresponding to the lower KIVC bands for $\downarrow$ and the bands for $\uparrow$. We note that the degeneracy of the lower bands is an artefact of the enhanced symmetry $\U(4)$ of the model when the single-particle dispersion is neglected and will be generally lifted when such term is added. For the upper bands relevant to electron doping, such terms have no qualitative effect.}
    \label{fig:IVCSp}
\end{figure}

\begin{figure}
    \centering
    \includegraphics[width = 0.6 \textwidth]{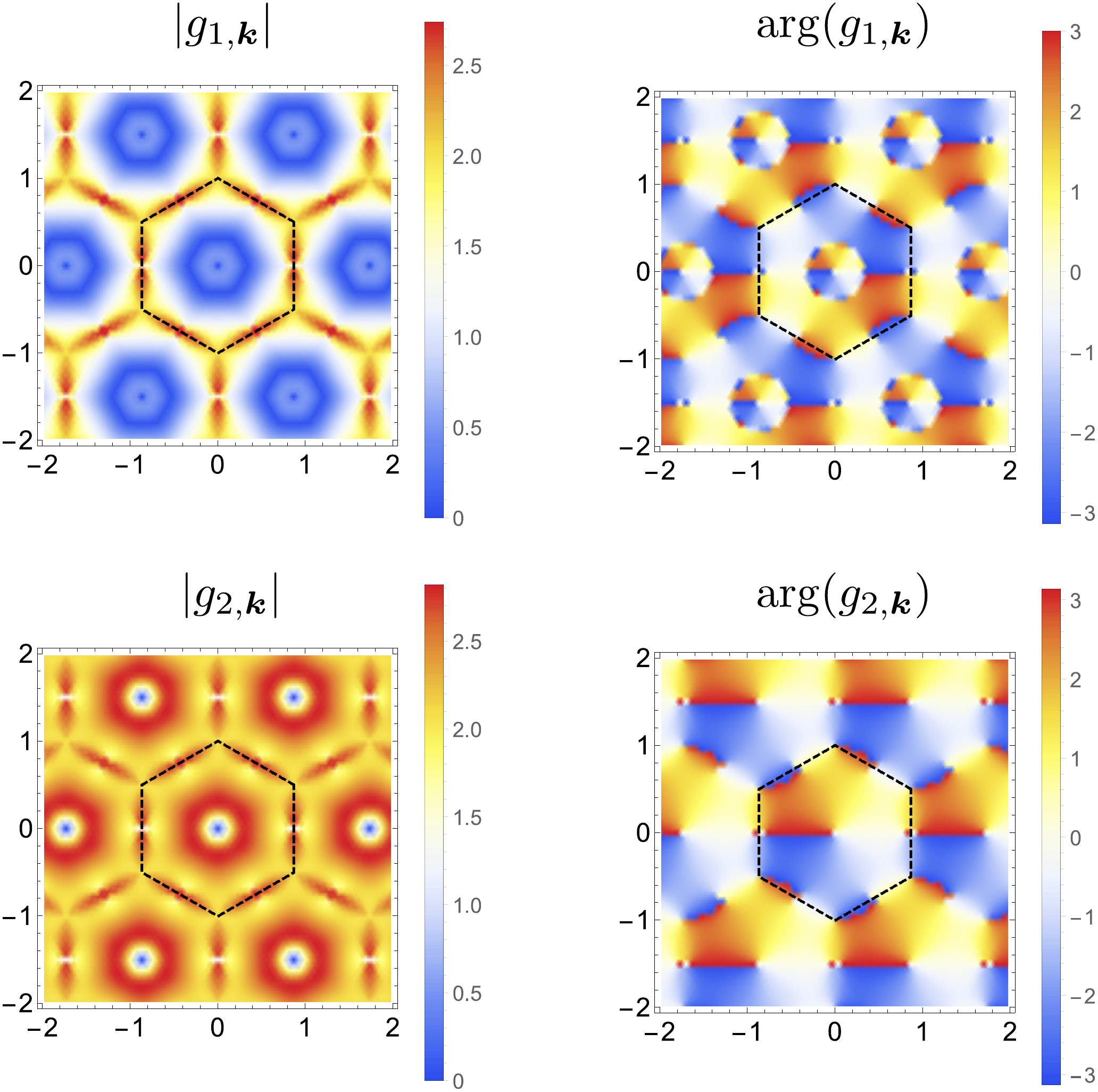}
    \caption{The magnitude and phase of the orbital $g$-factor $g_{n,\bk} = g^x_{n,\bk} + i g^y_{n,\bk}$ (Eq.~\ref{gnk} and \ref{gnkS}) for the KIVC bands $n = 1, 2$.}
    \label{fig:g12}
\end{figure}

In this appendix, we provide details for the weak coupling theory of pair-breaking. We start by considering the band structure for adding electrons to the KIVC state at $\nu = 2$. First, we note that if we neglect the single-particle dispersion term, there is still dispersion generated due to interaction on top of the insulating ground state at any integer filling which is given exactly by the Hartree-Fock Hamiltonian \cite{TBGV, VafekBernevig}. The insulating ground states at integer filling $\nu$ are described by $Q_\bk$ which is $\bk$-independent and satisfies $\tr Q = 2\nu$ and $[Q, \Lambda_\bq(\bk)] = 0$ for all $\bk$ and $\bq$. The Hartree-Fock Hamiltonian for a state specified by the matrix $Q_\bk$ is given by \cite{ShangNematic, KIVCpaper}
\begin{equation}
    H_{\rm HF}[Q](\bk) = \frac{1}{2A} \sum_\bG V_\bG \Lambda_\bG(\bk) \sum_{\bk'}\tr \Lambda_{-\bG}(\bk') Q_\bk - \frac{1}{2A} \sum_\bq V_\bq \Lambda_\bq(\bk) Q_{\bk + \bq} \Lambda_\bq(\bk)^\dagger
    \label{HHF}
\end{equation}
 We are interested in the vicinity of $\nu = 2$ where we can take the ground state to be a spin-polarized KIVC state (other possibilities like spin-valley locked states are related to it by applying $\U(2) \times \U(2)$ rotations) \cite{KIVCpaper}. The corresponding $Q$ can be taken to be $P_\uparrow \sigma_0 \tau_0 + P_\downarrow \sigma_y \tau_y$ \cite{KIVCpaper}, where $P_{\uparrow/\downarrow}$ denotes the projector onto the $\uparrow/\downarrow$ sector. Substituting in the Hartree-Fock Hamiltonian (\ref{HHF}) yields the dispersion shown in Fig.~\ref{fig:IVCSp} \cite{VafekBernevig}. We are going to focus on the case of electron doping where charge carriers enter the upper two bands which can be labelled by a band index $n = 1,2$. The effect of the in-plane field is obtained by projecting the term $\frac{g_{\text{orb}}}{2} \mu_B \mu_z \hat z \times \bB \cdot \bsigma$ onto the KIVC bands:
\beq
H_\bB = \mu_B \bg_{n,\bk} \cdot \bB, \qquad \bg_{n,\bk} = \frac{g_{\text{orb}}}{2} \langle u_{n,\bk}| \mu_z (\sigma_y, -\sigma_x \tau_z) | u_{n, \bk} \rangle
\label{gnkS}
\eeq
The two components $g^{x,y}_{n,\bk}$ for the two bands $n=1,2$ are plotted in Fig.~\ref{fig:g12}.

In the following, we will restrict ourselves to intraband pairing and drop the band index $n$. All the expressions that follow apply for either KIVC band $n = 1,2$. Given a KIVC band, we are going to assume there is some attractive interaction which gives rise to pairing in the KIVC bands and write the imaginary time Lagrangian 
\begin{equation}
    \L = \sum_\bk \psi_\bk^\dagger [\partial_\tau + \xi_\bk + \mu_B \bg_\bk \cdot \bB] \psi_\bk - \frac{1}{2} \sum_{\bk,\bk',\bq} v_{\bk,\bk'}(\bq) \psi^\dagger_{\bk + \bq/2} \psi^\dagger_{-\bk + \bq/2} \psi_{-\bk' + \bq/2} \psi_{\bk' + \bq/2} 
\end{equation}
Here, $v_{\bk,\bk'}(\bq)$ represents some attractive interaction such that $v_{\bk,\bk'}(\bq) \geq 0$. Note that the assumption of purely attractive interaction is not necessary here. It suffices to have some channel where the interaction is attractive, in which case we can understand $v_{\bk,\bk'}(\bq)$ as the projection of the full interaction onto the attractive channels. In the following, we will not make any assumption about the origin of the attraction and will assume its effective for an energy range of $2\Lambda$ around the Fermi energy which will provide the cutoff for the momentum summation (note that we have ignored umklapp processes). $\xi_\bk$ is the single particle KIVC energy spectrum.

To determine the effect of $\bB$ on the superconducting $T_c$, we will proceed in the standard fashion by introducing the Hubbard-Stratonovich field $\Delta_\bk(\bq)$ leading to
\begin{equation}
    \L = \sum_\bk \psi_\bk^\dagger [\partial_\tau + \xi_\bk + \mu_B \bg_\bk \cdot \bB] \psi_\bk + \frac{1}{2} \sum_{\bk,\bq} [\Delta_\bk^\dagger(\bq) \psi_{-\bk + \bq/2} \psi_{\bk + \bq/2} + \Delta_\bk(\bq) \psi^\dagger_{\bk + \bq/2} \psi^\dagger_{-\bk + \bq/2}] + \frac{1}{2} \sum_{\bk,\bk',\bq} \Delta^\dagger_\bk(\bq) [v(\bq)]^{-1}_{\bk,\bk'} \Delta_{\bk'}(\bq) 
\end{equation}
The action at finite temperature is given by the standard expression $S = \int_0^\beta d\tau \L(\tau)$. The order parameter field $\Delta_\bq(\bk)$ is generally assumed to be slowly varying in space (which means it is only non-zero for small $\bq$) and time. Here, we will completely ignoring the $\tau$ dependence and introduce the Fourier transform for $\psi(\tau)$ in terms of Matsubara frequency
\begin{equation}
    \psi(\tau) = \frac{1}{\sqrt{\beta}} \sum_{\omega_n} e^{i \omega_n \tau} \psi_n, \qquad \omega_n = \frac{2n+1}{\beta} \pi
\end{equation}
The action then becomes
\begin{equation}
    S = \frac{1}{2} \sum_{\omega_n,\bk,\bq} (\psi^\dagger_{n,\bk + \bq/2} \quad \psi_{-n,-\bk + \bq/2})
    \left(\begin{array}{cc}
       G^{-1}_{n,\bk + \bq/2} & \Delta^\dagger_\bk(\bq)  \\
        \Delta_\bk(\bq) & -G^{-1}_{-n,-\bk+\bq/2}
    \end{array} \right) \left(\begin{array}{c}
       \psi_{n,\bk + \bq/2} \\ \psi^\dagger_{-n,-\bk + \bq/2}
    \end{array} \right) + \frac{\beta}{2} \sum_{\bk,\bk',\bq} \Delta^\dagger_\bk(\bq) [v(\bq)]^{-1}_{\bk,\bk'} \Delta_{\bk'}(\bq) 
\end{equation}
where we introduced the single particle Green's function
\begin{equation}
    G_{n,\bk} = \frac{1}{i\omega_n + \xi_\bk + \mu_B \bg_\bk \cdot \bB}
\end{equation}
The fermions can be integrated out leading to a Pfafian which can be written in the exponential as the logarithm of the trace of some operator. The resulting free energy per unit cell can be expanded in powers of $\Delta$ as
\beq
F = F_0 + \frac{1}{2\beta N} \sum_{m,n,\bk,\bq} \frac{1}{m} \tr \left[\left(\begin{array}{cc}
       0 & G_{n,\bk + \bq/2} \Delta^\dagger_\bk(\bq)  \\
        -G_{-n,-\bk+\bq/2} \Delta_\bk(\bq) & 0
    \end{array} \right) \right]^m + \frac{1}{2} \sum_{\bk,\bk',\bq} \Delta^\dagger_\bk(\bq) [v(\bq)]^{-1}_{\bk,\bk'} \Delta_{\bk'}(\bq) 
\eeq
Taking the variation of $F$ relative to $\Delta_\bk(\bq)$ gives the BCS self-consistency equation which incorporates  the effect of the orbital field. Since we are interested in only identify $T_c$ for the different pairing channels and how it changes with in-plane field, we can make focus on the vicinity of $T = T_c$ where $\Delta$ is small small and restrict ourselves to the leading term in $\Delta^\dagger \Delta$. In addition, we assume the order parameter $\Delta_\bk(\bq)$ is spatially uniform which means that
\begin{equation}
    \Delta_\bk(\bq) = \delta_{\bq,0} \Delta_\bk
\end{equation}
Expanding to quadratic order in $\Delta$, we get
\begin{equation}
    F = F_0 + \frac{1}{2} \sum_{\bk,\bk'} \Delta_\bk^\dagger \left\{ [v(0)]^{-1}_{\bk,\bk'} - \frac{\delta_{\bk,\bk'}}{N} \Gamma_\bk \right\} \Delta_{\bk'}, \qquad \Gamma_\bk = \frac{1}{\beta} \sum_n G_{n,\bk} G_{-n,-\bk}
    \label{FE}
\end{equation}
$\Gamma_\bk$ can be evaluated using the standard Matsubara summation trick
\begin{align}
    \Gamma_\bk &= \frac{1}{\beta} \sum_n \frac{1}{(i \omega_n + \xi_\bk + g_\bk \cdot \bB)(-i \omega_n + \xi_\bk - \mu_B g_\bk \cdot \bB)} = \frac{\sinh \beta \xi_\bk}{2 \xi_\bk (\cosh \beta \xi_\bk + \cosh (\mu_B g_\bk \cdot \bB))} \nonumber \\
    &\simeq \frac{\tanh \frac{\beta \xi_\bk}{2}}{2 \xi_\bk} - (\mu_B g_\bk \cdot \bB)^2 \frac{\beta^2 \tanh \frac{\beta \xi_\bk}{2} \sech^2 \frac{\beta \xi_\bk}{2}}{8 \xi_\bk}
\end{align}
The expression in (\ref{FE}) can be written more compactly in the matrix form $\frac{1}{2} \hat \Delta^\dagger \left\{\hat v(0)^{-1} - \frac{1}{N} \hat \Gamma \right\} \hat \Delta$ where $\hat \Delta$, $\hat v(0)$, $\hat \Gamma$ are , respectively, an $N$-component vector, an $N \times N$ matrix, and an $N \times N$ diagonal matrix in the momentum index. The different pairing channels correspond to the eigenvectors of $\hat v(0)^{-1} - \frac{1}{N} \hat \Gamma$ 
% \begin{equation}
%     (\hat v(0)^{-1} - \frac{1}{N} \hat \Gamma) \hat \Delta_l = \lambda_l \hat \Delta_l
% \end{equation}
 with the corresponding $T_c$ obtained as the value of $T$ for which the corresponding eigenvalue vanishes. This condition is equivalent to solving the linearized BCS equation which can be written compactly as 
 \begin{equation}
     \det \Xi(T_c) = 0, \qquad \hat \Xi(T) = \hat 1 - \frac{1}{N} \hat v(0) \hat \Gamma(T), \quad \leftrightarrow \quad \Xi_{\bk,\bk'} = \delta_{\bk,\bk'} - \frac{1}{N} [v(0)]_{\bk,\bk'} \Gamma_{\bk'} 
 \end{equation} 
 To simplify further, we make the standard assumption of the weak coupling theory that the relevant temperatures as well as the energy cutoff $\Lambda$ are much smaller than the Fermi energy. In this case, we can reduce the momentum integrals to integrals over the Fermi surface by writing
 \begin{equation}
     \frac{1}{N} \sum_\bk = \int \frac{d^2 \bk}{(2\pi)^2} = \int_{-\Lambda}^{\Lambda} d\xi N(\xi) \int \frac{d\phi_\bk}{2\pi} = N(0) \int_{-\Lambda}^{\Lambda} d\xi \int \frac{d\phi_\bk}{2\pi}
 \end{equation}
 where $\phi_\bk$ is an angular variable labeling a point at the FS defined as $\phi_\bk = \arg(k_x + ik_y)$, $N(\xi)$ is the density of states and $N(0)$ is the density of states at the Fermi energy. In the last equality, we used the assumption that the cutoff $\Lambda$ is much smaller than the Fermi energy such that the variations of the density of states for a window of $2\Lambda$ around the Fermi energy can be neglected.
 
 We now make use of the rotation symmetry to express the matrix $\Xi$ in angular momentum basis. To make $C_6$ rotation symmetry manifest, we will write the angular momentum $l$ which is an integer as $m + 6 M$ where $m = -2,\dots, 3$ and $M$ an integer. $C_6$ rotation symmetry implies the conservation of $l \mod 6 = m$. The angular momentum matrix elements are
  \begin{align}
      \langle m,M|\Xi|m',M' \rangle &=
     \Xi_{m,m';M,M'} = \frac{1}{2 \Lambda N N(0)} \sum_{\bk,\bk'} \Xi_{\bk,\bk'} e^{i (m + 6M) \phi_\bk - i (m' + 6M') \phi_{\bk'}} \nonumber \\
     &= \delta_{m,m'} \delta_{M,M'} - \sum_{M''} V_{m;M,M''}  \Gamma_{m - m',M'' - M'}
  \end{align}
 where $V_{m;M,M'}$ and $\Gamma_{m;M}$ are defined as
 \begin{equation}
     \Gamma_{m,M} = \frac{1}{N} \sum_\bk  \Gamma_\bk e^{i (m + 6M) \phi_\bk} \qquad V_{m;M,M'} = \int \frac{d\phi_\bk}{2\pi} \int \frac{d\phi_{\bk'}}{2\pi} [v(0)]_{\bk,\bk'} e^{i [m (\phi_\bk - \phi_{\bk'}) + 6 (M \phi_\bk - M' \phi_{\bk'})]}
 \end{equation}
 Here, we used the rotation symmetry of $v[0]$: $[v(0)]_{\phi_\bk + \frac{2\pi}{6},\phi_{\bk'} + \frac{2\pi}{6}} = [v(0)]_{\phi_\bk, \phi_{\bk'}}$. Following the discussion in the main text, we can restrict ourselves to even angular momenta $m = 0, \pm 2$ since these are the only possible pairing channels for a single Fermi surface with $\T'^2 = -1$. 
 
 We now further assume that $V_{m;M,M'}$ is only non-zero for $M = M' = 0$:  $V_{m;M,M'} = \Delta_{M,0} \delta_{M',0} V_m$. This amounts to the assumption that the interaction does not change very rapidly as we go around the FS. Thus, in the following, we will drop the $M$ index. We note that time-reversal symmetry $\T'$ implies $V_{-2} = V_2$. The matrix $\Xi$ is now a $3 \time 3$ matrix in the $m = 0, \pm 2$ space given explicitly by
 \begin{equation}
     \Xi = \left(\begin{array}{ccc}
        1 - V_0 \Gamma_0  & V_0 \Gamma_2 & V_0 \Gamma_{-2} \\
        V_2 \Gamma_{2} & 1 - V_2 \Gamma_0 & V_2 \Gamma_4 \\
        V_2 \Gamma_{-2} & V_2 \Gamma_{-4} & 1 - V_2 \Gamma_0
     \end{array}\right)
     \label{XiM}
 \end{equation}
 
 Let us start by considering the case of zero field first. Since the Fermi energy is approximately circular, we can take $\xi_\bk$ to be only a function of $|\bk|$. At zero field, this leads to the simplification
 \begin{equation}
     \Gamma_{m} = \delta_{m,0}  N(0) \int_{-\Lambda}^\Lambda d\xi \frac{\tanh \frac{\beta \xi}{2}}{2 \xi} = \delta_{m,0} N(0) \log \beta \tilde \Lambda, \qquad \tilde \Lambda = \frac{2 e^{\gamma_E}}{\pi} \Lambda 
 \end{equation}
 where $\gamma_E \approx 0.577$ is the Euler constant. 
 
%  We now further assume that $V_{m;M,M'}$ is only non-zero for $M = M' = 0$:  $V_{m;M,M'} = \Delta_{M,0} \delta_{M',0} V_m$. This amounts to the assumption that the interaction does not change very rapidly as we go around the FS. Thus, in the following, we will drop the $M$ index. We note that time-reversal symmetry $\T'$ implies $V_{-2} = V_2$. In this limit, we find
% \begin{equation}
%     \Xi_{m,m'} = \delta_{m,m'}[1 - V_m N(0) \log \beta \tilde \Lambda]
% \end{equation}
Thus, the eigenvalues of $\Xi$ are given by
\begin{equation}
    \lambda_m = 1 - V_m N(0) \log \beta \tilde \Lambda, \quad \implies \quad T_m = \tilde \Lambda e^{-\frac{1}{N(0) V_m}}
\end{equation}
This yields the famous BCS expression $T_c = \tilde \Lambda e^{-\frac{1}{N(0) {\rm max}(V_2, V_0)}}$.

 For non-zero magnetic field, $\Gamma_{m}(\bB)$, we start by introducing the notation $B = |\bB| e^{i \arg(B_x + i B_y)}$ and $g_{\bk} = |\bg_{\bk}| e^{i \arg(g^x_{\bk} + i g^y_{\bk})}$ leading to
 \begin{gather}
     \Gamma_{m}(\bB) = \delta_{m,0} N(0) \log \beta \tilde \Lambda - \frac{1}{4} c \beta^2 N(0) \mu_B^2 (B^2 \bar \chi_{-m} + \bar B^2 \chi_m + 2 B \bar B \gamma_m),  \\ c = 
     \int dx \frac{\tanh \frac{x}{2} \sech^2 \frac{x}{2}}{8 x} = -7 \zeta'(-2) \approx 0.21314, \qquad 
     \gamma_m = \int \frac{d\phi_\bk}{2\pi} |g_{\bk}|^2 e^{i m \phi_\bk}, \qquad 
     \chi_m =  \int \frac{d\phi_\bk}{2\pi} g_{\bk}^2 e^{i m \phi_\bk}
     \label{GammaB}
 \end{gather}
 We now note that $g_\bk$ transforms under $C_6$ rotations as $g_{C_6 \bk} = e^{\frac{2\pi i}{6}} g_\bk$ (cf.~Fig.~\ref{fig:g12}). This implies that $\gamma_m = 0$ unless $m = 0 \mod 6$ and $\chi_m = 0$ unless $m = -2 \mod 6$. Substituting in (\ref{XiM}) yields
  \begin{equation}
     \Xi = \left(\begin{array}{ccc}
        1 - V_0 N(0) [\log \beta \tilde \Lambda - \frac{1}{2} c \beta^2 \mu_B^2 |B|^2 \gamma_0]  & -\frac{1}{4} c \beta^2 V_0 N(0) \mu_B^2 B^2 \bar \chi_{-2}  & -\frac{1}{4} c \beta^2 V_0 N(0) \mu_B^2 \bar B^2  \chi_{-2} \\
        -\frac{1}{4} c \beta^2 V_2 N(0) \mu_B^2 B^2 \bar \chi_{-2} & 1 - V_2 N(0) [\log \beta \tilde \Lambda - \frac{1}{2} c \beta^2 \mu_B^2 |B|^2 \gamma_0] & 0 \\
        -\frac{1}{4} c \beta^2 V_2 N(0) \mu_B^2 \bar B^2  \chi_{-2} & 0 & 1 - V_2 N(0) [\log \beta \tilde \Lambda - \frac{1}{2} c \beta^2 \mu_B^2 |B|^2 \gamma_0]
     \end{array}\right)
 \end{equation}
 The leading correction to the eigenvalues $\lambda_m$ of $\Xi$ can be obtained perturbatively. Notice that the off-diagonal terms will only affect the eigenvalues to order $\bB^4$ unless the diagonal terms are degenerate i.e. $V_0 \approx V_2$ which can only be realized through fine tuning since these channels are not related by any symmetry. Thus, assuming we can neglect the off-diagonal terms, we find that the eigenvalues of $\Xi$ are given by
 \begin{equation}
     \lambda_m = 1 - V_m N(0)[\log \beta \tilde \Lambda - \frac{1}{2} c \beta^2 \mu_B^2 \gamma_0 \bB^2] + O(\bB^4) \quad \implies \quad T_m(\bB) = T_m[1 - \frac{c \mu_B^2 \gamma_0}{T_m^2} \bB^2] + O(\bB^4)
 \end{equation}
 Thus, we get the expression for the $T_c$ as a function of field
 \begin{equation}
     \frac{T_c(\bB)}{T_c} =  1 - \frac{c \mu_B^2 \gamma_0}{2 T_c^2} \bB^2
 \end{equation}
 Rather remarkabely, this expression does not depend explicitly on the pairing channel. This can be solved for the critical field:
 \begin{equation}
     \frac{B_c}{T_c} = \frac{2}{\mu_B \sqrt{c \gamma_0}}
 \end{equation}
 We can use this relation to relate the critical field due to in-plane orbital effect to the Pauli field given by $B_p = [1.86 T/K] T_c$. This leads to the simple relation between the orbital in-plane critical field and the Pauli field
 \begin{equation}
     \frac{B_c}{B_p} = 3.5  \frac{1}{\sqrt{ \gamma_0}} .
 \end{equation}

\end{document}